\documentclass[%
 twocolumn,
superscriptaddress,
 amsmath,amssymb,
 pre,
 longbibliography
]{revtex4-1}

\usepackage{graphicx}
\usepackage{dcolumn}
\usepackage{bm}

\usepackage{xcolor}
\newcommand{\rt}[1]{\textcolor{black}{#1}}                            
\usepackage{soul}
\usepackage[normalem]{ulem}
\newcommand{\rsmath}[1]{\bgroup\markoverwith{\textcolor{red}{\rule[0.5ex]{2pt}{0.4pt}}}\ULon {\textcolor{red}{#1}}}                                           

\usepackage{eso-pic}
\AddToShipoutPictureBG*{%
\AtPageUpperLeft{%
\hspace{\paperwidth}%
\raisebox{-\baselineskip}{%
}}}%

\begin{document}

\title{Non-equilibrium dynamic hyperuniform states}
\author{Yusheng Lei}
 \affiliation{School of Chemistry, Chemical Engineering and Biotechnology, Nanyang Technological University, \\62 Nanyang Drive, 637459, Singapore}
\author{Ran Ni}\email{r.ni@ntu.edu.sg}
 \affiliation{School of Chemistry, Chemical Engineering and Biotechnology, Nanyang Technological University, \\62 Nanyang Drive, 637459, Singapore}

\begin{abstract}
Disordered hyperuniform structures are an exotic state of matter having suppressed density fluctuations at large length-scale similar to perfect crystals and quasicrystals but without any long range orientational order. In the past decade, an increasing number of non-equilibrium systems were found to have dynamic hyperuniform states, which have emerged as a new research direction coupling both non-equilibrium physics and hyperuniformity. Here we review the recent progress in understanding dynamic hyperuniform states found in various non-equilibrium systems, including the critical hyperuniformity in absorbing phase transitions, non-equilibrium hyperuniform fluids and the hyperuniform structures in phase separating systems via spinodal decomposition.
\end{abstract}
\maketitle


\tableofcontents


\section{Introduction}
The concept of hyperuniformity was introduced by Torquato and Stillinger in 2003~\cite{torquato2003local}, while its importance was brought to the fore only in recent years. A hyperuniform structure is defined when the structure factor $S({\bf k}) \rightarrow 0$ with the wavevector $|{\bf k}| \rightarrow 0$, and ordered structures like crystals and quasicrystals are hyperuniform with $S({\bf k}) = 0$ at small $|{\bf k}|$. Beside those ordered hyperuniform structures, one of the major research focuses of hyperuniformity has been on the disordered hyperuniform structures without any broken rotational symmetry, which were observed in various systems including maximally random jammed packings \cite{donev2005unexpected}, avian photoreceptor patterns \cite{kram2010avian, jiao2014avian} and even early universe fluctuations \cite{gabrielli2002glass}, etc. This is partially due to the fact that disordered hyperuniform structures were found to have special properties, e.g., the isotropic robust photonic bandgaps opened at low dielectric contrast \cite{florescu2009designer, man2013isotropic}, and abnormal transparency \cite{leseur2016high}. 

Hyperuniformity was originally described in terms of density fluctuations in a point pattern. 
Consider a many-point-particle system in $d$-dimensional Euclidean space, and let $N_{\Omega}(R)$ be the amount of particles inside the sampling window $\Omega$ within a specific length-scale $R$. The particle number fluctuation at length-scale $R$ is given by $\sigma_{N}^{2}(R)=\left\langle N_{\Omega}^{2}(R)\right\rangle-\left\langle N_{\Omega}(R)\right\rangle^{2}$ . 
If the point pattern is random, e.g., a Poisson point distribution, we can have $\sigma_{N}^{2}(R) \propto R^d$ with $R \rightarrow \infty$, while for ordered point patterns, like a crystal lattice, the number of particle fluctuation $\sigma_{N}^{2}(R) \propto R^{d-1}$ with $R \rightarrow \infty$. Hyperuniform point pattern is defined if the number of particle fluctuation 
\begin{equation}\label{eq1}
\sigma_{N}^{2}(R \rightarrow \infty) \propto R^{2d-\lambda}, \quad d < \lambda \leq d+1.
\end{equation}
When $\lambda=d+1$, the system is denoted as the \textit{maximally hyperuniform} state, which possess the same power-law scaling of the particle number fluctuation in ordered structures like perfect crystals. Beside characterising hyperuniformity using the particle number fluctuation $\sigma_{N}^{2}(R)$, one can also check the structure factor of the system defined as
\begin{equation}
S(\mathbf{k})=1+\rho \tilde{h}(\mathbf{k}),
\end{equation}
where $\tilde{h}(\mathbf{k})$ is the Fourier transform of the total correlation function ${h}(\mathbf{r})=g(\mathbf{r})-1$, and $g(\mathbf{r})$ is the radial distribution function with $\rho$ the density of the system. One can see that, for a random particle point pattern of the Poisson distribution, $S(\mathbf{k})=1$ for all $\mathbf{k}$. In the thermodynamic limit, from the vanishing fluctuation in long-wavelength in hyperuniform systems, we have \cite{hansen2013theory}
\begin{equation}
\lim _{\mathbf{k} \rightarrow \mathbf{0}} S(\mathbf{k})=\lim _{v_1(R) \rightarrow \infty} \frac{\sigma_{N}^{2}(R)}{\langle N_{\Omega}(R)\rangle} = 0,
\end{equation}
where $\langle N_{\Omega}(R)\rangle=\rho v_1(R)$ with $v_1(R)$ the volume of the sampling window $\Omega$ at the length-scale $R$. Besides, the structure factor $S(\mathbf{k})$ follows a power-law scaling
\begin{equation}\label{eq4}
S({\bf k}) \sim k^\alpha, \quad \mathbf{k} \rightarrow \mathbf{0}.
\end{equation}
Based on the asymptotic analysis on the expansion of particle number fluctuation at large length-scales, the exponent $\alpha$ here can be related to the number fluctuation in Eq.~\ref{eq1} by classifying the degree of hyperuniformity into three classes~\cite{torquato2018hyperuniform}
\begin{equation}
    \sigma_N^2(R \rightarrow \infty) \sim\left\{\begin{array}{l}R^{d-1}, \qquad\quad \alpha>1 \quad \text { (CLASS I) } \\ R^{d-1} \ln R, \quad \alpha=1 \quad \text { (CLASS II) } \\ R^{d-\alpha}, \quad 0<\alpha<1 \quad \text { (CLASS III) }\end{array}\right.
\end{equation}
If the system is anisotropic, one can also have directional hyperuniformity in different directions~\cite{torquato2016hyperuniformity}.
In addition to particle systems, the concept of hyperuniformity can be also generalized in two phase medium~\cite{zachary2009hyperuniformity}. 
For a scalar concentration field $\rho({\bf r})$, by considering the autocovariance function $\psi({\bf r})$, one can perform the Fourier transform on $\psi({\bf r})$ to get the spectral density function $\tilde{\psi}({\bf k})$. In analogy to general many-particle systems, the concept of hyperuniformity for a scalar field is defined as the vanishing spectral density function at small wave-number $\tilde{\psi}( {\bf |k|} \rightarrow 0) \rightarrow 0$~\cite{torquato2016hyperuniformity}.

In equilibrium disordered systems, to ensure a hyperuniform structure factor scaling $S(k \rightarrow 0)\sim k^\alpha$ with $\alpha>0$, as suggested in Ref.~\cite{lomba2018disordered,torquato2018hyperuniform}, the direct correlation function in Fourier space $\tilde{c}(k \rightarrow 0) \sim -\beta \tilde{v}(k\rightarrow 0)$, and the effective pair potential in Fourier space $\tilde{v}(k)$ should have a power-law scaling $\tilde{v}(k \rightarrow 0) \sim  -\beta^{-1} k^{-\alpha}$, which is the repulsive interaction in ‘‘like-charged’’ particles, e.g., hyperuniform binary plasmas~\cite{lomba2017disordered,lomba2018disordered}. 
Here $\beta = 1/k_BT$ with $k_B$ and $T$ the Boltzmann constant and temperature of the system, respectively.
Differently, in non-equilibrium systems, the ways to achieve hyperuniformity are not restricted by equilibrium constraints, and long-range interactions are not necessary. 
In recent decades, a growing number of non-equilibrium systems of short-range interacting particles have been found to form disordered hyperuniform states. Examples include jammed particles~\cite{zachary2011hyperuniform,ricouvier2017optimizing,dreyfus2015diagnosing,berthier2011suppressed}, perfect glass~\cite{zhang2016perfect}, sheared particles~\cite{weijs2015emergent,bartolo2017,tjhung2015hyperuniform,wilken2020hyperuniform,mitra2021hyperuniformity}, self-propelled particles~\cite{oppenheimer2022hyperuniformity,huang2021circular,lei2019nonequilibrium}, interacting-diffusing particles~\cite{grassberger2016oslo,bertrand2019nonlinear}, \rt{sedimenting non-Brownian irregularly-shaped paricles~\cite{goldfriend2017screening}}, non-reciprocally interacting robots~\cite{hepeng2024}, \rt{driven-diffusive particles with exclusion process~\cite{jack2015hyperuniformity,karevski2017conformal,garrahan2024topological} and quantum systems~\cite{torquato2008point,carollo2017fluctuating,sakai2022quantum}}
etc. This research direction has attracted significant scientific attention lately.
\rt{Here we review the recent progress of various approaches to obtain non-equilibrium dynamic hyperuniform states in classical many-particle systems theoretically, numerically and experimentally.} The review is organized as follows: in Sec. II, we briefly introduce the random organization dynamics and relavent modified models, and review the hyperuniform states in the absorbing phase transition; in Sec. III, we review the hyperuniform fluids in various active particle systems, e.g., non-equilibrium hyperuniform fluids, chiral active fluids and circularly swimming algae systems, etc.; in Sec. IV, we show another type of approaches to achieve dynamical hyperuniform structures, i.e., the field patterns generated from phase separating systems via spinodal decomposition, like the Model-B, Model-H and active field theories;  lastly, some concluding remarks on these dynamic hyperuniform states are given in Sec. V.

\begin{figure*}[htbp]
    \centering 
    \includegraphics[width=0.9\textwidth] {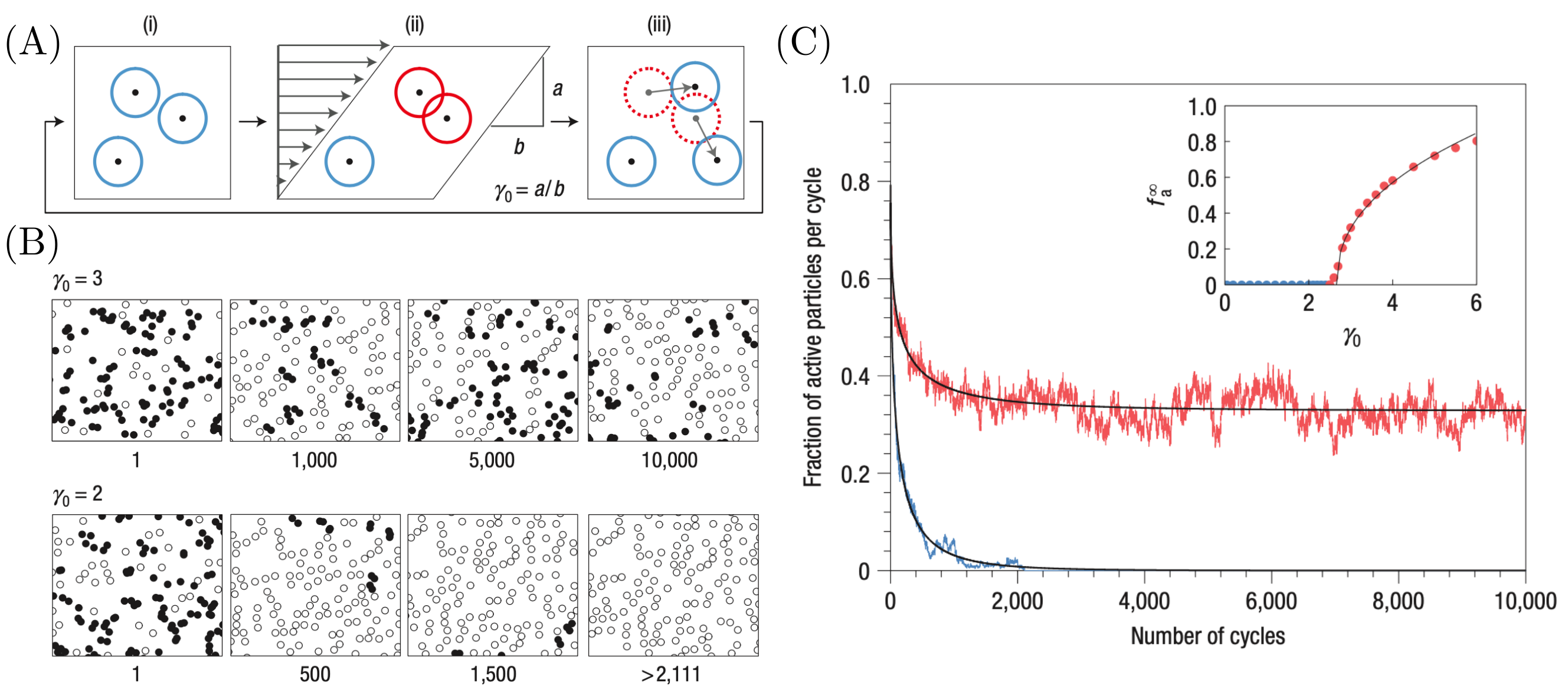}
    \caption{\textbf{(A)} The schematic diagram illustrates the dynamics of random organization. \textbf{(B)} Snapshots depict the evolution of particle distributions for two typical strain amplitudes: $\gamma_0 = 3$ in the active state and $\gamma_0 = 2$ in the absorbing state. The simulation step number is shown below each snapshot, with filled black circles indicating active particles. \textbf{(C)} The time evolution of the active particle fraction $f_a$ is presented for two strain amplitudes: $\gamma_0 = 3$ (red) and $\gamma_0 = 2$ (blue), corresponding to the system evolving into the active state and absorbing state, respectively. The inset plots the dependence of the steady-state active particle fraction, serving as the order parameter of the absorbing phase transition, on the strain amplitude $\gamma_0$.\cite{corte2008random} Reproduced with permission.\cite{corte2008random} Copyright 2008, Springer Nature Limited.}
    \label{Fig.1}
\end{figure*}

\begin{figure*}[htb]
    \centering
    \includegraphics[width=0.95\textwidth] {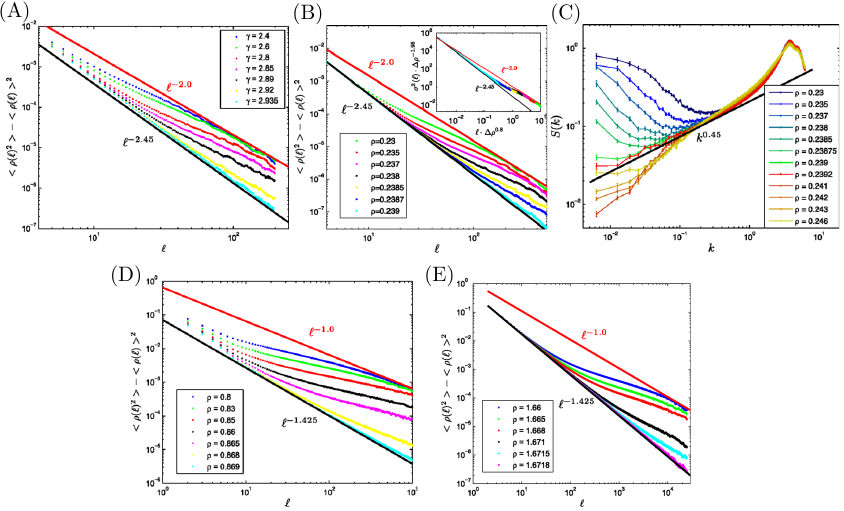}
    \caption{\textbf{(A)} Mean squared density fluctuation $\sigma^2(l)\equiv \langle \rho(l)^2 \rangle - \langle \rho(l) \rangle^2$ for the 2D random organization model as a function of length $l$ at various values of strain amplitude $\gamma<\gamma_c \approx$ 2.935. The black solid line indicates the hyperuniform fluctuation $\sigma^2(l) \sim l^{-2.45}$ and the red solid line indicates the number fluctuation in a 2D random distribution. \textbf{(B)} Mean squared density fluctuations for the 2D conserved lattice gas (CLG) model for different densities $\rho<\rho_c\approx 0.2391$, and the inset shows the data collapse. \textbf{(C)} Averaged structure factor $S(k)$ at small $k$ for the 2D conserved lattice gas (CLG) model for different densities, and $\rho_c\approx 0.2391$. \textbf{(D)} Density fluctuations in the 1D random organization model with various density $\rho<\rho_c \approx 0.8692$. \textbf{(E)} Density fluctuations in the 1D Manna model with various density $\rho<\rho_c \approx 1.6718$.\cite{hexner2015hyperuniformity} Reproduced with permission.\cite{hexner2015hyperuniformity} Copyright 2015, American Physical Society.}
    \label{Fig.2}
\end{figure*}

\section{Hyperuniformity from random organization}
\subsection{Hyperuniformity in critical absorbing state of random organization model}
In 2005, Pine \textit{et al.} experimentally investigated suspensions of periodically strained viscous spherical polymethylmethacrylate (PMMA) particles subjected to circular Couette flow. It was found that particle motion in slowly sheared non-Brownian suspensions at low Reynolds numbers is highly irreversible and chaotic~\cite{pine2005chaos}. By measuring particle trajectories, it was discovered that when the strain amplitude $\gamma$ exceeds a certain concentration-dependent threshold $\gamma_c$, the particles do not return to their initial positions after a shearing cycle, indicating irreversible and chaotic dynamics resulting from hydrodynamic interactions.
\begin{figure*}[htb]
    \centering
    \includegraphics[width=0.9\textwidth] {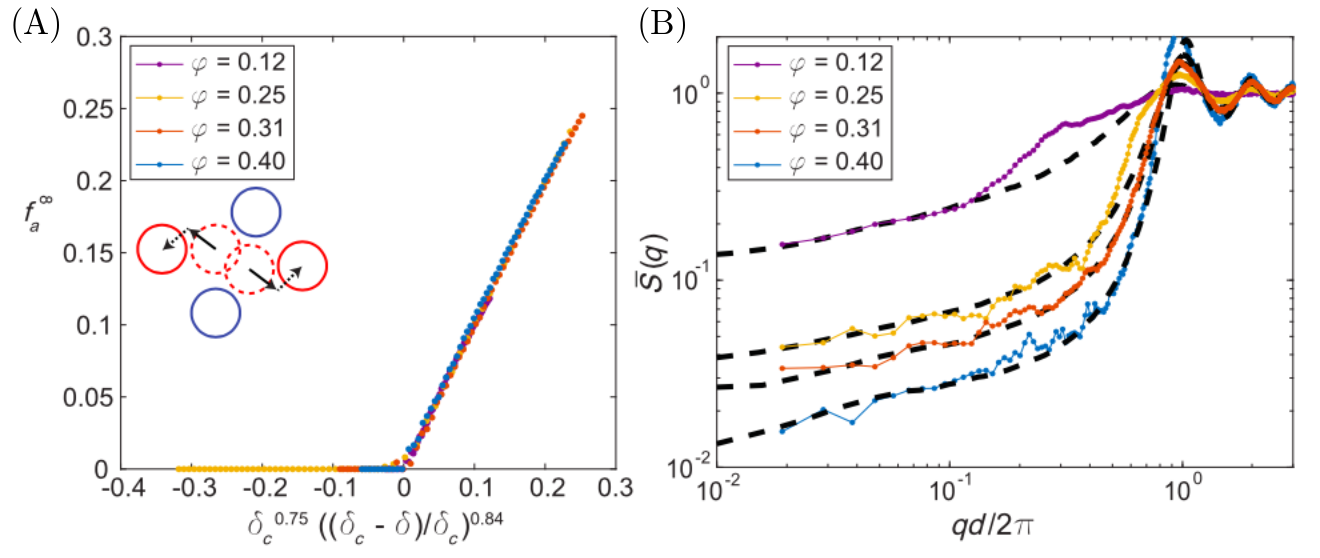}
    \caption{\textbf{(A)} The active particle fraction in steady state is plotted against the reduced control parameter for the biased random organization model. The inset illustrates a schematic diagram of the biased random organization model. \textbf{(B)} The structure factor at the critical strain amplitude $\gamma_c$ is shown for several volume fractions $\phi$, with experimental data represented by dotted curves, and simulation results from the biased random organization model at criticality for the same volume fractions indicated by dashed curves.\cite{wilken2020hyperuniform} Reproduced with permission.\cite{wilken2020hyperuniform} Copyright 2020, American Physical Society.} 
    \label{Fig.3}
\end{figure*}

To investigate the dynamics of the reversible-irreversible transition, Cort\'{e} \textit{et al.} proposed a model system known as the random organization (RO)~\cite{corte2008random}. In this model setup, a rectangular box containing a certain number of particles is periodically sheared to a specified strain amplitude and then returned to its initial rectangular form, as depicted in the schematic diagram shown in Fig.~\ref{Fig.1}(A), to simulate the dynamics of periodically sheared suspensions. The dynamics of the model involves three main steps: (i) identifying the initial positions of the particles before the shearing operation; (ii) shearing the system by a strain amplitude $\gamma_0$ and detecting particle pairs that collide, which are defined as active particles with a center-to-center distance smaller than their diameter; (iii) resetting the system to its initial positions and randomly displacing all active particles (from dashed red circles to blue circles) by a random distance in a random direction. This process mimics the chaotic collisions occurring in sheared suspensions, which was shown to be not ergodic~\cite{frenkelnonergodic}.
As shown in Fig.~\ref{Fig.1}(B), for small strain amplitudes $\gamma_0<\gamma_c$, the number of active particles vanishes after a certain number of shearing cycles, leading to reversible particle trajectories. On the contrary, for large strain amplitudes $\gamma_0>\gamma_c$, active particles persist in the system, maintaining chaotic irreversibility in steady states. In the context of absorbing phase transition, the steady-state active particle fraction $f_a^\infty$ serves as the order parameter, and systems with $f_a^\infty=0$ are denoted as in the absorbing state, while those with $f_a^\infty>0$ are denoted as in the active state. Throughout the absorbing phase transition, as $\gamma_0$ increases, the parameter $f_a^\infty$ undergoes a smooth continuous transition from the absorbing state to the active state, as illustrated in the inset of Fig.~\ref{Fig.1}(C).

Subsequently, it was found that structures at the critical absorbing transition in the random organization model exhibit abnormal fluctuations suppressed at large scales, indicating the presence of hyperuniformity~\cite{tjhung2015hyperuniform,hexner2015hyperuniformity}. As shown in Fig.~\ref{Fig.2}(A), for the random organization model considered in Ref.~\cite{hexner2015hyperuniformity}, as the strain amplitude $\gamma$ increases from the absorbing state towards the critical transition point $\gamma_c$, the hyperuniform scaling of density fluctuation $\sigma^2(l \rightarrow \infty) = \langle \rho(l)^2 \rangle - \langle \rho(l) \rangle ^2 \sim l^{-2.45}$ gradually emerges at large length-scale $l$.  \rt{Besides,  in Ref.~\cite{menon2009universality}, it was confirmed that the absorbing state transition for random organization model belongs to the conserved directed percolation (CDP) universality class, which is a class of models sharing similar critical behaviors with universal critical exponents~\cite{rossi2000universality,henkel2008non}. In order to confirm whether other models in the same universality class exhibit the same hyperuniform structural characteristics at the critical point, Ref.~\cite{hexner2015hyperuniformity} also investigated two other models within the CDP universality class, namely, the Manna model and the conserved lattice gas (CLG) model.} As shown in Fig.~\ref{Fig.2}(B) and (C), for the CLG model in 2D, they calculated both the particle density fluctuation $\sigma^2(l)$ and the structure factor $S(k)$, and found the same hyperuniform fluctuation scaling $\sigma^2(l \rightarrow \infty )\sim l^{-2.45}$ as observed in the 2D RO model. Additionally, they also examined the structure factor near the critical point of the 2D conserved lattice gas model in a large system and found the scaling $S(k \rightarrow 0) \sim k^{0.45}$ at the critical absorbing transition point. Furthermore, as shown in Fig.~\ref{Fig.2}(D) and (E), the authors analyzed 1D systems and found that the density fluctuation of the 1D RO and Manna models exhibited the same hyperuniform scaling $\sigma^2(l \rightarrow \infty)\sim l^{-1.425}$ at the critical point. This suggests that the hyperuniform scaling observed at the critical point of the CDP universality class depends on the dimensionality of the system, and the critical hyperuniform scaling at the absorbing transition was found to be $S(k \rightarrow 0) \sim k^{0.25}$ in 3D CLG model~\cite{hexner2015hyperuniformity}.
Later, it was confirmed that the critical hyperuniformity observed at the absorbing transition is robust against the change of particle shape in the RO model~\cite{ma2019hyperuniformity}. 
Below the critical point of absorbing transition $\rho_c$, the system eventually evolves into an absorbing state, which is hyperuniform up to a length-scale that diverges at $\rho_c$. 
\rt{Besides, Hexner \textit{et al.} found that by adding extra weak noises to structures in the absorbing state, for example, reactivating the system by giving small thermal-noise-like displacements to all particles and waiting for the noise-induced activity to vanish, one can enhance the hyperuniformity of the absorbing state~\cite{hexner2017enhanced}. Moreover, Mari \textit{et al.} modified the original RO dynamics with mediated interactions where passive particles undergo total-activity related diffusions to mimic the hydrodynamic effect in realistic systems, and observed that the hyperuniformity at critical absorbing state is destroyed at large length-scales~\cite{mari2022absorbing}.} 


To verify the hyperuniformity found at the critical absorbing transition point in 3D, Wilken \textit{et al.} experimentally investigated the structural change during the absorbing phase transition of periodically sheared colloidal suspensions consisting of monodisperse copolymer particles~\cite{wilken2020hyperuniform}, similar to the setup in Ref.~\cite{pine2005chaos}. They determined the critical strain amplitude $\gamma_c$ at several volume fractions $\phi$ by measuring the particle mean squared displacement per strain cycle and subsequently calculated the structure factor $S(q)$, as shown in Fig.~\ref{Fig.3}(B). At the critical point, they observed a hyperuniform scaling of the structure factor $S(q) \sim q^{0.25}$ with $q\rightarrow 0$. Additionally, they also compared the experimental results with simulations from a modified version of the original RO model. The authors adjusted the interactions in the RO model, as illustrated in the inset of Fig.~\ref{Fig.3}(A): at each strain cycle, active particle pairs were not only subjected to a random displacement with magnitude $\epsilon_r$ chosen from a Gaussian distribution of width $\epsilon_{r0}$, but also experienced an equal and opposite repulsive bias displacement with an amplitude ranging in $[0,\epsilon_d]$ along the direction connecting their centers. As shown in Fig.~\ref{Fig.3}(B), the simulation results closely matched the experimental findings with predetermined prefactors. 

Moreover, in a subsequent study~\cite{wilken2021random}, it was found that this modified RO model can effectively establish a dynamical approach towards the random close packing (RCP). Comparing to the original random organization model proposed in Ref.~\cite{corte2008random}, they investigated the zero-strain limit ($\gamma=0$) and found that the biased random organization dynamics resulted in a higher critical packing fraction. Remarkably, they found that this critical packing fraction is very close to the volume fraction of random close packing $\phi_{\rm RCP}$~\cite{ni_rcp_2013}. Furthermore, they demonstrated that the configurations obtained from the biased random organization critical point appeared to be structurally identical to RCP configurations obtained by other protocols. These similarities included isostatic coordination ($Z=6$), a cusp in the pair correlation function $g(r)$, and the structure factor scaling of $S(k \rightarrow 0) \sim k^{0.25}$ in 3D. Afterwards, it was found that the RCP structures obtained by this dynamic approach are hyperuniform only in $d<4$ dimensions~\cite{rcpjamming}.

To further understand the origin of critical hyperuniform structures in the CDP universality class, renormalization group (RG) analysis is necessary. Comparing to the simple directed percolation (DP) model, the treatment on the coupling becomes more complex due to the presence of additional conservation laws in CDP models~\cite{van2002universality,wijland2003infinitely,le2015exact}. 
In Ref.~\cite{ma2023theory}, Ma \textit{et al.} performed the renormalization group analysis on a CDP toy model by applying one loop expansion on the Doi-Peliti field theory, and they found that the hyperuniform scaling $S(k \rightarrow 0)\sim k^{2\epsilon/9+O(\epsilon^2)}$ with $\epsilon=4-d$ in dimension $d < 4$, and $S(k \rightarrow 0) \sim k^0$ when $d \ge 4$. 
However, Wiese proposed another approach by mapping the CDP model to the depinning model and argued that the higher order loop expansion, i.e., 2-loop or 3-loop, in functional RG is crucial to describe the nontrivial behavior in CDP models~\cite{joerg2024hyperuniformity}.
This approach allows to relate the hyperuniformity exponent $\alpha$ to the roughness exponent $\zeta$ at depinning, 
\begin{equation}
\alpha = (4-d)- 2 \zeta, \quad \epsilon = 4-d.
\end{equation}
This relation can be used in two different ways to obtain predictions for $\alpha$ in the CDP class: 
use field theory, with $\zeta= \epsilon/3 + 0.04777 \epsilon^2 - 0.06835 \epsilon^3$, and Padé-Borel resummation supplemented 
by the knowledge of $\zeta(d=0)=2$~\cite{Wiese_2022}, and $\zeta(d=1)=5/4$~\cite{grassberger2016oslo,Shapira_2023}.
This leads to 
$\alpha = 1/2 $ in $d=1$, $\alpha=0.4964$ in $d=2$ and $\alpha = 0.2868$ in $d=3$, or use the best available simulation results to get 
$\alpha = 1/2 $ in $d=1$~\cite{grassberger2016oslo,Shapira_2023} 
$\alpha=0.494(4)$ in $d=2$ and $\alpha = 0.29(2)$ in $d=3$~\cite{rosso2003}.

\begin{figure*}[htbp]
    \centering
    \includegraphics[width=0.9\textwidth] {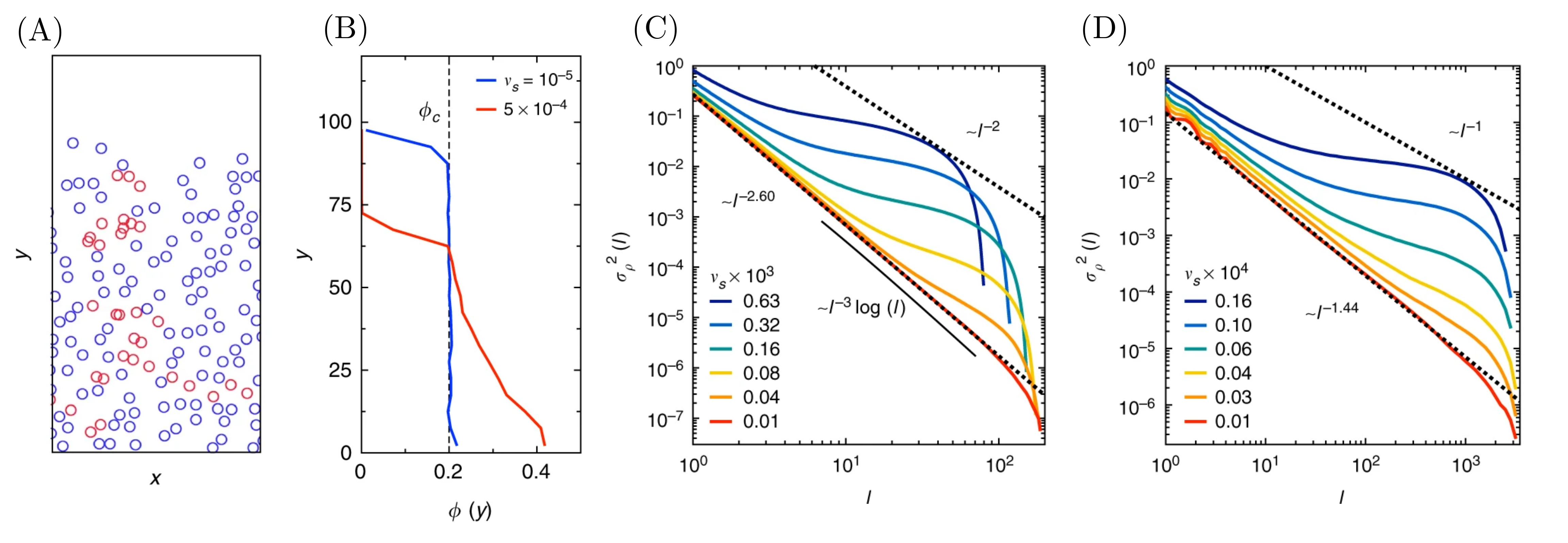}
    \caption{\textbf{(A)} Snapshot of a system in the steady state, with red particles indicating active particles. \textbf{(B)} Particle concentration versus vertical coordinate $y$ at low and high sedimentation velocities $v_s$. \textbf{(C)} Mean squared density fluctuation $\sigma^2(l)\equiv \langle \rho^2(l) \rangle - \langle \rho(l) \rangle ^2$ in 2D simulations, where the $l^{-2}$ scaling is from a 2D random distribution. \textbf{(D)} Mean squared density fluctuation $\sigma^2(l)$ in 1D simulations, where the $l^{-1}$ scaling is from a 1D random distribution.\cite{wang2018hyperuniformity} Reproduced with permission.\cite{wang2018hyperuniformity} Copyright 2018, Springer Nature Limited.}
    \label{Fig.4}
\end{figure*}

\subsection{Hyperuniformity in self-organized criticality based on random organization model}
In the previous section, we have mentioned recent studies on the hyperuniformity at the critical point of absorbing phase transitions. In Ref.~\cite{wang2018hyperuniformity}, Wang \textit{et al.} numerically investigated an intriguing non-fine-tuned hyperuniform system based on self-organized critical (SOC) states in sheared non-Brownian suspension experiments proposed by Cort\'{e} {\it et al.} in Ref.~\cite{corte2009self}. Specifically, based on the random organization dynamics in Fig.~\ref{Fig.1}(A), the SOC random organization model introduced an additional sedimentation velocity $v_s$ acting on all particles towards a specific direction to mimic the gravity-induced sedimentation in cyclically sheared non-Brownian suspensions. The authors defined the horizontal $x$-axis as the shear flow direction and the vertical $y$-axis as the sedimentation direction, as shown in Fig.~\ref{Fig.4}(A). In computer simulations, $N$ particles of diameter $d$ were placed in a square box of width $L$, with an area fraction $\phi = N\pi (d/2)^2/L^2$. 
Periodic boundary condition is applied in $x$-axis, and a hard impenetrable wall is placed at the bottom of the box ($y=0$).
Additionally, a linear density of particles was defined as $\kappa \equiv N/L$ along the $x$-axis. In this system, two competing dynamics are at play: sedimentation dynamics and shear-induced random organization dynamics. \rt{Sedimentation dynamics tend to concentrate particles towards the bottom of the box in $y$ direction, while the effective repulsive pair-interactions via random organization dynamics tend to push particles away from each other against sedimenting to the bottom.} As demonstrated in Ref.~\cite{corte2009self,wang2018hyperuniformity}, due to these competing dynamics, the system automatically self-organizes into a uniform or non-uniform steady state, as shown in Fig.~\ref{Fig.4}(B), depending on the sedimentation speed $v_s$. At low sedimentation speed $v_s$, particles are more likely to self-organize into a uniform distribution in the $y$ direction with a steady-state height $h^{\infty}$, while this is not observed at high sedimentation speed where the sedimentation effect dominates so that more particles tend to stay near the bottom. Remarkably, in the low $v_s$ case, it was found that the constant mean area fraction, defined as $\bar{\phi}^{\infty} \equiv \pi d^2 \kappa/4 \bar{h}^{\infty}$, converges to the critical volume fraction \rt{$\phi_c$} of the random organization model at the same strain amplitude $\gamma_0$.

In the absence of thermal diffusion, the area fraction at the bottom wall would increase due to sedimentation from $\phi_c$ to $2\phi_c$ in a time $\tau_s=\xi/v_s$, where $\xi=\sqrt{\pi d^2/4\phi_c}$ is the mean interparticle spacing. Meanwhile, shearing would also induce diffusion and redistribute the increased $\phi$ due to sedimentation to a uniform steady state with a suspension height $h^{\infty}$, characterized by a typical time scale \rt{$\tau_D = {h^{\infty}}^2/4D$}, where $D$ is the diffusivity under the strain amplitude $\gamma_0$ and the area fraction $2\phi_c$. Thus, it was found that the competing dynamics can be well captured by a dimensionless factor $A=\tau_D/\tau_s=\left(\pi/\phi_c\right)^{3 / 2} d^3 \kappa^2 v_s/(4 D)$~\cite{corte2009self}. \rt{In the limit $A \ll 1$, the authors found that the system evolves into a uniform steady state with a self-organized volume fraction $\phi_\infty=\phi_c$, and $\phi_\infty/\phi_c$ can be collapsed as a function of $A$ for various linear densities $\kappa$ and strain amplitudes $\gamma_0$ at small $A$. Later, in Ref.~\cite{wang2018hyperuniformity}, Wang \textit{et al.} found the dimensionless factor $A$ can not be used to describe high sedimentation speed case ($A\gg1$), as it results in poor data collapse for different linear densities $\kappa$ with increasing $A$. For better data collapsing, the authors proposed that the timescales for diffusion and sedimentation should instead be considered over the same length-scale, and defined a new dimensionless factor $\bar{A}$. By taking $\tau'_D$ and $\tau'_s$ as the new timescales for particle transport over the critical height of the bed of particles, $h_c$, Wang \textit{et al.} defined a new dimensionless factor}
\begin{equation}
    \bar{A}=\frac{\tau'_D}{\tau'_s}=\frac{\pi d^2 \kappa v_s}{16 \phi_c D},
\end{equation}
which served as a dimensionless sedimentation speed and captured the uniform steady state at $\bar{A}\ll 1$. Furthermore, they measured the number density fluctuation in the uniform steady states ($\bar{A}\ll 1$) from the 2D and 1D simulations, as shown in Fig.~\ref{Fig.4}(C) and (D). As the sedimentation velocity $v_s$ decreases, the number fluctuation exhibits a hyperuniform scaling: $\sigma_\rho^2(l\rightarrow \infty) \sim l^{-3}\log(l)$ in 2D and $\sigma_\rho^2(l\rightarrow \infty) \sim l^{-1.44}$ in 1D. These values are close to the hyperuniform scaling observed at the CDP critical point: $\sigma^2(l \rightarrow \infty) \sim l^{-2.45}$ in 2D and $\sim l^{-1.42}$ in 1D. Besides, they also found that the hyperuniform scaling only appeared below a length-scale in 2D 
\begin{equation}
l_{H} \approx 0.22(\bar{A}\sqrt{\phi_c})^{-1.54}.
\end{equation}
Thus, for low enough sedimentation velocity $v_s \rightarrow 0$ with the dimensionless factor $\bar{A} \ll 1$, the length-scale $l_H$ can be very large.

\subsection{Hyperuniformity in active states of random organization model}

\begin{figure*}[htb]
    \centering
    \includegraphics[width=0.9\textwidth] {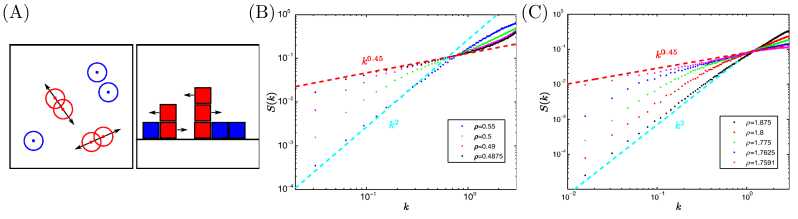}
    \caption{\textbf{(A)} Schematic diagram of random organization model and Manna model with center-of-mass conservation (CMC). \textbf{(B)} Structure factor $S(k)$ of  the random organization model with CMC in active states ($\rho > \rho_c$).  \textbf{(C)} Structure factor $S(k)$ of the active states ($\rho > \rho_c$) of the Manna model with CMC.\cite{hexner2017noise} Reproduced with permission.\cite{hexner2017noise} Copyright 2017, American Physical Society.
}
    \label{Fig.5}
\end{figure*}

In Ref.~\cite{hexner2017noise}, Hexner and Levine studied variants of models based on the random organization and the Manna model. Comparing to the original model, they introduced the center-of-mass conservation (CMC) into the system and found that this modification led to new dynamic hyperuniform states. As shown in the left panel of Fig.~\ref{Fig.5}(A), pairs of active particles are given displacements along the axis connecting their centers, with the same amplitudes chosen from a uniform distribution in the range $[0,\sigma]$, where $\sigma$ is the particle diameter, ensuring the center-of-mass conservation.
\rt{Comparing to the other modified random organization model we have discussed in Section II.A~\cite{wilken2020hyperuniform}, the CMC version of the RO model can be regarded as setting $\epsilon_r=0$ only retaining the reciprocal repulsions. This adjustment enforces the system to exhibit less randomness in active particle diffusion as the direction of each collision is deterministic rather than randomly chosen.} In the study of the 2D zero-strain case, the structure factors of systems at several particle densities $\rho$ are shown in Fig.~\ref{Fig.5}(B). They observed that the structure factor exhibited the same scaling $S(k \rightarrow 0) \sim k^{0.45}$ as the original RO model at the critical point of the absorbing-state phase transition. Moreover, they found that the active states exhibited a new dimension-irrelevant structure factor scaling $S(k \rightarrow 0) \sim k^2$, which indicates that the center-of-mass conservation suppresses long-range fluctuations in the active state and induces strong long-ranged hyperuniform correlations. Additionally, they studied a CMC version of the Manna model, where particles in active sites could redistribute in opposite directions to adjacent neighboring sites, and the structure factor shows the same hyperuniform scaling in the corresponding critical state and active states, as shown in Fig.~\ref{Fig.5}(C).

\begin{figure}[htb]
    \centering
    \includegraphics[width=0.45\textwidth] {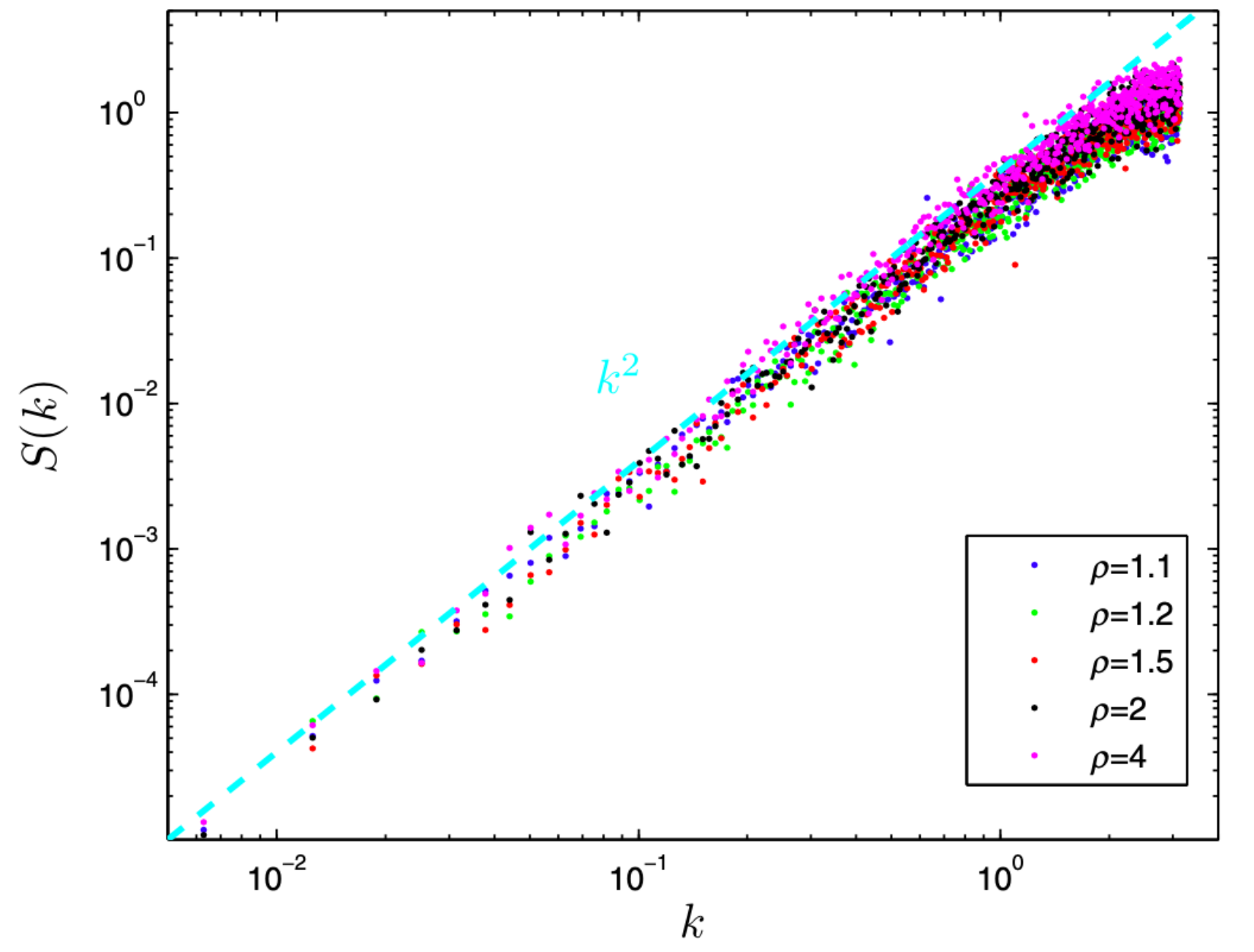}
    \caption{The structure factor, $S(k)\sim k^2$, of COMCon model for density $\rho > \rho_c \approx 1$ in the active states.\cite{hexner2017noise} Reproduced with permission.\cite{hexner2017noise} Copyright 2017, American Physical Society.
}
    \label{Fig.6}
\end{figure}

\begin{figure*}[htb]
    \centering
    \includegraphics[width=0.9\textwidth] {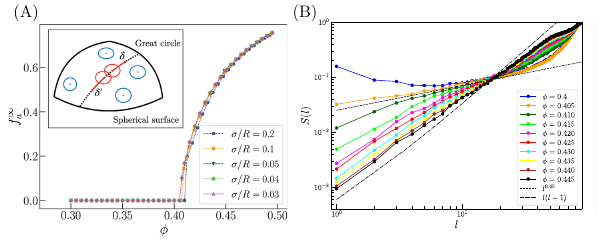}
    \caption{\textbf{(A)} The active particle fraction for several relative spherical surface size $\sigma/R$, with particle diameter $\sigma$ and spherical surface radius $R$. The inset shows the schematic diagram of center-of-mass conserved random organization model on a spherical surface. \textbf{(B)} Spherical structure factor $S(l)$ for several packing fractions on sphere for both absorbing state $\phi<\phi_c$, critical point $\phi_c=0.405$ and active state $\phi>\phi_c$.\cite{Lei2023Spherical} Reproduced with permission.\cite{Lei2023Spherical} Copyright 2023, AIP Publishing.}
    \label{Fig.7}
\end{figure*}

To further understand the origin of hyperuniformity induced by the center-of-mass conservation, the authors studied a 1D minimal lattice model called COMCon, which allows to establish an analytic description of the active state. In the COMCon model, similar to the Manna model, there are $n_i$ particles at site $i$, and the site with $n_i>1$ is regarded as an active site. Pairs of particles on the active site $i$ move to sites $i-1$ and $i+1$ at a rate $\omega(n_i-1)$, while particles at sites with $n_i \le 1$ do not move. This construction of particle dynamics ensures the conservation of the center-of-mass, and the COMCon model also undergoes an absorbing phase transition from absorbing states to active diffusive states at a critical density $\rho_c \approx 1$. Then, the authors solved the COMCon model numerically and found that the structure factor exhibited a hyperuniform quadratic scaling $S(k \rightarrow 0) \sim k^{2}$ in active states, as shown in Fig.~\ref{Fig.6}. Moreover, by using the direct stochastic process analysis~\cite{hexner2017noise}, a coarse-grained field-level Langevin equation could be obtained analytically
\begin{equation} \label{eq:COMCon}
\partial_t n=D \partial_{x x} n+A \partial_{x x}(\sqrt{n-1} \eta),
\end{equation}
where  $D=\omega a^2$ is the intrinsic diffusion constant and $a$ is the lattice constant, the Gaussian white noise $\left\langle\eta(x, t) \eta\left(x^{\prime}, t^{\prime}\right)\right\rangle=\delta\left(x-x^{\prime}\right) \delta\left(t-t^{\prime}\right)$ with the noise amplitude \rt{$A=\sqrt{\omega} a^2$}.
To understand this quadratic noise term $\partial_{xx}\eta$, the authors considered the system in the continuum limit, and the particle number conservation could be written as
\begin{equation} \label{eq:density_consrv}
\partial_t n =- \nabla \cdot J,
\end{equation}
where $J$ is the density flux. Also, by defining the global center-of-mass quantity at chosen freedom $\alpha$ as $R_\alpha \equiv \int {\rm d}^d r n(r)r_\alpha$, the center-of-mass conservation could be represented by $\partial_t R_\alpha=0$. Using Eq.~\ref{eq:density_consrv} and the definition of $\alpha$, the center-of-mass conservation can be represented as follows 
\begin{equation}\label{eq:CMC}
\partial_t R_\alpha  =-\int {\rm d}^d r r_\alpha \nabla \cdot J  =-\int {\rm d} S \cdot\left(J r_\alpha\right)+\int {\rm d}^d r J_\alpha.
\end{equation}
To ensure \rt{$\partial_t R_\alpha=0$}, $J_\alpha$ in the last term of Eq.~\ref{eq:CMC} should be written in the divergence form as $J_\alpha=-\nabla \cdot \sigma_\alpha$, which leads Eq.~\ref{eq:density_consrv} to $\delta_t n = \nabla \cdot \nabla \cdot \sigma_\alpha$. In 1D, $\sigma$ could be written as $\sigma=f(n)+g(n)\eta$, where $f(n)$ is deterministic and $g(n)\eta$ with Gaussian white noise $\eta$ represents the stochastic motion. Thus, in Eq.~\ref{eq:density_consrv}, the term accounting for stochastic motion should be written in the quadratic divergence form $\partial_{xx}\eta$.
This center-of-mass conserved active state leads to a new dynamic hyperuniform state and further inspires a number of non-equilibrium hyperuniform fluid systems, e.g., chiral active fluids and non-equilibrium driven-dissipative hyperuniform fluids~\cite{lei2019nonequilibrium,lei2019hydrodynamics,tjhung2017discontinuous}, which are discussed in the next section.

\begin{figure*}[htb]
    \centering
    \includegraphics[width=0.9\textwidth] {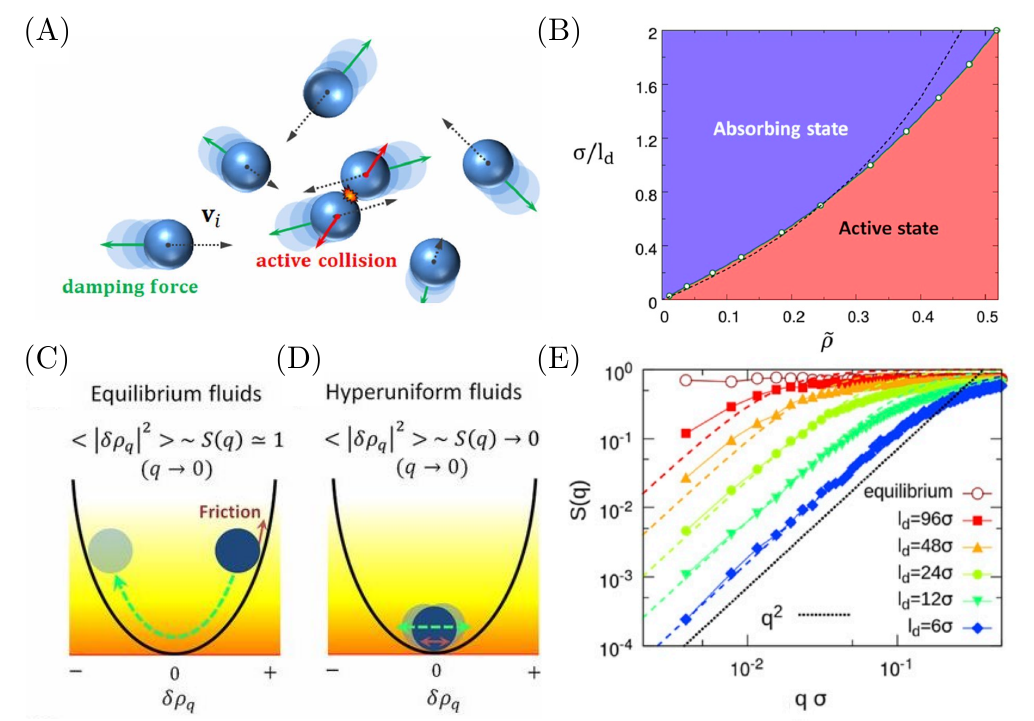}
    \caption{\textbf{(A)} Schematic illustration of a reactive hard-sphere fluid. \textbf{(B)} The phase diagram of reactive hard spheres. \textbf{(C and D)} The density fluctuation $\delta \rho_q$ represented as damped stochastic harmonic oscillators for (C) equilibrium fluids and (D) non-equilibrium hyperuniform fluids. \textbf{(E)} Hyperuniform scaling $S(q \rightarrow 0)\sim q^2$ for various dissipation lengths $l_d$ at density $\tilde{\rho}=0.1$, with the hollow red dotted line indicating the structure factor of the equilibrium liquid at $\tilde{\rho}=0.1$. The dashed curves represent theoretical predictions from the generalized Navier-Stokes equations of hyperuniform fluids.\cite{lei2019hydrodynamics} Reproduced with permission.\cite{lei2019hydrodynamics} Copyright 2019, National Academy of Sciences.}
    \label{Fig.8}
\end{figure*}

Next, to study the effect of topology on the dynamic hyperuniform state, Lei \textit{et al.} extended the CMC version of the random organization model to a 2D curved space by preserving similar local random organizing dynamics~\cite{Lei2023Spherical}. As shown in the inset of Fig.~\ref{Fig.7}(A), at each simulation step, active particles (depicted as red circles in the figure, i.e., overlapped particles) are assigned random displacements of the same magnitude with opposite directions along the great circle crossing the centers of two particles. The authors investigated the absorbing phase transition and demonstrated that the transition area fraction on the sphere $\phi$ is not significantly affected by the spherical surface size $\sigma/R$ ($\sigma$ is the particle diameter and $R$ is the spherical surface radius). To study the hyperuniformity of the particle structure on a spherical surface, the authors calculated the spherical structure defined as ~\cite{brauchart2019hyperuniform,bovzivc2019spherical}
\begin{equation}
S(l) = \frac{1}{N} \sum_{i,j = 1}^{N} P_l \left[ \cos \left( \frac{d_{ij}}{R}\right)\right],
\end{equation}
where $d_{ij}$ is the (great circle) spherical distance between particles $i$ and $j$, and $P_l(\cdot)$ is the $l$th order Legendre polynomial. Here $l$ is the wave number, which plays the role of wave vector as in Euclidean spaces, and $S(l \rightarrow 0) \rightarrow 0$ indicates the hyperuniformity. As shown in Fig.~\ref{Fig.7}(B), it was found that $S(l \rightarrow 0) \sim l^{0.45}$. In active states, the spherical structure factor shows a hyperuniform scaling $S(l \rightarrow 0) \sim l(l+1)$, and these hyperuniform structure factor scalings do not change with spherical surface size $R$. The authors proposed a effective dynamical field theory for the active state in the Fourier space $(\omega, l)$ 
\begin{equation}
i \omega \delta \rho_{l, \omega}R^2=-D_0 l(l+1) \delta \rho_{l, \omega}-l(l+1) \sqrt{\bar{\rho}} \, \eta_{l, \omega},
\end{equation}
and obtained the hyperuniform scaling $S(l \rightarrow 0) \sim l(l+1)/R^2$. This suggests a way to produce dynamically self-organized hyperuniform structures on a closed manifold without introducing interfaces, which normally has significant influence in long-range correlated states.

\section{Non-equilibrium hyperuniform fluids}
\subsection{Stable and metastable non-equilibrium hyperuniform fluids}

In the previous section, we mentioned that by modifying the noise in the random organization model to conserve the center-of-mass of the system, one can obtain an active hyperuniform dynamic state with $S(k \rightarrow 0) \sim k^2$, and this special noise was recently also shown to enhance the quasi-long range translational order in two-dimensional crystals~\cite{berthier2d,2dquasilong}. However, we note that such center-of-mass conserved noise is unphysical, and it can not be realized in any realistic particle system, because of the inertia of particles. Therefore, in Ref.~\cite{lei2019hydrodynamics}, as shown in Fig.~\ref{Fig.8}(A),  an athermal dynamical model consisting of $N$ reactive hard spheres in $d$-dimensions with mass $m$ and \rt{diameter $\sigma$} at dimensionless particle number density $\tilde{\rho}\equiv \rho \sigma^d$ was proposed. In this model, particles undergo reciprocal elastic collisions with an additional kinetic energy $\Delta E$ injected during each collision (reaction). Between collisions, particles with velocity $\mathbf{v}_i$ follow the equation of motion $m \frac{d \mathbf{v}_i (t)}{dt}=-\gamma  \mathbf{v}_i (t)$, where $\gamma$ is the linear damping coefficient. 
There are two characteristic length-scales in the system controlling its phase behaviour: (i) the mean-free path $l_m \propto \sigma/\tilde{\rho}$ at low density, which represents the typical distance between two subsequent collisions for a particle; (ii) the dissipation length $l_d \equiv \sqrt{m\Delta E}/\gamma$, which represents the typical distance that an isolated active particle can travel before stop. It was found that when $l_d<l_m$, the system eventually gets trapped in an absorbing state with $T_k\equiv m \overline{v^2}/k_Bd=0$. On the other hand,  when $l_d > l_m$, the system self-organizes into a homogeneous active fluid state with a positive kinetic temperature $T_k >0$. To understand this, the authors considered the balance between the energy injection $W_{\rm driv} \simeq f_a\Delta E$ and dissipation $W_{\rm disp} \simeq \bar{v}^2/\gamma$ per particle in the system, with the average collision frequency per particle $f_a \simeq \bar{v}/2l_m$ and $\bar{v}$ is the average particle speed. For the active fluid state in 2D, a driving-dissipation balance $W_{\rm driv}=W_{\rm disp}$ is achieved, and the kinetic temperature could be written as $k_BT_k/\Delta E= A(l_d/l_m)^2/8$ with the prefactor $A$ obtained from data collapsing of simulations. The phase diagram of the system is shown in Fig.~\ref{Fig.8}(B), and the dash curve is the mean-field theory prediction for the phase boundary.

Then, it was observed that the critical point and active fluid state of the random organizing fluid in both 2D and 3D are hyperuniform, with the structure factor scaling $S(q \rightarrow 0)\sim q^{0.45}$ at the CDP class critical point for 2D, and $S(q \rightarrow 0) \sim q^2$ in the active state for both systems in 2D and 3D. To explain the hyperuniformity in the active fluid state, an athermal fluctuating hydrodynamic theory based on generalized Navier–Stokes equations was proposed
\begin{equation}\label{eq:HUfluid_NS}
\begin{aligned}
\frac{\partial \rho}{\partial t} & =-\nabla \cdot(\rho \mathbf{u}), \\
\frac{\partial(\rho \mathbf{u})}{\partial t}+\nabla \cdot(\rho \mathbf{u u}) & =-\tilde{\gamma} \rho \mathbf{u}-\nabla p+\nabla \cdot\left(\boldsymbol{\sigma}^v+\boldsymbol{\sigma}^r\right),
\end{aligned}
\end{equation}
where $\rho$  and $\mathbf{u}$ are the density and velocity field of the fluid, respectively.
$\tilde{\gamma}\equiv\gamma/m$, and the local pressure $p$ is assumed to follow $p=c_s^2\rho$, where $c_s$ represents the speed of sound. $\boldsymbol{\sigma}^v$ and $\boldsymbol{\sigma}^r$ are the classical momentum-conserved viscous stress tensor~\cite{hansen2013theory} and the random (noise) stress tensor~\cite{landau1992hydrodynamic}, respectively. By performing the hydrodynamics linearization and Helmholtz decomposition on Eq.~\ref{eq:HUfluid_NS}, and transforming the equation in Fourier $q$ space, a Langevin equation for density fluctuation $\delta \rho_q$ can be written as 
\begin{equation} \label{eq:HUfluid_langevin}
q^{-2} \frac{\partial^2 \delta \rho_{\mathbf{q}}}{\partial t^2}=-\left(\tilde{\gamma} q^{-2}+\nu^{\|}\right) \frac{\partial \delta \rho_{\mathbf{q}}}{\partial t}-c_s^2 \delta \rho_{\mathbf{q}}+\sigma_{\|, \mathbf{q}}^r,
\end{equation}
where, $\nu^{\parallel}$ represents the longitudinal kinematic viscosity, and $\sigma^r_{\parallel,q}$ denotes the longitudinal component of random noise in $q$ space. This noise follows the relationship $\left\langle\sigma_{\|, \mathbf{q}}^r(t) \sigma_{\|, \mathbf{q}}^r\left(t^{\prime}\right)\right\rangle=2 \rho_0 \nu^{\|} k_B T_k V \delta\left(t-t^{\prime}\right)$, where $\rho_0$ is the average density with $V$ the volume of the system. Here, Eq.~\ref{eq:HUfluid_langevin} formally represents an equation for a damped stochastic harmonic oscillator with the effective mass $q^{-2}$, the displacement $\delta \rho_{\mathbf{q}}$, the damping force $\left(\tilde{\gamma} q^{-2}+\nu^{\|}\right)$, the restoring force $c_s^2 \delta \rho_{\mathbf{q}}$, and the driving random noise $\sigma_{\|, \mathbf{q}}^r$. Then one can obtain the magnitude of oscillation or the static structure factor of the system:
\begin{equation}\label{eq:HUfluid_Sq}
S(q)=\frac{q^2}{(B l_d^{-2}+q^2)},
\end{equation}
where $B=\sqrt{8/A}$ in 2D. When $\tilde{\gamma} = 0$, according to Eq.~\ref{eq:HUfluid_Sq}, the system behaves like a low-density equilibrium fluid with $S(q)=1$, consistent with the fact that the average potential of the oscillator at a fixed temperature is independent of its effective mass, as illustrated in Fig.~\ref{Fig.8}(C). For a non-equilibrium fluid system with $\tilde{\gamma}>0$, when $q \ll l_d^{-1}$, a hyperuniform scaling regime $S(q) \sim q^2$ is observed, as shown in Fig.~\ref{Fig.8}(E). This regime can be interpreted as a stochastic oscillator with suppressed oscillation, where $\left< |\delta \rho_q|^2 \right> \rightarrow 0$, resulting from the infinitely large effective damping coefficient, as shown in Fig.~\ref{Fig.8}(D). This also implies that based on the equipartition theorem~\cite{ept}, there is no well defined temperature in non-equilibrium hyperuniform fluids.

\begin{figure}[htb!]
    \centering
    \includegraphics[width=0.45\textwidth] {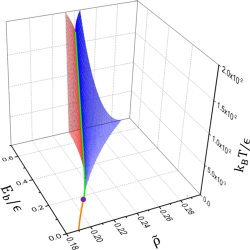}
    \caption{Mean-field theory predicted phase diagram of barrier-controlled reactive hard spheres.\cite{lei2021barrier} Reproduced with permission.\cite{lei2021barrier} Copyright 2021, American Physical Society.}
    \label{Fig.9}
\end{figure}

\begin{figure*}[htb]
    \centering
    \includegraphics[width=0.95\textwidth] {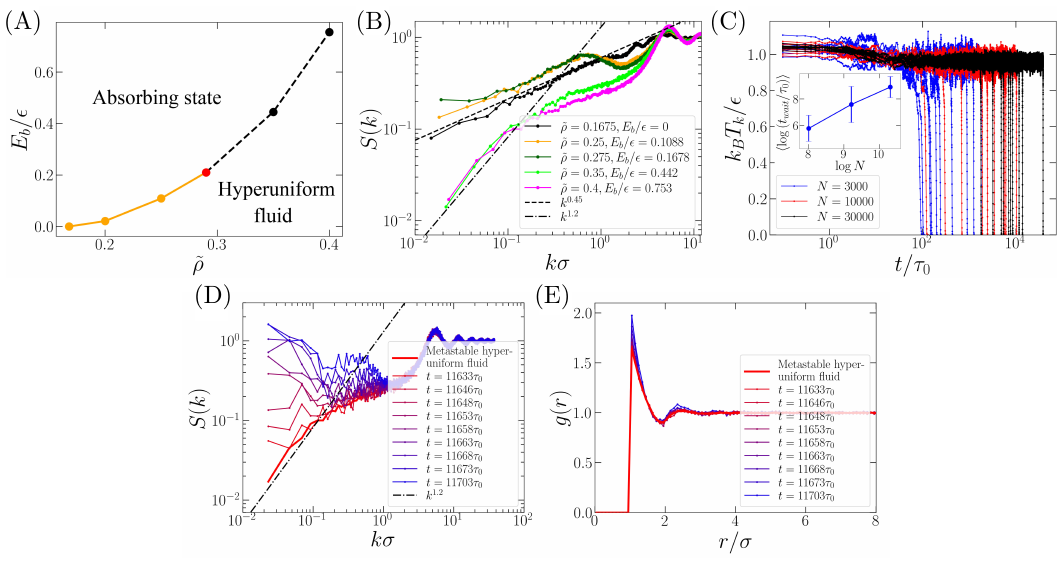}
    \caption{\textbf{(A)} Phase behavior of athermal barrier-controlled reactive hard spheres from simulation. \textbf{(B)} The structure factor near the two typical critical points of continuous phase transition (black and orange data curves), the tricritical point (dark green data curve), and two typical points selected near the stability limits of the hyperuniform fluid in the discontinuous phase transition (green and pink data curves). \textbf{(C)} The time evolution of the kinetic temperature $T_k$ in the metastable hyperuniform fluid near the stability limit of different sizes. The inset shows $\left <t_{wait}\right >$ as a function of the system size. \textbf{(D and E)} The structure factor $S(k)$ (D) and radial distribution function $g(r)$ (E) for the system at various times during the discontinuous phase transition from a metastable hyperuniform fluid to an absorbing state.\cite{lei2023howHUfreeze} Reproduced with permission.\cite{lei2023howHUfreeze} Copyright 2023, National Academy of Sciences.}
    \label{Fig.10} 
\end{figure*}

To further study the non-equilibrium phase behavior of the driven-dissipative hard spheres~\cite{lei2019hydrodynamics}, Lei {\textit{et al.}} proposed a modified model by introducing an additional activation barrier to the system~\cite{lei2021barrier,lei2023howHUfreeze}. In the modified model, an active collision occurs only when the relative kinetic energy of the two colliding particles along the center-to-center direction surpasses an activation barrier $E_b$. Otherwise, the colliding particles undergo an elastic collision with no injection of extra energy. By selecting the kinetic temperature $T_k$ as the order parameter and employing the mean-field approximation, the barrier controlled reactive hard-sphere fluid at finite temperature $T$, which represents the strength of thermal noise, can be approximated by a cubic equation
\begin{equation}\label{eq:BarrierHUFluid_mf}
\frac{\partial  T_k}{\partial t}=a T_k-b T_k^2-c T_k^3+h,
\end{equation}
where $a=\frac{\tilde{\rho}}{2\tau_0}[(1-B E_b/\Delta E)-(\tau_0/\tau_d)]$, $b=-\frac{dk_B\tilde{\rho}}{2 \tau_0 \Delta E^2 }[(1+A) B E_b - A\Delta E]$, $c=\frac{d^2k_B^2}{2\tau_0 \Delta E^3}\tilde{\rho}[4 \Delta E A^2+(3 A-4 A^2) B E_b]$, $h=\gamma T/m$, and $\tau_d=m/\gamma$, with prefactors $A$ and $B$. The typical excitation speed is $v_0=\sqrt{\Delta E/m}$, and  $\tau_0=\sigma/v_0$ is the time unit in the system. A mean-field phase diagram is obtained by solving Eq.~\ref{eq:BarrierHUFluid_mf} as shown in Fig.~\ref{Fig.9}, the purple point on $T=0$ plane is the tricritical point which is the crossover point from the CDP class absorbing phase transition (orange curve) to the Ising class phase transition (green curve). The red and blue surfaces are the discontinuous phase transition boundary separating the absorbing steady state and active steady state, in between which is the metastable regime.

Subsequently, the kinetic pathway of the discontinuous absorbing transition from the metastable non-equilibrium hyperuniform fluid was investigated in Ref.~\cite{lei2023howHUfreeze}, which focused on the barrier controlled athermal driven-dissipative hard-sphere fluid at $T=0$ in 2D.
The phase diagram is shown in Fig.~\ref{Fig.10}(A), where the orange curve represents the continuous absorbing transition, the red dot marks the tricritical point, and the black dashed curve represents the stability limit of the active state. As shown in Fig.~\ref{Fig.10}(B), the critical hyperuniformity at the continuous absorbing transition with the structure factor scaling $S(k \rightarrow 0)\sim k^{0.45}$ was observed, which is in agreement with the hyperuniform scaling at the CDP universality class critical point. Additionally, they found a new hyperuniform scaling $S(k \rightarrow 0 )\sim k^{1.2}$ in the structure of metastable states near the stability limit of the discontinuous absorbing transition, of which the physical origin remains unknown.

\begin{figure}[htb]
    \centering
    \includegraphics[width=0.45\textwidth] {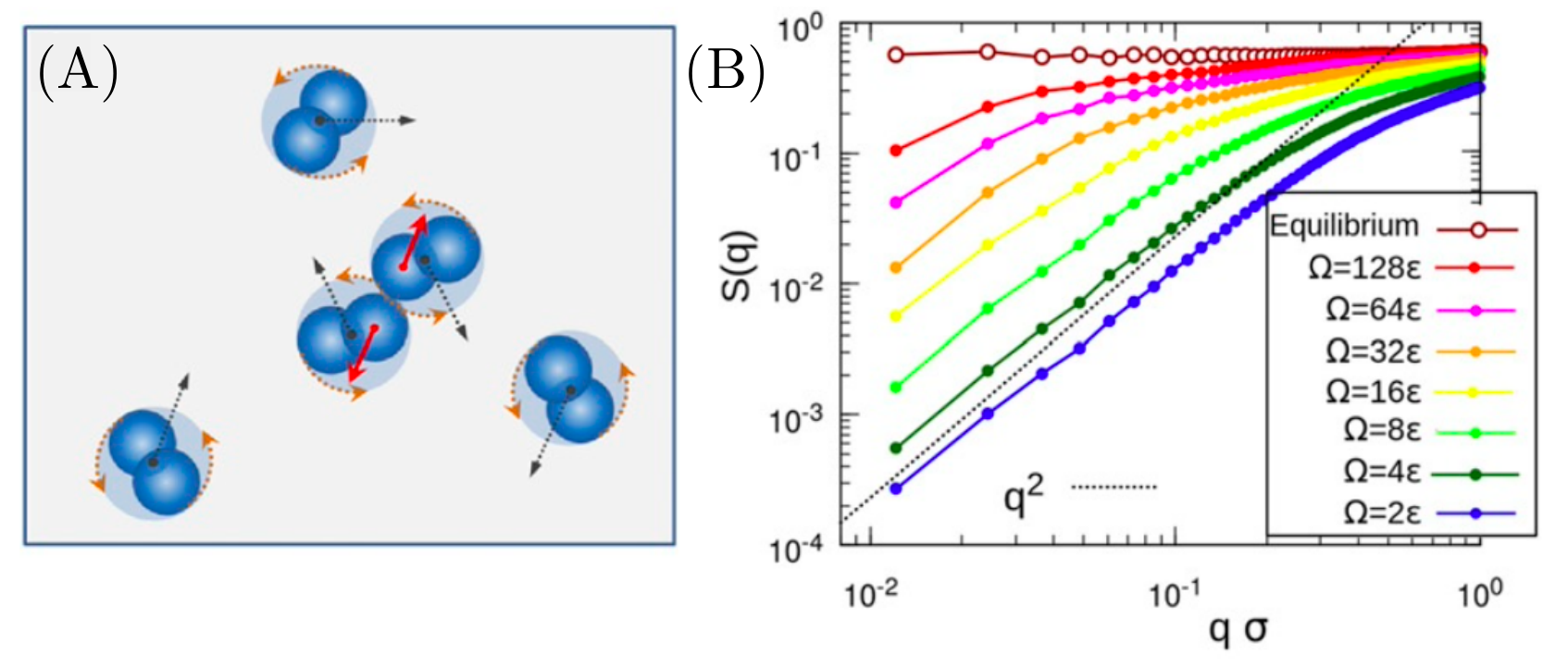}
    \caption{\textbf{(A)} Schematic diagram of the active spinner model. \textbf{(B)} Hyperuniform structure factor scaling at various driven torque $\Omega$.\cite{lei2019hydrodynamics} Reproduced with permission.\cite{lei2019hydrodynamics} Copyright 2019, National Academy of Sciences.}
    \label{Fig.11}
\end{figure}

\begin{figure*}[htb]
    \centering
    \includegraphics[width=0.95\textwidth] {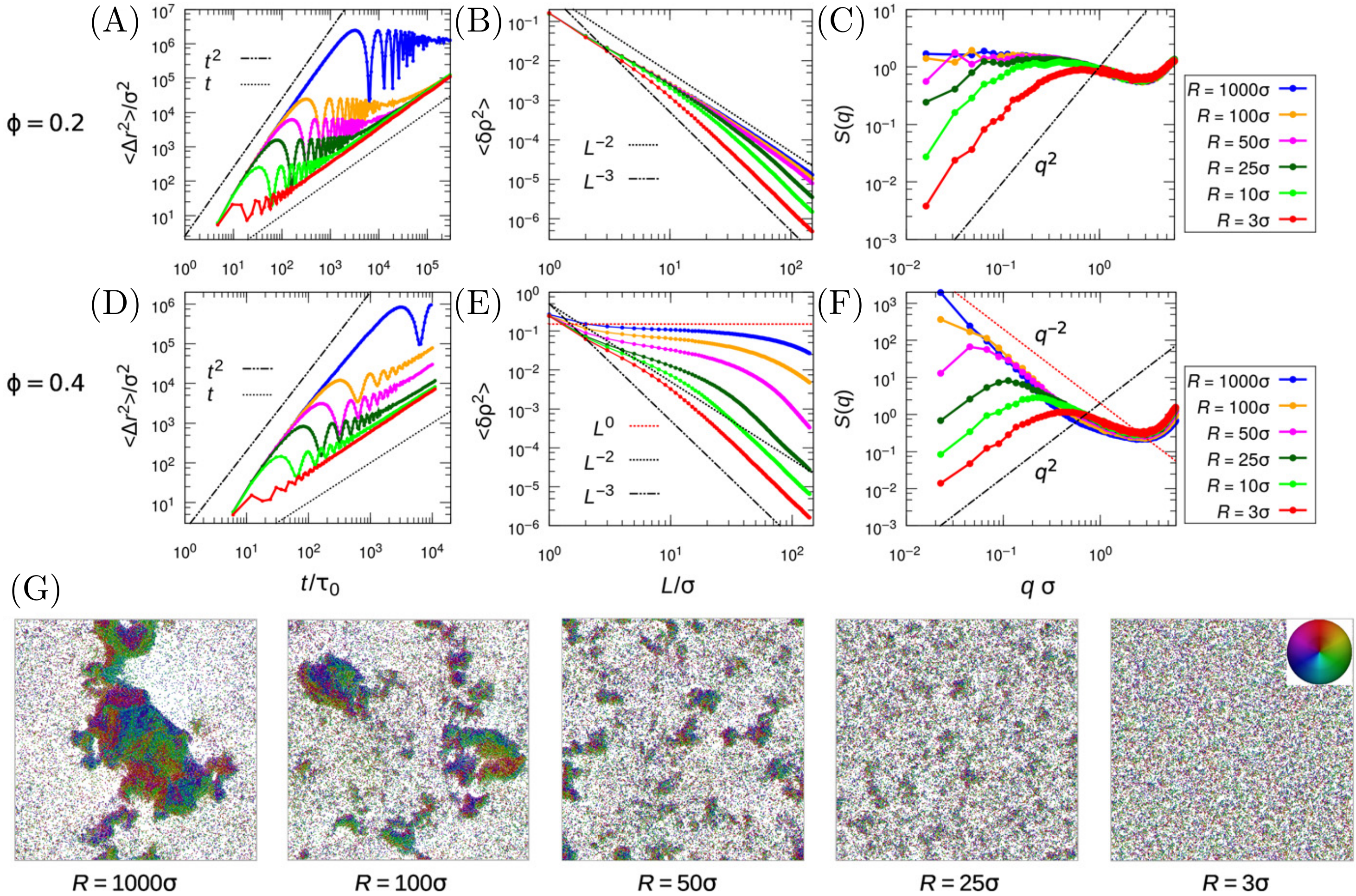}
    \caption{\textbf{(A and D)} MSD as functions of $t$ for various driven circular motion radius $R$. \textbf{(B and E)} Density variances $\left <\delta \rho^2 \right >$  as functions of window size $L$ for various $R$.  \textbf{(C and F)} Hyperuniform scaling of structure factor $S(q) \sim q^2$ for various $R$ at small $R$ at two typical area fraction $0.2$ and $0.4$. \textbf{(G)} Typical snapshots for systems at the area fraction $0.4$ with various $R$.\cite{lei2019nonequilibrium} Reproduced with permission.\cite{lei2019nonequilibrium} Copyright 2019, American Association for the Advancement of Science.}
    \label{Fig.12}
\end{figure*}

Furthermore, while studying the discontinuous absorbing transition in the system, the authors found that the metastable dynamic hyperuniform state could be kinetically stable at the thermodynamic limit, as the waiting time for phase transition from the metastable active state to the absorbing state increases with the system size, as shown in the Fig.~\ref{Fig.10}(C). The kinetic pathway of the discontinuous absorbing transition indicates that the transition is triggered by the large-scale fluctuations (i.e. small $k$ correlation in $S(k)$ in Fig.~\ref{Fig.10}(D)) while the local structure (i.e. small $r$ regime in $g(r)$) remains intact (Fig.~\ref{Fig.10}(E)). This suggests that the hyperuniformity found near the absorbing transition stability limit  is metastable yet kinetically stable for infinitely large systems, and this challenges the common understanding of metastability in equilibrium, where discontinuous phase transitions can always be triggered by localized fluctuations, i.e., nucleation.

\begin{figure*}[htb]
    \centering 
    \includegraphics[width=0.95\textwidth] {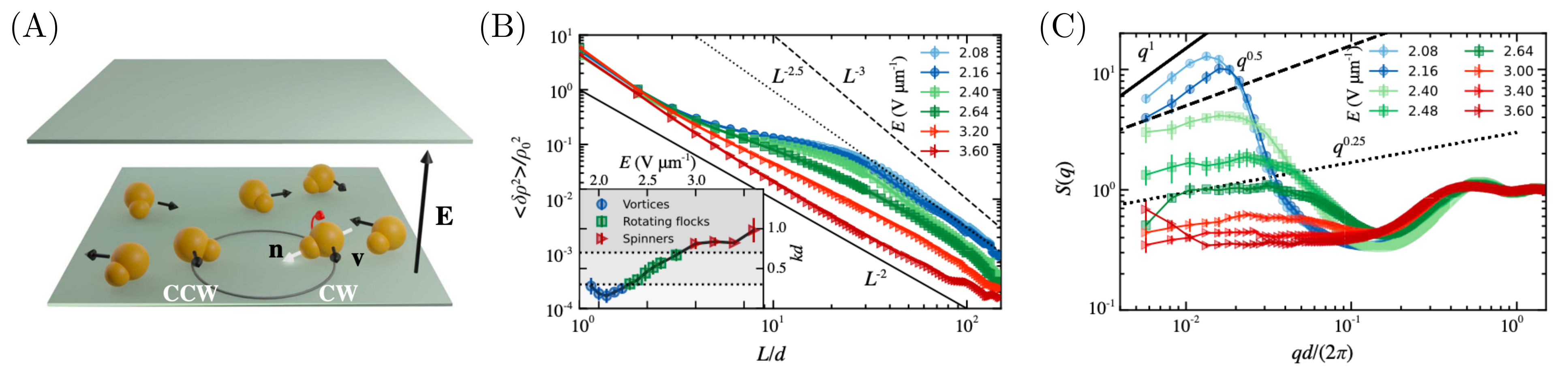}
    \caption{\textbf{(A)} A sketch of experimental setup. \textbf{(B)} Density fluctuations $\left < \delta \rho^2(L) \right >$ of pear-shaped rollers as a function of the window size $L$ at several electric field strengths. \textbf{(C)} Structure factors $S(q)$ of circle rollers at different dynamic states of various electric field strengths.\cite{zhang2022hyperuniform} Reproduced with permission.\cite{zhang2022hyperuniform} Copyright 2022, American Physical Society.}
    \label{Fig.13}
\end{figure*}

\subsection{Hyperuniformity in chiral active particles}

In the previous section, we have reviewed a hyperuniform fluid model with energy dissipation via damping and energy driven via reciprocal excitation, and the system is found to be hyperuniform in the active state shown in Fig.~\ref{Fig.8}. In Ref.~\cite{lei2019hydrodynamics}, the authors also investigated this hyperuniform mechanism in a more realistic system consisting of $N$ active spinners with the same chirality on a frictional substrate. As shown in Fig.~\ref{Fig.11}(A), each spinner consists of two spherical monomers of mass $m/2$ and diameter $\sigma$ connected by a center-to-center rigid bond with length $l_B$. The motion of the spinner driven by a constant torque $\Omega$ is described by the underdamped Langevin equation
\begin{equation}\label{eq:cABPs_Langevin}
\begin{aligned}
m \ddot{\mathbf{r}}_i(t) & =-\gamma \dot{\mathbf{r}}_i(t)-\nabla_i U(t), \\
I \ddot{\theta}_i(t) & =\Omega-\gamma_r \dot{\theta}_i(t)-\frac{\partial}{\partial \theta_i} U(t),
\end{aligned}
\end{equation}
where the moment of inertia of each spinner is $I=ml_B^2/4$, $\gamma$ is the translational friction coefficient, $\gamma_r=\gamma l_B^2/4$ is the rotational friction coefficient, and the spinners interact through the  Weeks-Chandler-Andersen (WCA) potential between their monomers. When two spinners collide, the rotational kinetic energy is transferred to their translational kinetic energy, which induces an active reciprocal collision similar to the driven-dissipative hard spheres in Ref.~\cite{lei2019hydrodynamics}. With increasing $\Omega$, the kinetic temperature $T_k$ of the system increases, and the system stays in an active state with positive $T_k$ at large $\Omega$. As shown in Fig.~\ref{Fig.11}, an quadratic hyperuniform scaling $S(q \rightarrow 0) \sim q^{2}$ at large torque $\Omega$ was observed, which is identical to the active hyperuniform fluid at high dissipation length $l_d$ case, i.e., high active energy $\Delta E$, in Fig.~\ref{Fig.8}.

Similarly, another chiral active particle system was investigated in Ref.~\cite{lei2019nonequilibrium}
using Brownian dynamics simulations, in which particles interact through a WCA potential and the motion of particles at temperature $T$ is described by the overdamped Langevin equations
\begin{equation}\label{eq:cABPs_Langevin}
\begin{aligned}
&\dot{\mathbf{r}}_i(t)=\gamma_t^{-1}\left[-\nabla_i U(t)+F^{{p}} e_i(t)\right]+\sqrt{2 k_{\mathrm{B}} T / \gamma_{\mathrm{t}}} \xi_i^t(t),\\
&\dot{\mathbf{e}}_i(t)=\left[\gamma_{\mathrm{r}}^{-1} \Omega+\sqrt{2 k_{\mathrm{B}} T / \gamma_{\mathrm{r}}} \xi_i^{\mathrm{r}}(t)\right] \times \mathbf{e}_i(t),
\end{aligned}
\end{equation}
where $\mathbf{r}_i$ and $\mathbf{e}_i$ are the position of particle $i$ and its self-propulsion orientation, respectively. $\gamma_t$ and $\gamma_r$ are the translational and rotational friction coefficients, respectively, with $\xi_i^t$ and $\xi_i^r$ the Gaussian noise terms. The self-propulsion speed of the particle is $v_0 = \gamma_t^{-1}F^p$, and the reduced noise strength is defined as $T_R = k_BT/(F^p\sigma)$ with $\sigma$ and $F^p$ the particle size and self-propulsion force, respectively. In the zero noise limit, i.e., $T_R = 0$, active particles perform circular motions with a fixed radius $R = F_p \sigma^2/\omega$ and period $\Gamma = 2\pi \gamma_r/\omega$. When $T_R = 0$, by increasing the particle area fraction $\phi$ or the circular motion radius $R$, it was found that the system undergoes a phase transition from a non-collisional absorbing state with a long time diffusion coefficient $D = 0$ to a colliding active phase with a diffusion coefficient $D > 0$, which was measured by using the mean squared displacement (MSD).

At a low area fraction of $0.2$ in 2D, particles exhibit the long-time diffusive motion with ${\rm MSD} \sim t$ in the active homogeneous state. A strong hyperuniformity at large length-scale was observed with the particle density fluctuation scaling $\left< \delta \rho(L \rightarrow \infty)^2 \right> \sim L^{-3}$ and the structure factor scaling $S(q \rightarrow 0) \sim q^2$, as shown in Fig.~\ref{Fig.12}(B) and (C). At a higher area fraction of $0.4$ in 2D active state, a critical scaling of $S(q) \sim q^{-2}$ appears at some intermediate length-scale, which is due to the giant density fluctuation induced by the motility-induced micro-phase separation~\cite{mazhan2022}, and the hyperuniform scaling $S(q \rightarrow 0) \sim q^{2}$ persists at large length-scale as shown in Fig.~\ref{Fig.12}(E) and (F). 
The co-existence of two seemingly different phenomena at different length-scales, i.e., the giant local density fluctuation and global hyperuniformity, is remarkable, and a similar phenomenon was previously only found in the early stage of universe, in which the enhanced local density fluctuation induced by gravity co-exists with the global hyperuniformity of the universe~\cite{gabrielli2002glass}. 

\begin{figure*}[htb]
    \centering
    \includegraphics[width=0.9\textwidth] {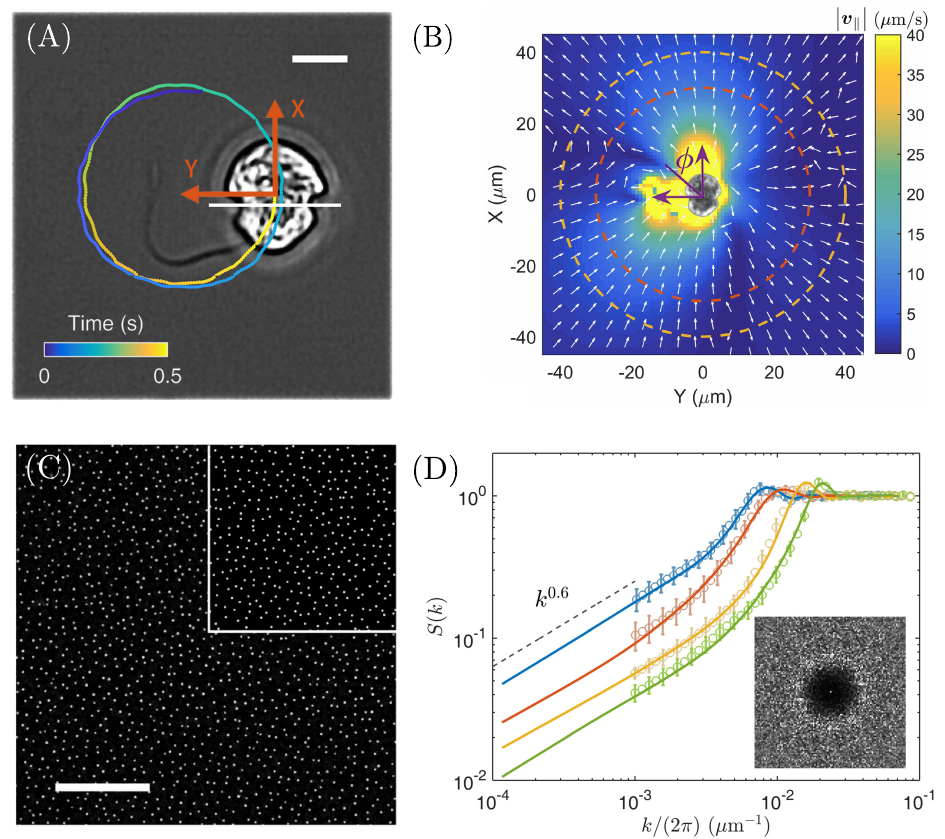}
    \caption{\textbf{(A)} Circular trajectory plotted on an optical image of an algae cell at the interface. \textbf{(B)} Mean in-plane flow field, $\mathbf{v}_{\parallel}$, calculated from regularized Stokeslet  model simulation. \textbf{(C)} Streak image of cell motion. Inset: Typical snapshot from simulation. \textbf{(D)} Structure factors at various densities (different colors indicates various cell number densities). Experimental and numerical results are shown by light symbols and dark curves, respectively.\cite{huang2021circular} Reproduced with permission.\cite{huang2021circular} Copyright 2021, National Academy of Sciences.}
    \label{Fig.14}
\end{figure*}

To explain the quadratic hyperuniform structure factor scaling in the active homogeneous state, by treating the particle circulating center $\mathbf{r}_i^o=\mathbf{r}_i-\sigma^2\left(\mathbf{F}_i^p \times \Omega\right) /|\Omega|^2$ as the effective particle, the authors proposed a dynamic equation for the density field of these effective particles $\rho^o$.
Inspired by the linear response theory for chiral active particles, at the large length-scale or small $q$ regime, there is only a diffusion mode in the density field of effective particles~\cite{lei2019nonequilibrium}. Therefore, at large length-scale $q \ll 2\pi/R$, the dynamic equation for the density field in $q$ space can be written as
\begin{equation} \label{eq:cABPs_Field}
\partial_t \rho^o_{\mathbf{q}} = -D_e^o  q^2 \rho^o_{\mathbf{q}}+\xi_{\mathbf{q}}(t),
\end{equation}
where $D^o_e$ is the diffusion coefficient of effective particles and $\xi_{\bf q}(t)$ is the Fourier transform of the noise $\xi(t)$ mimicking collisions between particles in the system. As the collisions between particles are reciprocal, $\xi(t)$ conserves the center-of-mass of the system~\cite{hexner2017noise}, and can be written as $\xi(t) = \sqrt{\bar{\rho}} \nabla^2 \eta(t)$ with $\langle \eta({{\bf r},t}) \eta({{\bf r}',t'}) \rangle = A^2 \delta ({\bf r - r}') \delta(t-t')$, where $A$ is the strength of the noise, and $\bar{\rho}$ is the average particle density in the system.
By solving the equation above in Fourier space, the hyperuniform scaling of effective particle can be obtained as $S^o(q \rightarrow 0) \sim A^2q^2/2D_e^o$.
Recently, Kuroda and Miyazaki proposed a microscopic theory~\cite{kuroda2023microscopic}, in which they used an effective fluctuating hydrodynamic field description for the active state of chiral active particles obtained from the Langevin equation (Eq.~\ref{eq:cABPs_Langevin}), and recovered the field equation (Eq.~\ref{eq:cABPs_Field}). This provides a microscopic analytical approach to the hyperuniform quadratic scaling in the active fluid state. Moreover, Kuroda \textit{et al.} recently also showed that the system can crystallize into a two-dimensional chiral active crystal at high density, which can be quantitatively explained by a linear elastic theory~\cite{kuroda2024}. 

The experimental realization of the predicted dynamic hyperuniform state of chiral active particles was recently done by Zhang and Snezhko in Ref.~\cite{zhang2022hyperuniform}, where they prepared a number of pear-shaped polystyrene particles between two glass slides in an AOT/hexadecane solution, and the particles were driven by the electrohydrodynamic Quincke rotation phenomenon in an electric field $\mathbf{E}$ applied perpendicular to the glass slides, as shown in Fig.~\ref{Fig.13}(A). This resulted in the half of the particles performing clockwise rotation while the other half rotating counterclockwisely in the experiment. As the external electric field $E$ increases, at a fixed particle area fraction, the system formed several emergent fluid patterns. These active chiral fluid patterns exhibited long-range fluctuation suppression as the field $E$ decreased. The authors found that the particle number fluctuation scaling transformed from $\sim L^{-2}$ in the spinner pattern phase to $\sim L^{-3}$ in the vortices pattern phase at the large length-scale $L$, and correspondingly the structure factor $S(q)$ scaling changed from $\sim q^{0}$ to $\sim q^1$, as shown in Fig.~\ref{Fig.13}(C) and (D). This qualitatively agrees with the theoretical prediction of dynamics hyperuniform states formed by chiral active particles~\cite{lei2019nonequilibrium}, while the exact exponent in the hyperuniform scaling is different, which may be due to the unwanted noises, e.g., hydrodynamic effects, in experiments not considered theoretically.

\begin{figure*}[htb]
    \centering
    \includegraphics[width=0.95\textwidth] {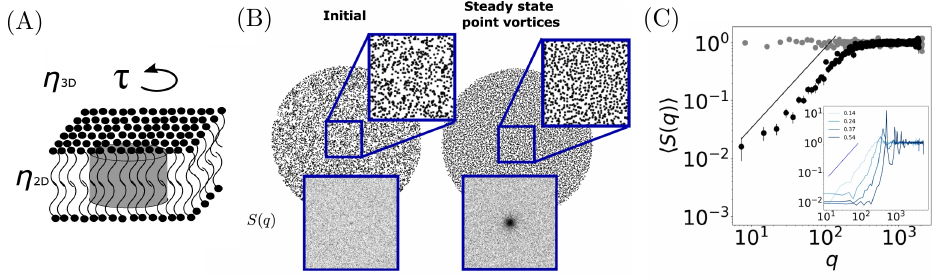}
    \caption{\textbf{(A)} A schematic diagram of a membrane rotor. \textbf{(B)} Snapshots of vortices initially (left) and at steady state (right). Insets show the structure factor $S(q_x,q_y)$. \textbf{(C)} The angular averaged structure factor $S(q)$ of initial and steady state vortices structure show in (B). Inset shows the $S(q)$ in rotors of various area fraction.\cite{oppenheimer2022hyperuniformity} Reproduced with permission.\cite{oppenheimer2022hyperuniformity} Copyright 2022, Springer Nature Limited.}
    \label{Fig.15}
\end{figure*}

\subsection{Hyperuniformity in algae systems intermediated by hydrodynamic flows}
In Ref.~\cite{huang2021circular}, Huang \textit{et al.} investigated the system of marine algae cells (Effrenium voratum), which swim circularly at the air–liquid interface, as shown in Fig.~\ref{Fig.14}(A). The authors experimentally prepared samples of interacting swimming cells at the air–liquid interface, which self-organized into a homogeneous steady state, as shown in Fig.~\ref{Fig.14}(C). They obtained the snapshots and the cell distributions, with which they found a hyperuniform structure factor scaling $S(k \rightarrow 0) \sim k^{0.6}$ for various cell number densities, as shown in Fig.~\ref{Fig.14}(D). Moreover, they also performed computer simulations to model the algae cell system, using a regularized Stokeslet model to obtain the fluid flow generated by cell swimming, as shown in Fig.~\ref{Fig.14}(B). \rt{Then, they proposed a particle based model to describe the motion of $i$-th algae cells the at time $n \tau$ during a time step $\tau$}
\begin{widetext}
\begin{equation}\label{eq:Algae}
\overline{\boldsymbol{r}}^{(i)}((n+1) \tau)-\overline{\boldsymbol{r}}^{(i)}(n \tau)= \sum_{j \neq i} \overline{\boldsymbol{V}}\left(\overline{\boldsymbol{r}}^{(i)}(n \tau)-\overline{\boldsymbol{r}}^{(j)}(n \tau); v_c^{(j)}\right) \tau +\boldsymbol{\eta}\left(D^{(i)}\right) \delta\left(\bmod (n, p), s^{(i)}\right),
\end{equation}
\end{widetext}
\rt{where $\overline{\boldsymbol{V}}(\mathbf{R};v_c)$ is the fluid flow from the Stokeslet model for cell circling with velocity $v_c$ and distance $\mathbf{R}$, and particles interact through hydrodynamic flow with a long cutoff length. In Eq.~\ref{eq:Algae}, the second term on the right-hand side indicates that particles undergo adjacent jumps with random temporal steps and random displacements. Specifically, the Kronecker delta function $\delta(\cdot)$ indicates that the $i$-th particle jumps at time $n\tau$ if $n$ satisfies modulo operation $\bmod(n, p)=s^{(i)}$ with $p$ a constant value, and $s^{(i)}$ the an integer constant between zero and $p-1$, which is randomly assigned to all particles. $\mathbf{\eta}\left( D^{i} \right )$ indicates that the jumping displacements are independently chosen from a normal distribution with deviation $\sqrt{2D^{(i)}p\tau}$, and $D^{(i)}$ is the diffusivity for the $i$-th particle randomly drawn from a Pareto distribution~\cite{huang2021circular}. The authors measured particle position distribution after randomly initialization and wait the system evolving into a steady state, as shown in the inset of Fig.~\ref{Fig.14}(C).} The structure factor agrees with the experimental data, as shown in Fig.~\ref{Fig.14}(D), indicating that the hyperuniform structure observed in the system is induced by the \rt{long-range} hydrodynamic interactions caused by the circular swimming of algae.

\begin{figure*}[htb]
    \centering
    \includegraphics[width=0.9\textwidth] {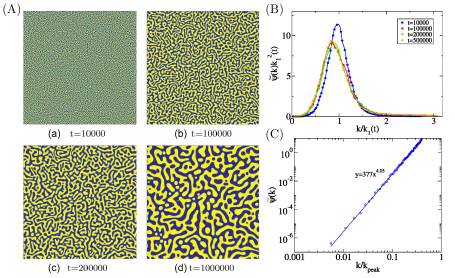}
    \caption{\textbf{(A)} The time evolution of the concentration field of a system evolving under the Cahn-Hilliard equation (Model-B) with critical quench. \textbf{(B)} The time-normalized spectral density $\tilde{\psi}(k/k_1(t))$. \textbf{(C)} The hyperuniform scaling of the spectral density $\tilde{\psi}(k)\sim k^{4.05}$.\cite{ma2017random} Reproduce with permission.\cite{ma2017random} Copyright 2017, AIP Publishing.}
    \label{Fig.16}
\end{figure*}

\subsection{Hyperuniformity in vortex systems induced by active rotors on membrane}
In Ref.~\cite{oppenheimer2022hyperuniformity}, Oppenheimer \textit{et al.} investigated a system of $N$ point vortices induced by membrane rotors, which are disks rotating due to torque $\tau$ in the plane of the membrane or the surface of the fluid, as shown in Fig.~\ref{Fig.15}(A). For a single vortex in an ideal inviscid fluid surface induced by a rotor, the stream function can be written as $\Psi(r) = \frac{1}{4\left(H_0\left(\frac{r}{\lambda}\right) - Y_0\left(\frac{r}{\lambda}\right)\right)}$ for $r \ll \lambda = \frac{\eta_{2D}}{2 \eta_{3D}}$, where $\lambda$ is the Shaffman-Delbrück length, $\eta_{2D}$ is the 2D viscosity on the membrane, and $\eta_{3D}$ is the viscosity of the outer bulk fluid. $H_0$ and $Y_0$ are zeroth-order Struve
function and Bessel function of the second kind, respectively. The dynamics of $N$ point vortices could be written as the Hamilton’s equation
\begin{equation}
\Gamma_{i} \mathbf{v}_{i}=\partial_{i}^{\perp} \mathit{H},\,\,{\rm with}\,\,\mathit{H}=\frac{1}{2} \sum_{i \neq j} \Gamma_{i} \Gamma_{j} \Psi(|\mathbf{r}_{i}-\mathbf{r}_{j}|),
\end{equation}
where $\mathbf{v}_i$ is the velocity of the $i$-th vortex, and the circulation $\Gamma_i = \frac{\tau_i}{\eta_{2 D}}$ is proportional to the magnitude of the torque on the rotors. $\Psi(\mathbf{r})$ denotes the stream function. By solving the Hamilton's equation in 2D with random initial configurations, the authors found that the vortices self-organize into a hyperuniform steady state at low rotor area fractions, as shown in Fig.~\ref{Fig.15}(C), where the black dots represent data points and the inset illustrates the structure factor scaling $S(q \rightarrow 0) \sim q^{1.3}$. Moreover, the authors explained the steady state hyperuniformity by considering the mathematical relationship between $S(q)$ and the Hamiltonian $\mathcal{H}$ written in the explicit form of  $\rho(\mathbf{r})$:
\begin{equation}
\mathcal{H}[\rho(\mathbf{r})] \sim \frac{\Gamma^2}{2} \int {\rm d} \mathbf{r} \int {\rm d} \mathbf{r}^{\prime} \rho(\mathbf{r}) \rho\left(\mathbf{r}^{\prime}\right) \Psi\left(\left|\mathbf{r}-\mathbf{r}^{\prime}\right|\right).
\end{equation}
By using the convolution theorem in Fourier space, a relationship between the Hamiltonian and the structure factor can be obtained as follows
\begin{equation}\label{eq:vortice_Hamiltonian}
\mathcal{H}[\rho(\mathbf{r})]=\frac{N \Gamma^2}{4 \pi} \int S(\mathbf{q}) \widetilde{\Psi}(\mathbf{q}) {\rm d} \mathbf{q}.
\end{equation}
As the Fourier transform of a single point vortice stream function is $\tilde{\Psi}(q)=1/q(q+\lambda^{-1})$, at the long-range limit $\tilde{\Psi}(q\rightarrow 0)=1/q^2$. Then, the Hamiltonian in Eq.~\ref{eq:vortice_Hamiltonian} could be written as
$\mathcal{H}[\rho(\mathbf{r})]=\frac{N \Gamma^2}{4 \pi} \int \frac{S({\bf q})}{q^2} {\rm d}{\bf q} = 
\frac{N \Gamma^2}{2} \int \frac{S(\mathbf{q})}{q} {\rm d} q$ with ${\rm d}{\bf q}= 2 \pi q {\rm d}q$. \rt{Thus, due to the conservation of Hamiltonian, the structure factor should have the scaling $S(q\rightarrow 0) \sim q^\alpha$ with $\alpha>0$, i.e., hyperuniformity, which ensures the convergence of Hamiltonian at large length scales.} Moreover, we note that the exponent $\alpha \approx 1.3$ found here is different from the value of 2 obtained in the theory and simulations of hyperuniform fluids of active spinners without considering hydrodynamic effects~\cite{lei2019hydrodynamics}, and this suggests that hydrodynamic effects can indeed change the exponent of the scaling of structure factor in dynamic hyperuniform states of chiral active particles.

\begin{figure*}[htb]
    \centering
    \includegraphics[width=0.95\textwidth] {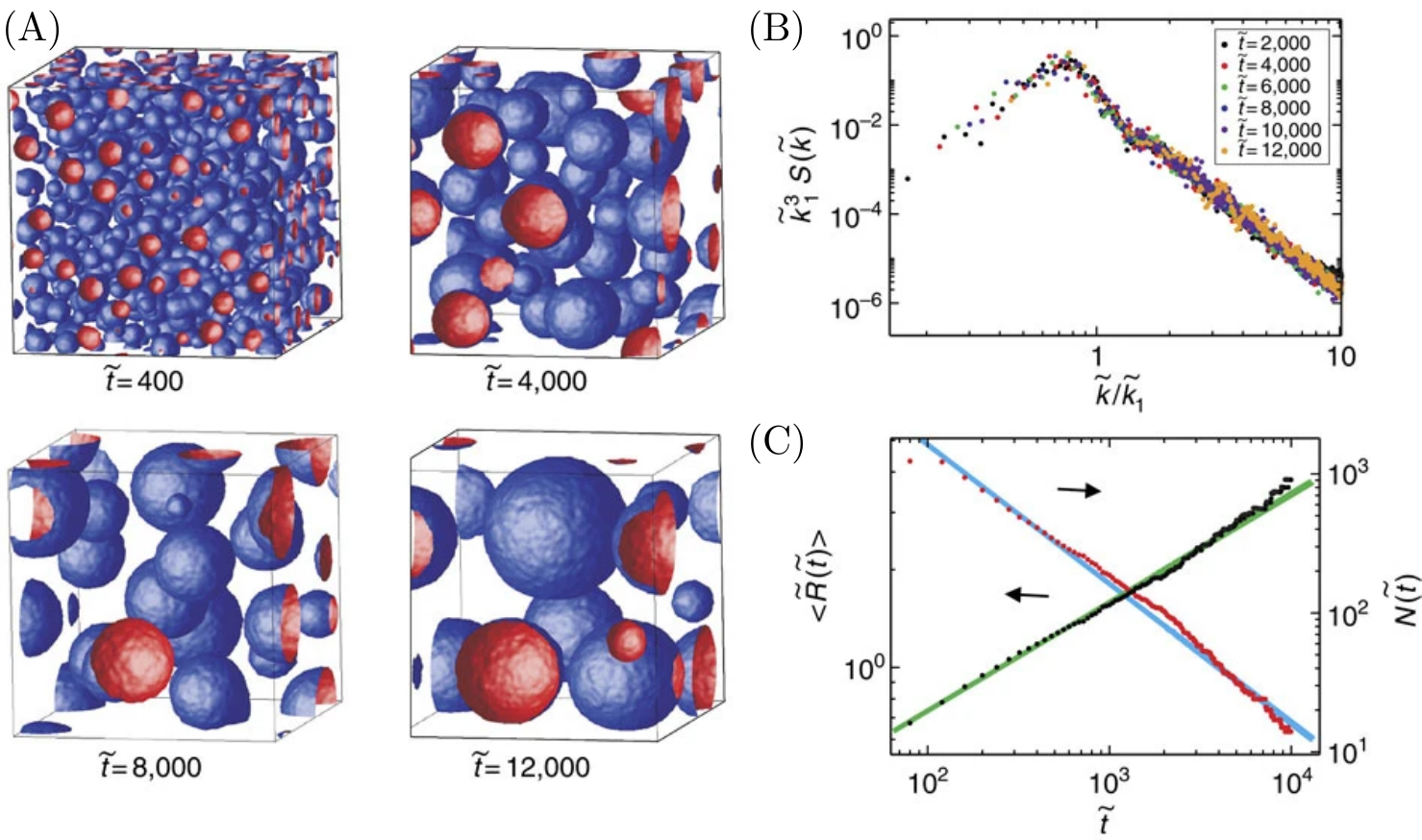}
    \caption{\textbf{(A)} The time evolution of the droplet structure in the late stages of the phase-separation process. \textbf{(B)} The time-normalized structure factor $S(\tilde{k})$. \textbf{(C)} Time evolution of the average droplet size $\left< \tilde{R}(\tilde{t}) \right>$ and the total number of droplets $N(\tilde{t})$ during phase separation.\cite{shimizu2015novel} Reproduced with permission.\cite{shimizu2015novel} Copyright 2015, Springer Nature Limited}
    \label{Fig.17}
\end{figure*}

\section{Hyperuniformity in spinodal decomposition}
\subsection{Hyperuniformity and Cahn-Hilliard Equations}
Spinodal decomposition is a well-known class of dynamics, in which systems undergo spontaneous phase separation without crossing any thermodynamic barrier. It has been investigated in various systems, such as binary alloys ~\cite{rundman1967early,langer1971theory,findik2012improvements}, polymer blends~\cite{de1980dynamics,bruder1992spinodal} and binary mixtures of molecular fluids~\cite{laradji1996spinodal,findik2012improvements,tanaka1996coarsening}, and the Cahn-Hilliard (CH) equation was proposed to describe the spinodal decomposition dynamics in those phase separating systems~\cite{cahn1958free}, which is also called the Model-B in the language of classifying critical phenomena~\cite{hohenberg1977theory}. In detail, the CH equation can be written as
\begin{widetext}
\begin{equation}\label{eq:CH}
\partial_t \phi(r, t)=- \nabla \cdot \mathbf{J}(r,t),\,{\rm with}\,\,\mathbf{J}(r,t)=-D\nabla\mu,\,{\rm and}\,\,\mu[\phi]=\frac{\delta\mathcal F[\phi]}{\delta \phi},
\end{equation}
\end{widetext}
where $\phi$ is a scalar field describing the phase ordering, e.g., the composition field in phase separating binary liquid, the density field in liquid-gas phase separation, or an indicator ranging from $-1$ to $1$ for different phase domains in the phase separation. $D$ is the diffusion coefficient, and $\mathbf{J}$ is the diffusive flux. $\mu[\phi]$ and $\mathcal F[\phi]$ are the chemical potential and free energy, respectively, which can be written in the functional form
\begin{equation}
\mathcal{F}[\phi]=\int {\rm d} r\left\{f[\phi]+\frac{\gamma_a}{2}|\nabla \phi|^2\right\},
\end{equation}
where the bulk free energy $f[\phi]$ is in the Landau-Ginzburg form of a forth order polynomial of $\phi$, e.g., $f[\phi]=\frac{1}{4}\left(\phi^2-1\right)^2$. By setting a random initial condition, e.g., set the initial concentration field $\phi(\mathbf{x},0)=0.5-a \delta$ with $\delta$ a random uniform distribution within $[0,1]$ and $a$ the initial fluctuation amplitude, Eq.~\ref{eq:CH} can be solved numerically to mimic a binary liquid after a rapid quench. When two phases separate, clusters grow through a coarsening process. After the Brownian diffusion at the early stage, time-evolved self-similar patterns appear and the domain size grows as $L\sim t^{1/3}$. A typical snapshots of the time-evolved pattern formation in Ref.~\cite{ma2017random} can be found in Fig.~\ref{Fig.16}. Besides, in the scaling regime, the spectral density $\tilde{\psi}(k)$ of field $\phi(\mathbf{x},t)$ could be collapsed by a time dependent scale $k_1=\int k\tilde{\psi}(k,t){\rm d} k/\int\tilde{\psi}(k,t){\rm d} k$ into a normalized $\tilde{\psi}(k) k_1^2(t)$ in 2D, as shown in Fig.~\ref{Fig.16}(B). As shown in Fig.~\ref{Fig.16}(C), one can find a strong two-media hyperuniform spectral density scaling $\tilde{\psi}({k \rightarrow 0})\sim k^4$ in the system, and similar phenomena of the converging spectral density $\tilde{\psi}(k)$ at small $k$ in spinodal decomposition was observed in a number of studies in past decades experimentally~\cite{komura1984dynamical,katano1984crossover,wiltzius1988spinodal,hashimoto1986late}, numerically~\cite{yeung1988scaling,shinozaki1993spinodal,kabrede2006spinodal,midya2020kinetics} and theoretically~\cite{yeung1988scaling,tomita1991preservation,mazenko1994growth,bray2002theory}.

\begin{figure*}[htb]
    \centering
    \includegraphics[width=0.9\textwidth] {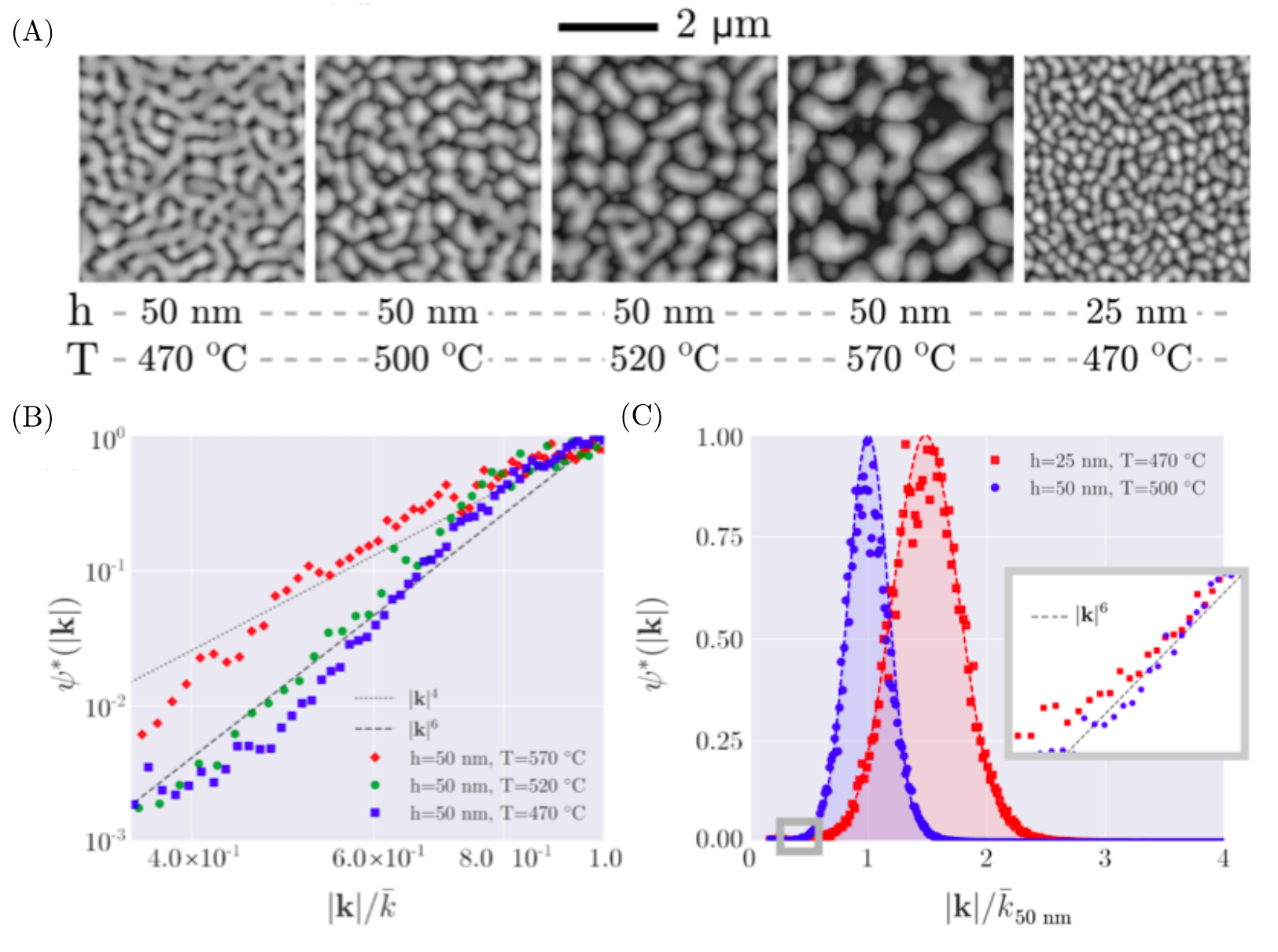}
    \caption{\textbf{(A)} AFM images of Ge film and annealed at different temperature. \textbf{(B)} The time-normalized spectral density $\psi^*(k)$ of height profile from experiments for similar samples annealed at different temperature. \textbf{(C)} Comparison of $\psi^*(k)$ of the patterns for the last and second sample of (A) with different thickness of the film.\cite{salvalaglio2020hyperuniform} Reproduced with permission.\cite{salvalaglio2020hyperuniform} Copyright 2020, American Physical Society.}
    \label{Fig.18}
\end{figure*}

\subsection{Beyond Cahn-Hilliard Equations}
\subsubsection{Hyperuniformity in Model-H with thermal noise}
In Ref.~\cite{shimizu2015novel}, Shimizu and Tanaka investigated the so-called Model-H phase-separation kinetics, which
comparing to the Model-B, considers not only the thermal diffusion but also hydrodynamic effects. Therefore, the Model-H has been commonly employed to characterize the spinodal decomposition phase separation dynamics in binary fluids.  
Generally, the Model-H for immiscible binary incompressible fluids mixture can be described by the equations
\begin{equation}\label{eq:Model-H}
\begin{aligned}
\frac{\partial \phi}{\partial t}&=-\mathbf{v} \cdot \nabla \phi+L \nabla^2 \frac{\delta(\beta \mathcal{H})}{\delta \phi}+\theta,\\
\rho \frac{\partial \mathbf{v}}{\partial t}&= - \rho( \mathbf{v}  \cdot \nabla) \mathbf{v}+\mathbf{F}_\phi-\nabla p+\eta \nabla^2 \mathbf{v}+\zeta,
\end{aligned}
\end{equation}
where $\phi$ is the composition, $v$ is the fluid velocity, $p$ is the pressure, $\rho$ stands for the mixture density, $\eta$ is the viscosity, $L$ is the kinetic coefficient. $\beta\mathcal{H}$ represents the free energy in the Landau-Ginzberg functional form, while $\theta$ and $\zeta$ are the Gaussian white noise.
By solving Eq.~\ref{eq:Model-H} with finite thermal noise and random initial conditions in 3D, typical snapshots are shown in Fig.~\ref{Fig.17}(A). In the scaling regime, where the domain size $\left< \tilde{R}\tilde{(t)}\right >\sim \tilde{t}^{1/3}$ as shown in Fig.~\ref{Fig.17}(C), the system exhibits time-evolved statistically self-similar patterns akin to the Model-B dynamics. The spectral density of $\phi$, normalized by wave number $\tilde{k}_1=\int \tilde{k} S(\tilde{k},t) {\rm d}k/\int\tilde{\psi}(\tilde{k},t){\rm d}\tilde{k}$, is plotted in Fig.~\ref{Fig.17}(B), revealing a strong hyperuniform spectral density scaling approximately $\sim \tilde{k}^4$ in the small $\tilde{k}$ regime.

\begin{figure*}[htb]
    \centering
    \includegraphics[width=0.95\textwidth] {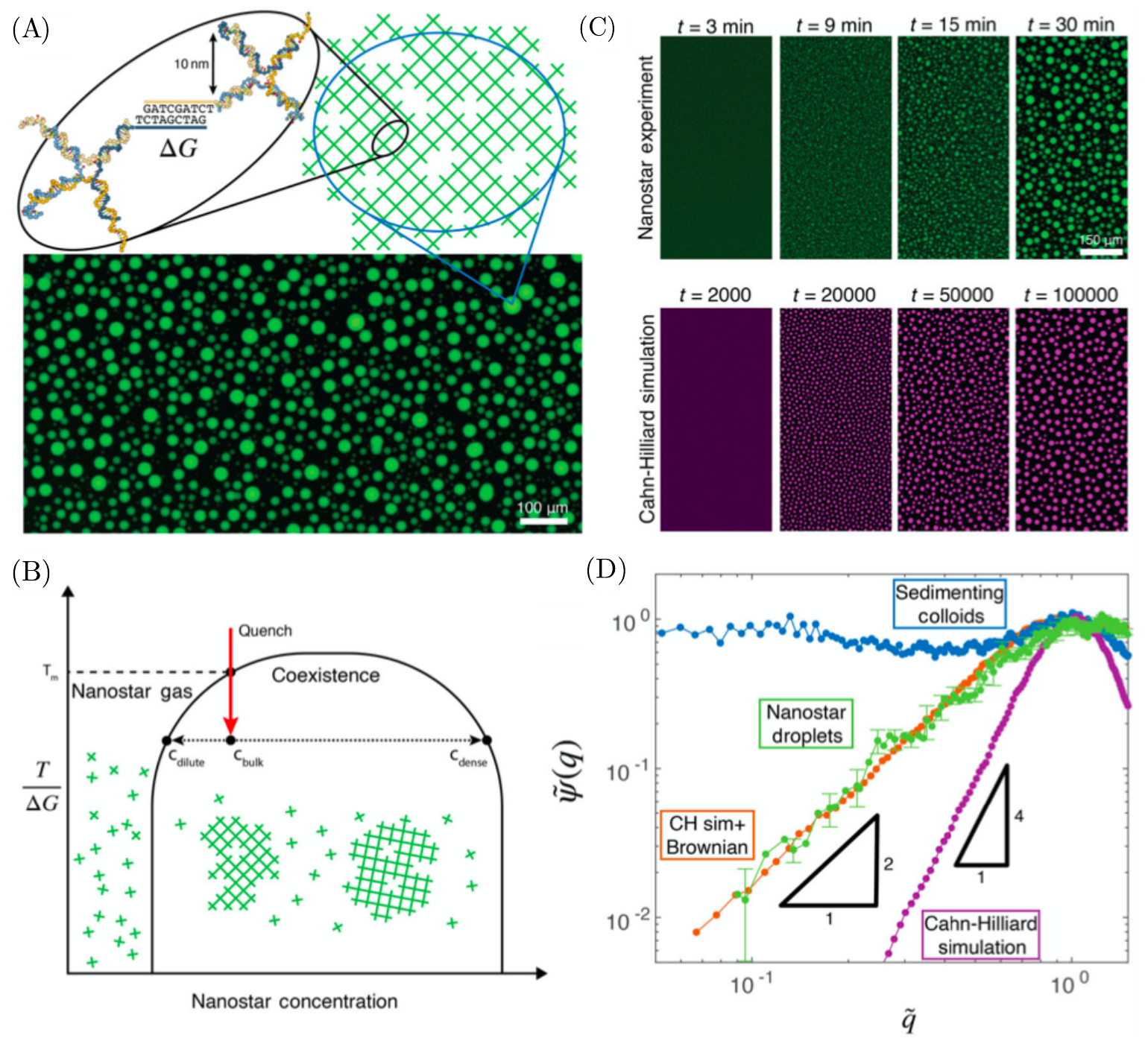}
    \caption{\textbf{(A)} Schematic diagram of DNA nanostar networks in the phase separated droplet and typical fluorescent image. \textbf{(B)} Schematic of the nanostar phase diagram. \textbf{(C)} Droplet formation dynamics from experiments on nanostars (top) and from Cahn-Hilliard simulations (bottom). \textbf{(D)} The scattering function $\tilde{\psi}(q)$ of the droplet intensity as a function of wave number $q$.\cite{wilken2023spatial} Reproduced with permission.\cite{wilken2023spatial} Copyright 2023, American Physical Society.}
    \label{Fig.19}
\end{figure*}

\subsubsection{Hyperuniformity in spinodal solid-state dewetting systems}

Physically similar to the spinodal decomposition in binary demixing fluids,
Salvalaglio \textit{et al.} experimentally investigated the spinodal solid-state dewetting system by preparing $\text{Si}_{1-x}\text{Ge}_x$ based thin films on ultrathin silicon and insulator substrates in a molecular beam epitaxy reactor under ultrahigh vacuum conditions~\cite{salvalaglio2020hyperuniform}. Upon annealing at high temperatures, the $\text{Si}_{1-x}\text{Ge}_x$ film undergoes a spinodal dewetting, and AFM images of various film thicknesses and annealing temperatures are shown in Fig.~\ref{Fig.18}(A). By analyzing the spectral density from the AFM images, a hyperuniform scaling is observed, as shown in Fig.~\ref{Fig.18}(B), and the hyperuniform scaling observed is different at different temperature during the dewetting process. 
\rt{The authors noted that dewetting at different annealing conditions may induce hyperuniform structures of different structure factor scalings. For example, the fourth sample shown in Fig.~\ref{Fig.18}(A) at a higher temperature $T=570 ^\circ {\rm C}$, exhibits the scaling $\psi^*(k) \sim k^4$ at small $k$, resembling the cases observed in Model-B and Model-H dynamics. In contrast, structures at lower temperatures, like those in the first ($T=470 ^\circ {\rm C}$) and third ($T=500 ^\circ {\rm C}$) samples shown in Fig.~\ref{Fig.18}(A), exhibit the scaling $\psi^*(k) \sim k^6$ at small $k$.}
 Additionally, the authors also considered the effect of the initial Ge film thickness, comparing spectral density plots as illustrated in Fig.~\ref{Fig.18}(C), while the peak positions are different, both structures exhibit the same scaling $\psi^*(k) \sim k^6$ at small $k$.

\subsubsection{Hyperuniformity in phase-separating DNA nanostar droplet systems}

Recently, experiments were performed to investigate the phase separating DNA nanostars by quenching a DNA nanostar solution from high temperature to below the DNA melting point, as illustrated in Fig.~\ref{Fig.19}(A) and (B)~\cite{wilken2023spatial}. 
Droplets formed through DNA sticky-end binding  sedimenting to the bottom of the sample chamber were observed via fluorescent imaging. In experiments, by quenching the DNA nanostar solution to different points in the coexisting region of the phase diagram shown in Fig.~\ref{Fig.19}(B), the authors observed slow condensation with droplet formation after a typical waiting time via nucleation and growth, and fast condensation with immediate droplet formation through spinodal decomposition.
By calculating the spectral density $\psi(q)$ from droplet configurations in fluorescent images, i.e., nanostar concentration field, a hyperuniform scaling $\psi(q\rightarrow 0)\sim q^2$ was observed during the fast condensation as shown in Fig.~\ref{Fig.19}(D), which was not observed during the slow condensation. For the fast condensation, similar to the Model-B dynamics, time-evolved self-similarity was observed, as shown in the upper panel in Fig.~\ref{Fig.19}(C). However, the obtained spectral density $\tilde{\psi}(q)$ scaling differed from the $\sim q^4$ scaling observed in Cahn-Hilliard simulations, as shown in Fig.~\ref{Fig.19}(D). The authors argued that this scaling difference is caused by the sedimentation process after droplets formation, which induced small Brownian displacements on the original structure produced by the spinodal decomposition.
To demonstrate this, the authors performed a modified Cahn-Hilliard simulation by displacing the droplets in random directions, with the displacement magnitudes $\delta$ proportional to the Stokes-Einstein mobility of the droplet $\delta \sim 1/R$ with $R$ the droplet radius. The simulation results shown in Fig.~\ref{Fig.19}(D) recovered the $q^2$ scaling, supporting their hypothesis that the scaling difference on spectral density $\tilde{\psi}(q)$ is caused by the noise in the sedimentation process.

\subsubsection{Hyperuniformity in active field theories for phase separation}
\begin{figure}[htb]
    \centering
    \includegraphics[width=0.45\textwidth] {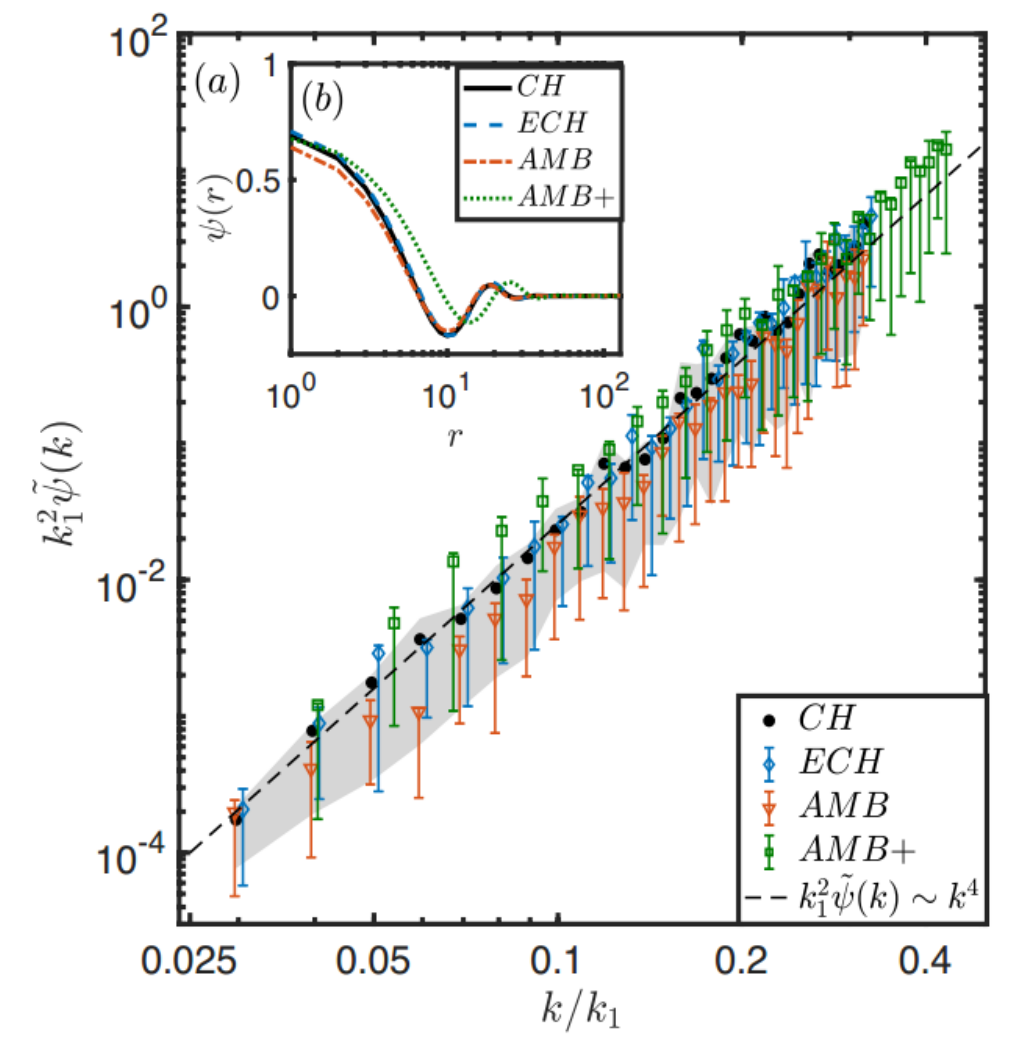}
    \caption{The average normalized scattering function $k_1^2\tilde{\psi}(k)$ of time-evolved patterns from CH equation (Model-B) and various active field theories (ECH, AMB and AMB+) as a function of wave number $k/k_1$. Inset: the auto-covariance function $\psi(r)$ for these patterns. $k_1 = 2\pi/l_1$, and $l_1$ characterizes the cluster length-scales, where $\psi(r)$ reaches its minimum.\cite{zheng2023universal} Reproduced with permission.\cite{zheng2023universal} Copyright 2023, Y. Zheng, M. A. Klatt, and H. L{\"o}wen.}
    \label{Fig.20}
\end{figure}
Beyond equilibrium systems, non-equilibrium field theories for active matter systems have received considerable scientific attention in past decades, given rich phenomena observed such as the motility-induced phase separation in active Brownian particles (ABPs)~\cite{cates_review,mazhan2020}. Over the last decade, several Cahn-Hilliard-like active field theories (AFTs) have been proposed to describe phase separation in systems like ABPs. Zheng \textit{et al.} recently investigated the spinodal decomposition in three recently developed AFTs~\cite{zheng2023universal}, which can be written in the general form:
\begin{equation}\label{eq:active_field}
\partial_t \phi(r, t)=D \nabla^2 \mu[\phi]+g_X \mathcal{A}_X[\phi],
\end{equation}
where $\phi$ is a scalar field, e.g., composition in phase separating binary liquid or density field in liquid-gas phase separation, and the functional term $g_X \mathcal{A}_X[\phi]$ signifies the non-equilibrium contribution from AFTs, while the other parts of the equation remain the same as in the Cahn-Hilliard equation. The authors studied three AFTs: the Effective Chan-Hilliard (ECH) model with $\mathcal{A}_\chi=\nabla \cdot(\phi \nabla \phi)$~\cite{speck2014effective}, the Active Model B (AMB) with $\mathcal{A}_\chi=\nabla^2\left[(\nabla \phi)^2\right]$~\cite{wittkowski2014scalar}, and the Active Model B+ (AMB+) with $\mathcal{A}_\chi=\nabla \cdot\left[\left(\nabla^2 \phi\right) \nabla \phi\right]$~\cite{tjhung2018cluster}.
By numerically solving Eq.~\ref{eq:active_field} for various AFTs with critical quenching in 2D, \rt{and normalizing the spectral density from the result with a time-dependent length-scale $k_1 = 2\pi/l_1$, where $l_1$ characterizes the cluster length-scales in the scaling regime}, all three active field models exhibit the same universal hyperuniform scaling $\sim k^4$ at $k \rightarrow 0$, as shown in Fig.~\ref{Fig.20}. This suggests that their large-scale structural properties are similar. The authors also considered the fluctuation $\sigma^2(R)$ and higher-order correlation moments (skewness and kurtosis) at the length-scale $R$ by coarse-graining $\phi(r)$. They found that the variance for the CH model and all active models decay with the same scaling $\sigma^2(R)\sim R^{-3}$ at large length-scale $R \rightarrow \infty$, consistent with the hyperuniform spectral density. However, they also found that the higher order moments (skewness and kurtosis) for these models do not exhibit any universal behavior, indicating activity-dependent higher-order correlations.

\section{Concluding remarks}
Disordered hyperuniformity has been introduced for more than twenty years, and it is known that in equilibrium systems to realize disordered hyperuniform states one has to use delicately designed long range interactions~\cite{torquato2018hyperuniform}, which is highly challenging experimentally. However, long range interactions are not necessary in disordered hyperuniform states in non-equilibrium systems, and in this article, we have briefly reviewed the progress on non-equilibrium dynamic hyperuniform states, which has become an emerging research direction on hyperuniformity in the past decade. We first introduced the background of random organization model, which was originally proposed to describe the chaotic dynamics and absorbing phase transition in sheared non-Brownian suspensions. Afterwards, we showed the critical absorbing transition point in random organization model and other CDP class models that were found to be hyperuniform. Moreover, we also mentioned that by adding a center-of-mass conservation to the random organization dynamics, the active state would have a stronger hyperuniformity, which could be understood as the unusually suppressed noise term in the dynamical mean-field equation. However, the center-of-mass conserved noise is unphysical in nature, as all particles have inertia. Therefore, we introduced the non-equilibrium hyperuniform fluid found in driven-dissipative particles with finite inertia in dry systems such as reactive hard-sphere fluids and chiral active fluids, which undergo similar absorbing phase transition as in the random organization, with critical point and active state hyperuniformity. Moreover, we reviewed driven particles in wet systems such as the circularly swimming algae system and driven motor systems, in which long range hydrodynamic interactions induced hyperuniformity. Lastly, we also recapped about some recent literature on  hyperuniform structures found in phase separating systems through spinodal decomposition, including numerical systems like the Model-B, Model-H, and several active field model systems, as well as experimental systems like phase-separating DNA nanostar droplets and dewetting Ge-based film systems. 

All those non-equilibrium systems offer new possibilities for realizing hyperuniform structures experimentally, and pave the way for fabricating hyperuniform functional materials. 
\rt{Future research may consider finding hyperuniform structures in other systems that share similar characteristics with those mentioned in this review. 
For example, just as hyperuniform critical states were found in various systems with CDP class absorbing phase transitions, it would be interesting to investigate systems with similar non-equilibrium transitions, such as general reversible-irreversible transitions~\cite{reichhardt2023reversible}, yielding transitions~\cite{jocteur2024yielding}, depinning transitions~\cite{reichhardt2009random,joerg2024hyperuniformity}, and even transitions involving time crystals~\cite{yao2020classical}, etc.
Additionally, as discussed in Section II.C and III, noises such as the center-of-mass conserved (CMC) noise or the momentum-conserved reciprocal excitations are known to be the origin of hyperuniform structures with $S(k\rightarrow 0)\sim k^2$ in various driven-dissipative steady states~\cite{hexner2017noise,lei2019nonequilibrium,lei2019hydrodynamics}. This idea of manipulating the noise form has recently been generalized in Model-A and Model-B field theories~\cite{ikeda2023correlated}, which appears to be a promising approach for designing hyperuniform systems. 
Furthermore, it is noteworthy that CMC noise was found to be related to the breaking of the Mermin-Wagner theorem in low dimensions, allowing the existence of 1D or 2D crystals~\cite{berthier2d,ikeda2023harmonic,ikeda2023correlated,kuroda2024}. A recent study discussed hyperuniformity and center-of-mass conservation in Cahn-Hilliard dynamics, suggesting that a stronger hyperuniform $k^4$ scaling for $k\rightarrow 0$ is enhanced by the unique CMC noise, other than $k^2$ scaling if there is conventional thermal noise ~\cite{de2024hyperuniformity}. Although most of the systems reviewed here are classical, dynamic hyperuniform states are also expected in quantum systems. For example, the vortex systems in super-conducting materials~\cite{le2017enhanced,rumi2019hyperuniform,llorens2020disordered,sanchez2023disordered}, as well as boundary-driven open quantum chains~\cite{carollo2017fluctuating} were found to have a dynamic hyperuniform state~\cite{jack2015hyperuniformity}. In addition, designing ``temporal hyperuniform'' systems rather than the spatial ones by performing time series analysis may also be an interesting future direction, which could be used for exploring hidden temporal orders in networks, economics, or social science data.}

\begin{acknowledgments}
We thank Kay Jörg Wiese and Salvatore Torquato for helpful discussions and constructive comments on the manuscript.
This work is supported by the Academic Research Fund from the Singapore Ministry of Education (RG59/21, RG151/23 and MOE2019-T2-2-010), and the National Research Foundation, Singapore, under its 29th Competitive Research Program (CRP) Call (Award No. NRF-CRP29-2022-0002).
\end{acknowledgments}

\bibliography{review}

\begin{thebibliography}{117}%
\makeatletter
\providecommand \@ifxundefined [1]{%
 \@ifx{#1\undefined}
}%
\providecommand \@ifnum [1]{%
 \ifnum #1\expandafter \@firstoftwo
 \else \expandafter \@secondoftwo
 \fi
}%
\providecommand \@ifx [1]{%
 \ifx #1\expandafter \@firstoftwo
 \else \expandafter \@secondoftwo
 \fi
}%
\providecommand \natexlab [1]{#1}%
\providecommand \enquote  [1]{``#1''}%
\providecommand \bibnamefont  [1]{#1}%
\providecommand \bibfnamefont [1]{#1}%
\providecommand \citenamefont [1]{#1}%
\providecommand \href@noop [0]{\@secondoftwo}%
\providecommand \href [0]{\begingroup \@sanitize@url \@href}%
\providecommand \@href[1]{\@@startlink{#1}\@@href}%
\providecommand \@@href[1]{\endgroup#1\@@endlink}%
\providecommand \@sanitize@url [0]{\catcode `\\12\catcode `\$12\catcode
  `\&12\catcode `\#12\catcode `\^12\catcode `\_12\catcode `\%12\relax}%
\providecommand \@@startlink[1]{}%
\providecommand \@@endlink[0]{}%
\providecommand \url  [0]{\begingroup\@sanitize@url \@url }%
\providecommand \@url [1]{\endgroup\@href {#1}{\urlprefix }}%
\providecommand \urlprefix  [0]{URL }%
\providecommand \Eprint [0]{\href }%
\providecommand \doibase [0]{http://dx.doi.org/}%
\providecommand \selectlanguage [0]{\@gobble}%
\providecommand \bibinfo  [0]{\@secondoftwo}%
\providecommand \bibfield  [0]{\@secondoftwo}%
\providecommand \translation [1]{[#1]}%
\providecommand \BibitemOpen [0]{}%
\providecommand \bibitemStop [0]{}%
\providecommand \bibitemNoStop [0]{.\EOS\space}%
\providecommand \EOS [0]{\spacefactor3000\relax}%
\providecommand \BibitemShut  [1]{\csname bibitem#1\endcsname}%
\let\auto@bib@innerbib\@empty
\bibitem [{\citenamefont {Torquato}\ and\ \citenamefont
  {Stillinger}(2003)}]{torquato2003local}%
  \BibitemOpen
  \bibfield  {author} {\bibinfo {author} {\bibfnamefont {S.}~\bibnamefont
  {Torquato}}\ and\ \bibinfo {author} {\bibfnamefont {F.~H.}\ \bibnamefont
  {Stillinger}},\ }\bibfield  {title} {\enquote {\bibinfo {title} {Local
  density fluctuations, hyperuniformity, and order metrics},}\ }\href@noop {}
  {\bibfield  {journal} {\bibinfo  {journal} {Phys. Rev. E}\ }\textbf {\bibinfo
  {volume} {68}},\ \bibinfo {pages} {041113} (\bibinfo {year}
  {2003})}\BibitemShut {NoStop}%
\bibitem [{\citenamefont {Donev}\ \emph {et~al.}(2005)\citenamefont {Donev},
  \citenamefont {Stillinger},\ and\ \citenamefont
  {Torquato}}]{donev2005unexpected}%
  \BibitemOpen
  \bibfield  {author} {\bibinfo {author} {\bibfnamefont {A.}~\bibnamefont
  {Donev}}, \bibinfo {author} {\bibfnamefont {F.~H.}\ \bibnamefont
  {Stillinger}}, \ and\ \bibinfo {author} {\bibfnamefont {S.}~\bibnamefont
  {Torquato}},\ }\bibfield  {title} {\enquote {\bibinfo {title} {Unexpected
  density fluctuations in jammed disordered sphere packings},}\ }\href@noop {}
  {\bibfield  {journal} {\bibinfo  {journal} {Phys. Rev. Lett.}\ }\textbf
  {\bibinfo {volume} {95}},\ \bibinfo {pages} {090604} (\bibinfo {year}
  {2005})}\BibitemShut {NoStop}%
\bibitem [{\citenamefont {Kram}\ \emph {et~al.}(2010)\citenamefont {Kram},
  \citenamefont {Mantey},\ and\ \citenamefont {Corbo}}]{kram2010avian}%
  \BibitemOpen
  \bibfield  {author} {\bibinfo {author} {\bibfnamefont {Y.~A.}\ \bibnamefont
  {Kram}}, \bibinfo {author} {\bibfnamefont {S.}~\bibnamefont {Mantey}}, \ and\
  \bibinfo {author} {\bibfnamefont {J.~C.}\ \bibnamefont {Corbo}},\ }\bibfield
  {title} {\enquote {\bibinfo {title} {Avian cone photoreceptors tile the
  retina as five independent, self-organizing mosaics},}\ }\href@noop {}
  {\bibfield  {journal} {\bibinfo  {journal} {PloS one}\ }\textbf {\bibinfo
  {volume} {5}},\ \bibinfo {pages} {e8992} (\bibinfo {year}
  {2010})}\BibitemShut {NoStop}%
\bibitem [{\citenamefont {Jiao}\ \emph {et~al.}(2014)\citenamefont {Jiao},
  \citenamefont {Lau}, \citenamefont {Hatzikirou}, \citenamefont
  {Meyer-Hermann}, \citenamefont {Corbo},\ and\ \citenamefont
  {Torquato}}]{jiao2014avian}%
  \BibitemOpen
  \bibfield  {author} {\bibinfo {author} {\bibfnamefont {Y.}~\bibnamefont
  {Jiao}}, \bibinfo {author} {\bibfnamefont {T.}~\bibnamefont {Lau}}, \bibinfo
  {author} {\bibfnamefont {H.}~\bibnamefont {Hatzikirou}}, \bibinfo {author}
  {\bibfnamefont {M.}~\bibnamefont {Meyer-Hermann}}, \bibinfo {author}
  {\bibfnamefont {J.~C.}\ \bibnamefont {Corbo}}, \ and\ \bibinfo {author}
  {\bibfnamefont {S.}~\bibnamefont {Torquato}},\ }\bibfield  {title} {\enquote
  {\bibinfo {title} {Avian photoreceptor patterns represent a disordered
  hyperuniform solution to a multiscale packing problem},}\ }\href@noop {}
  {\bibfield  {journal} {\bibinfo  {journal} {Phys. Rev. E}\ }\textbf {\bibinfo
  {volume} {89}},\ \bibinfo {pages} {022721} (\bibinfo {year}
  {2014})}\BibitemShut {NoStop}%
\bibitem [{\citenamefont {Gabrielli}\ \emph {et~al.}(2002)\citenamefont
  {Gabrielli}, \citenamefont {Joyce},\ and\ \citenamefont
  {Labini}}]{gabrielli2002glass}%
  \BibitemOpen
  \bibfield  {author} {\bibinfo {author} {\bibfnamefont {A.}~\bibnamefont
  {Gabrielli}}, \bibinfo {author} {\bibfnamefont {M.}~\bibnamefont {Joyce}}, \
  and\ \bibinfo {author} {\bibfnamefont {F.~S.}\ \bibnamefont {Labini}},\
  }\bibfield  {title} {\enquote {\bibinfo {title} {Glass-like universe:
  Real-space correlation properties of standard cosmological models},}\
  }\href@noop {} {\bibfield  {journal} {\bibinfo  {journal} {Phys. Rev. D}\
  }\textbf {\bibinfo {volume} {65}},\ \bibinfo {pages} {083523} (\bibinfo
  {year} {2002})}\BibitemShut {NoStop}%
\bibitem [{\citenamefont {Florescu}\ \emph {et~al.}(2009)\citenamefont
  {Florescu}, \citenamefont {Torquato},\ and\ \citenamefont
  {Steinhardt}}]{florescu2009designer}%
  \BibitemOpen
  \bibfield  {author} {\bibinfo {author} {\bibfnamefont {M.}~\bibnamefont
  {Florescu}}, \bibinfo {author} {\bibfnamefont {S.}~\bibnamefont {Torquato}},
  \ and\ \bibinfo {author} {\bibfnamefont {P.~J.}\ \bibnamefont {Steinhardt}},\
  }\bibfield  {title} {\enquote {\bibinfo {title} {Designer disordered
  materials with large, complete photonic band gaps},}\ }\href@noop {}
  {\bibfield  {journal} {\bibinfo  {journal} {Proc. Natl. Acad. Sci. U.S.A.}\
  }\textbf {\bibinfo {volume} {106}},\ \bibinfo {pages} {20658--20663}
  (\bibinfo {year} {2009})}\BibitemShut {NoStop}%
\bibitem [{\citenamefont {Man}\ \emph {et~al.}(2013)\citenamefont {Man},
  \citenamefont {Florescu}, \citenamefont {Williamson}, \citenamefont {He},
  \citenamefont {Hashemizad}, \citenamefont {Leung}, \citenamefont {Liner},
  \citenamefont {Torquato}, \citenamefont {Chaikin},\ and\ \citenamefont
  {Steinhardt}}]{man2013isotropic}%
  \BibitemOpen
  \bibfield  {author} {\bibinfo {author} {\bibfnamefont {W.}~\bibnamefont
  {Man}}, \bibinfo {author} {\bibfnamefont {M.}~\bibnamefont {Florescu}},
  \bibinfo {author} {\bibfnamefont {E.~P.}\ \bibnamefont {Williamson}},
  \bibinfo {author} {\bibfnamefont {Y.}~\bibnamefont {He}}, \bibinfo {author}
  {\bibfnamefont {S.~R.}\ \bibnamefont {Hashemizad}}, \bibinfo {author}
  {\bibfnamefont {B.~Y.}\ \bibnamefont {Leung}}, \bibinfo {author}
  {\bibfnamefont {D.~R.}\ \bibnamefont {Liner}}, \bibinfo {author}
  {\bibfnamefont {S.}~\bibnamefont {Torquato}}, \bibinfo {author}
  {\bibfnamefont {P.~M.}\ \bibnamefont {Chaikin}}, \ and\ \bibinfo {author}
  {\bibfnamefont {P.~J.}\ \bibnamefont {Steinhardt}},\ }\bibfield  {title}
  {\enquote {\bibinfo {title} {Isotropic band gaps and freeform waveguides
  observed in hyperuniform disordered photonic solids},}\ }\href@noop {}
  {\bibfield  {journal} {\bibinfo  {journal} {Proc. Natl. Acad. Sci. U.S.A.}\
  }\textbf {\bibinfo {volume} {110}},\ \bibinfo {pages} {15886--15891}
  (\bibinfo {year} {2013})}\BibitemShut {NoStop}%
\bibitem [{\citenamefont {Leseur}\ \emph {et~al.}(2016)\citenamefont {Leseur},
  \citenamefont {Pierrat},\ and\ \citenamefont {Carminati}}]{leseur2016high}%
  \BibitemOpen
  \bibfield  {author} {\bibinfo {author} {\bibfnamefont {O.}~\bibnamefont
  {Leseur}}, \bibinfo {author} {\bibfnamefont {R.}~\bibnamefont {Pierrat}}, \
  and\ \bibinfo {author} {\bibfnamefont {R.}~\bibnamefont {Carminati}},\
  }\bibfield  {title} {\enquote {\bibinfo {title} {High-density hyperuniform
  materials can be transparent},}\ }\href@noop {} {\bibfield  {journal}
  {\bibinfo  {journal} {Optica}\ }\textbf {\bibinfo {volume} {3}},\ \bibinfo
  {pages} {763--767} (\bibinfo {year} {2016})}\BibitemShut {NoStop}%
\bibitem [{\citenamefont {Hansen}\ and\ \citenamefont
  {McDonald}(2013)}]{hansen2013theory}%
  \BibitemOpen
  \bibfield  {author} {\bibinfo {author} {\bibfnamefont {J.~P.}\ \bibnamefont
  {Hansen}}\ and\ \bibinfo {author} {\bibfnamefont {I.~R.}\ \bibnamefont
  {McDonald}},\ }\href@noop {} {\emph {\bibinfo {title} {Theory of simple
  liquids: with applications to soft matter}}}\ (\bibinfo  {publisher}
  {Academic press},\ \bibinfo {year} {2013})\BibitemShut {NoStop}%
\bibitem [{\citenamefont {Torquato}(2018)}]{torquato2018hyperuniform}%
  \BibitemOpen
  \bibfield  {author} {\bibinfo {author} {\bibfnamefont {S.}~\bibnamefont
  {Torquato}},\ }\bibfield  {title} {\enquote {\bibinfo {title} {Hyperuniform
  states of matter},}\ }\href@noop {} {\bibfield  {journal} {\bibinfo
  {journal} {Phys. Rep.}\ }\textbf {\bibinfo {volume} {745}},\ \bibinfo {pages}
  {1--95} (\bibinfo {year} {2018})}\BibitemShut {NoStop}%
\bibitem [{\citenamefont {Torquato}(2016)}]{torquato2016hyperuniformity}%
  \BibitemOpen
  \bibfield  {author} {\bibinfo {author} {\bibfnamefont {S.}~\bibnamefont
  {Torquato}},\ }\bibfield  {title} {\enquote {\bibinfo {title}
  {Hyperuniformity and its generalizations},}\ }\href@noop {} {\bibfield
  {journal} {\bibinfo  {journal} {Phys. Rev. E}\ }\textbf {\bibinfo {volume}
  {94}},\ \bibinfo {pages} {022122} (\bibinfo {year} {2016})}\BibitemShut
  {NoStop}%
\bibitem [{\citenamefont {Zachary}\ and\ \citenamefont
  {Torquato}(2009)}]{zachary2009hyperuniformity}%
  \BibitemOpen
  \bibfield  {author} {\bibinfo {author} {\bibfnamefont {C.~E.}\ \bibnamefont
  {Zachary}}\ and\ \bibinfo {author} {\bibfnamefont {S.}~\bibnamefont
  {Torquato}},\ }\bibfield  {title} {\enquote {\bibinfo {title}
  {Hyperuniformity in point patterns and two-phase random heterogeneous
  media},}\ }\href@noop {} {\bibfield  {journal} {\bibinfo  {journal} {J. Stat.
  Mech.: Theory Exp.}\ }\textbf {\bibinfo {volume} {2009}},\ \bibinfo {pages}
  {P12015} (\bibinfo {year} {2009})}\BibitemShut {NoStop}%
\bibitem [{\citenamefont {Lomba}\ \emph {et~al.}(2018)\citenamefont {Lomba},
  \citenamefont {Weis},\ and\ \citenamefont {Torquato}}]{lomba2018disordered}%
  \BibitemOpen
  \bibfield  {author} {\bibinfo {author} {\bibfnamefont {E.}~\bibnamefont
  {Lomba}}, \bibinfo {author} {\bibfnamefont {J.~J.}\ \bibnamefont {Weis}}, \
  and\ \bibinfo {author} {\bibfnamefont {S.}~\bibnamefont {Torquato}},\
  }\bibfield  {title} {\enquote {\bibinfo {title} {Disordered
  multihyperuniformity derived from binary plasmas},}\ }\href@noop {}
  {\bibfield  {journal} {\bibinfo  {journal} {Phys. Rev. E}\ }\textbf {\bibinfo
  {volume} {97}},\ \bibinfo {pages} {010102} (\bibinfo {year}
  {2018})}\BibitemShut {NoStop}%
\bibitem [{\citenamefont {Lomba}\ \emph {et~al.}(2017)\citenamefont {Lomba},
  \citenamefont {Weis},\ and\ \citenamefont {Torquato}}]{lomba2017disordered}%
  \BibitemOpen
  \bibfield  {author} {\bibinfo {author} {\bibfnamefont {E.}~\bibnamefont
  {Lomba}}, \bibinfo {author} {\bibfnamefont {J.~J.}\ \bibnamefont {Weis}}, \
  and\ \bibinfo {author} {\bibfnamefont {S.}~\bibnamefont {Torquato}},\
  }\bibfield  {title} {\enquote {\bibinfo {title} {Disordered hyperuniformity
  in two-component nonadditive hard-disk plasmas},}\ }\href@noop {} {\bibfield
  {journal} {\bibinfo  {journal} {Phys. Rev. E}\ }\textbf {\bibinfo {volume}
  {96}},\ \bibinfo {pages} {062126} (\bibinfo {year} {2017})}\BibitemShut
  {NoStop}%
\bibitem [{\citenamefont {Zachary}\ \emph {et~al.}(2011)\citenamefont
  {Zachary}, \citenamefont {Jiao},\ and\ \citenamefont
  {Torquato}}]{zachary2011hyperuniform}%
  \BibitemOpen
  \bibfield  {author} {\bibinfo {author} {\bibfnamefont {C.~E.}\ \bibnamefont
  {Zachary}}, \bibinfo {author} {\bibfnamefont {Y.}~\bibnamefont {Jiao}}, \
  and\ \bibinfo {author} {\bibfnamefont {S.}~\bibnamefont {Torquato}},\
  }\bibfield  {title} {\enquote {\bibinfo {title} {Hyperuniform long-range
  correlations are a signature of disordered jammed hard-particle packings},}\
  }\href@noop {} {\bibfield  {journal} {\bibinfo  {journal} {Phys. Rev. Lett.}\
  }\textbf {\bibinfo {volume} {106}},\ \bibinfo {pages} {178001} (\bibinfo
  {year} {2011})}\BibitemShut {NoStop}%
\bibitem [{\citenamefont {Ricouvier}\ \emph {et~al.}(2017)\citenamefont
  {Ricouvier}, \citenamefont {Pierrat}, \citenamefont {Carminati},
  \citenamefont {Tabeling},\ and\ \citenamefont
  {Yazhgur}}]{ricouvier2017optimizing}%
  \BibitemOpen
  \bibfield  {author} {\bibinfo {author} {\bibfnamefont {J.}~\bibnamefont
  {Ricouvier}}, \bibinfo {author} {\bibfnamefont {R.}~\bibnamefont {Pierrat}},
  \bibinfo {author} {\bibfnamefont {R.}~\bibnamefont {Carminati}}, \bibinfo
  {author} {\bibfnamefont {P.}~\bibnamefont {Tabeling}}, \ and\ \bibinfo
  {author} {\bibfnamefont {P.}~\bibnamefont {Yazhgur}},\ }\bibfield  {title}
  {\enquote {\bibinfo {title} {Optimizing hyperuniformity in self-assembled
  bidisperse emulsions},}\ }\href@noop {} {\bibfield  {journal} {\bibinfo
  {journal} {Phys. Rev. Lett.}\ }\textbf {\bibinfo {volume} {119}},\ \bibinfo
  {pages} {208001} (\bibinfo {year} {2017})}\BibitemShut {NoStop}%
\bibitem [{\citenamefont {Dreyfus}\ \emph {et~al.}(2015)\citenamefont
  {Dreyfus}, \citenamefont {Xu}, \citenamefont {Still}, \citenamefont {Hough},
  \citenamefont {Yodh},\ and\ \citenamefont
  {Torquato}}]{dreyfus2015diagnosing}%
  \BibitemOpen
  \bibfield  {author} {\bibinfo {author} {\bibfnamefont {R.}~\bibnamefont
  {Dreyfus}}, \bibinfo {author} {\bibfnamefont {Y.}~\bibnamefont {Xu}},
  \bibinfo {author} {\bibfnamefont {T.}~\bibnamefont {Still}}, \bibinfo
  {author} {\bibfnamefont {L.~A.}\ \bibnamefont {Hough}}, \bibinfo {author}
  {\bibfnamefont {A.}~\bibnamefont {Yodh}}, \ and\ \bibinfo {author}
  {\bibfnamefont {S.}~\bibnamefont {Torquato}},\ }\bibfield  {title} {\enquote
  {\bibinfo {title} {Diagnosing hyperuniformity in two-dimensional, disordered,
  jammed packings of soft spheres},}\ }\href@noop {} {\bibfield  {journal}
  {\bibinfo  {journal} {Phys. Rev. E}\ }\textbf {\bibinfo {volume} {91}},\
  \bibinfo {pages} {012302} (\bibinfo {year} {2015})}\BibitemShut {NoStop}%
\bibitem [{\citenamefont {Berthier}\ \emph {et~al.}(2011)\citenamefont
  {Berthier}, \citenamefont {Chaudhuri}, \citenamefont {Coulais}, \citenamefont
  {Dauchot},\ and\ \citenamefont {Sollich}}]{berthier2011suppressed}%
  \BibitemOpen
  \bibfield  {author} {\bibinfo {author} {\bibfnamefont {L.}~\bibnamefont
  {Berthier}}, \bibinfo {author} {\bibfnamefont {P.}~\bibnamefont {Chaudhuri}},
  \bibinfo {author} {\bibfnamefont {C.}~\bibnamefont {Coulais}}, \bibinfo
  {author} {\bibfnamefont {O.}~\bibnamefont {Dauchot}}, \ and\ \bibinfo
  {author} {\bibfnamefont {P.}~\bibnamefont {Sollich}},\ }\bibfield  {title}
  {\enquote {\bibinfo {title} {Suppressed compressibility at large scale in
  jammed packings of size-disperse spheres},}\ }\href@noop {} {\bibfield
  {journal} {\bibinfo  {journal} {Phys. Rev. Lett.}\ }\textbf {\bibinfo
  {volume} {106}},\ \bibinfo {pages} {120601} (\bibinfo {year}
  {2011})}\BibitemShut {NoStop}%
\bibitem [{\citenamefont {Zhang}\ \emph {et~al.}(2016)\citenamefont {Zhang},
  \citenamefont {Stillinger},\ and\ \citenamefont
  {Torquato}}]{zhang2016perfect}%
  \BibitemOpen
  \bibfield  {author} {\bibinfo {author} {\bibfnamefont {G.}~\bibnamefont
  {Zhang}}, \bibinfo {author} {\bibfnamefont {F.~H.}\ \bibnamefont
  {Stillinger}}, \ and\ \bibinfo {author} {\bibfnamefont {S.}~\bibnamefont
  {Torquato}},\ }\bibfield  {title} {\enquote {\bibinfo {title} {The perfect
  glass paradigm: Disordered hyperuniform glasses down to absolute zero},}\
  }\href@noop {} {\bibfield  {journal} {\bibinfo  {journal} {Sci. Rep.}\
  }\textbf {\bibinfo {volume} {6}},\ \bibinfo {pages} {36963} (\bibinfo {year}
  {2016})}\BibitemShut {NoStop}%
\bibitem [{\citenamefont {Weijs}\ \emph {et~al.}(2015)\citenamefont {Weijs},
  \citenamefont {Jeanneret}, \citenamefont {Dreyfus},\ and\ \citenamefont
  {Bartolo}}]{weijs2015emergent}%
  \BibitemOpen
  \bibfield  {author} {\bibinfo {author} {\bibfnamefont {J.~H.}\ \bibnamefont
  {Weijs}}, \bibinfo {author} {\bibfnamefont {R.}~\bibnamefont {Jeanneret}},
  \bibinfo {author} {\bibfnamefont {R.}~\bibnamefont {Dreyfus}}, \ and\
  \bibinfo {author} {\bibfnamefont {D.}~\bibnamefont {Bartolo}},\ }\bibfield
  {title} {\enquote {\bibinfo {title} {Emergent hyperuniformity in periodically
  driven emulsions},}\ }\href@noop {} {\bibfield  {journal} {\bibinfo
  {journal} {Phys. Rev. Lett.}\ }\textbf {\bibinfo {volume} {115}},\ \bibinfo
  {pages} {108301} (\bibinfo {year} {2015})}\BibitemShut {NoStop}%
\bibitem [{\citenamefont {Weijs}\ and\ \citenamefont
  {Bartolo}(2017)}]{bartolo2017}%
  \BibitemOpen
  \bibfield  {author} {\bibinfo {author} {\bibfnamefont {J.~H.}\ \bibnamefont
  {Weijs}}\ and\ \bibinfo {author} {\bibfnamefont {D.}~\bibnamefont
  {Bartolo}},\ }\bibfield  {title} {\enquote {\bibinfo {title} {Mixing by
  unstirring: Hyperuniform dispersion of interacting particles upon chaotic
  advection},}\ }\href@noop {} {\bibfield  {journal} {\bibinfo  {journal}
  {Phys. Rev. Lett.}\ }\textbf {\bibinfo {volume} {119}},\ \bibinfo {pages}
  {048002} (\bibinfo {year} {2017})}\BibitemShut {NoStop}%
\bibitem [{\citenamefont {Tjhung}\ and\ \citenamefont
  {Berthier}(2015)}]{tjhung2015hyperuniform}%
  \BibitemOpen
  \bibfield  {author} {\bibinfo {author} {\bibfnamefont {E.}~\bibnamefont
  {Tjhung}}\ and\ \bibinfo {author} {\bibfnamefont {L.}~\bibnamefont
  {Berthier}},\ }\bibfield  {title} {\enquote {\bibinfo {title} {Hyperuniform
  density fluctuations and diverging dynamic correlations in periodically
  driven colloidal suspensions},}\ }\href@noop {} {\bibfield  {journal}
  {\bibinfo  {journal} {Phys. Rev. Lett.}\ }\textbf {\bibinfo {volume} {114}},\
  \bibinfo {pages} {148301} (\bibinfo {year} {2015})}\BibitemShut {NoStop}%
\bibitem [{\citenamefont {Wilken}\ \emph {et~al.}(2020)\citenamefont {Wilken},
  \citenamefont {Guerra}, \citenamefont {Pine},\ and\ \citenamefont
  {Chaikin}}]{wilken2020hyperuniform}%
  \BibitemOpen
  \bibfield  {author} {\bibinfo {author} {\bibfnamefont {S.}~\bibnamefont
  {Wilken}}, \bibinfo {author} {\bibfnamefont {R.~E.}\ \bibnamefont {Guerra}},
  \bibinfo {author} {\bibfnamefont {D.~J.}\ \bibnamefont {Pine}}, \ and\
  \bibinfo {author} {\bibfnamefont {P.~M.}\ \bibnamefont {Chaikin}},\
  }\bibfield  {title} {\enquote {\bibinfo {title} {Hyperuniform structures
  formed by shearing colloidal suspensions},}\ }\href@noop {} {\bibfield
  {journal} {\bibinfo  {journal} {Phys. Rev. Lett.}\ }\textbf {\bibinfo
  {volume} {125}},\ \bibinfo {pages} {148001} (\bibinfo {year}
  {2020})}\BibitemShut {NoStop}%
\bibitem [{\citenamefont {Mitra}\ \emph {et~al.}(2021)\citenamefont {Mitra},
  \citenamefont {Parmar}, \citenamefont {Leishangthem}, \citenamefont
  {Sastry},\ and\ \citenamefont {Foffi}}]{mitra2021hyperuniformity}%
  \BibitemOpen
  \bibfield  {author} {\bibinfo {author} {\bibfnamefont {S.}~\bibnamefont
  {Mitra}}, \bibinfo {author} {\bibfnamefont {A.~D.}\ \bibnamefont {Parmar}},
  \bibinfo {author} {\bibfnamefont {P.}~\bibnamefont {Leishangthem}}, \bibinfo
  {author} {\bibfnamefont {S.}~\bibnamefont {Sastry}}, \ and\ \bibinfo {author}
  {\bibfnamefont {G.}~\bibnamefont {Foffi}},\ }\bibfield  {title} {\enquote
  {\bibinfo {title} {Hyperuniformity in cyclically driven glasses},}\
  }\href@noop {} {\bibfield  {journal} {\bibinfo  {journal} {J. Stat. Mech.:
  Theor. Exp.}\ }\textbf {\bibinfo {volume} {2021}},\ \bibinfo {pages} {033203}
  (\bibinfo {year} {2021})}\BibitemShut {NoStop}%
\bibitem [{\citenamefont {Oppenheimer}\ \emph {et~al.}(2022)\citenamefont
  {Oppenheimer}, \citenamefont {Stein}, \citenamefont {Zion},\ and\
  \citenamefont {Shelley}}]{oppenheimer2022hyperuniformity}%
  \BibitemOpen
  \bibfield  {author} {\bibinfo {author} {\bibfnamefont {N.}~\bibnamefont
  {Oppenheimer}}, \bibinfo {author} {\bibfnamefont {D.~B.}\ \bibnamefont
  {Stein}}, \bibinfo {author} {\bibfnamefont {M.~Y.~B.}\ \bibnamefont {Zion}},
  \ and\ \bibinfo {author} {\bibfnamefont {M.~J.}\ \bibnamefont {Shelley}},\
  }\bibfield  {title} {\enquote {\bibinfo {title} {Hyperuniformity and phase
  enrichment in vortex and rotor assemblies},}\ }\href@noop {} {\bibfield
  {journal} {\bibinfo  {journal} {Nat. Comm.}\ }\textbf {\bibinfo {volume}
  {13}},\ \bibinfo {pages} {804} (\bibinfo {year} {2022})}\BibitemShut
  {NoStop}%
\bibitem [{\citenamefont {Huang}\ \emph {et~al.}(2021)\citenamefont {Huang},
  \citenamefont {Hu}, \citenamefont {Yang}, \citenamefont {Liu},\ and\
  \citenamefont {Zhang}}]{huang2021circular}%
  \BibitemOpen
  \bibfield  {author} {\bibinfo {author} {\bibfnamefont {M.}~\bibnamefont
  {Huang}}, \bibinfo {author} {\bibfnamefont {W.}~\bibnamefont {Hu}}, \bibinfo
  {author} {\bibfnamefont {S.}~\bibnamefont {Yang}}, \bibinfo {author}
  {\bibfnamefont {Q-X.}\ \bibnamefont {Liu}}, \ and\ \bibinfo {author}
  {\bibfnamefont {H.P.}\ \bibnamefont {Zhang}},\ }\bibfield  {title} {\enquote
  {\bibinfo {title} {Circular swimming motility and disordered hyperuniform
  state in an algae system},}\ }\href@noop {} {\bibfield  {journal} {\bibinfo
  {journal} {Proc. Natl. Acad. Sci. U.S.A.}\ }\textbf {\bibinfo {volume}
  {118}},\ \bibinfo {pages} {e2100493118} (\bibinfo {year} {2021})}\BibitemShut
  {NoStop}%
\bibitem [{\citenamefont {Lei}\ \emph {et~al.}(2019)\citenamefont {Lei},
  \citenamefont {Ciamarra},\ and\ \citenamefont {Ni}}]{lei2019nonequilibrium}%
  \BibitemOpen
  \bibfield  {author} {\bibinfo {author} {\bibfnamefont {Q.-L.}\ \bibnamefont
  {Lei}}, \bibinfo {author} {\bibfnamefont {M.~P.}\ \bibnamefont {Ciamarra}}, \
  and\ \bibinfo {author} {\bibfnamefont {R.}~\bibnamefont {Ni}},\ }\bibfield
  {title} {\enquote {\bibinfo {title} {Nonequilibrium strongly hyperuniform
  fluids of circle active particles with large local density fluctuations},}\
  }\href@noop {} {\bibfield  {journal} {\bibinfo  {journal} {Sci. Adv.}\
  }\textbf {\bibinfo {volume} {5}},\ \bibinfo {pages} {eaau7423} (\bibinfo
  {year} {2019})}\BibitemShut {NoStop}%
\bibitem [{\citenamefont {Grassberger}\ \emph {et~al.}(2016)\citenamefont
  {Grassberger}, \citenamefont {Dhar},\ and\ \citenamefont
  {Mohanty}}]{grassberger2016oslo}%
  \BibitemOpen
  \bibfield  {author} {\bibinfo {author} {\bibfnamefont {P.}~\bibnamefont
  {Grassberger}}, \bibinfo {author} {\bibfnamefont {D.}~\bibnamefont {Dhar}}, \
  and\ \bibinfo {author} {\bibfnamefont {P.}~\bibnamefont {Mohanty}},\
  }\bibfield  {title} {\enquote {\bibinfo {title} {Oslo model, hyperuniformity,
  and the quenched edwards-wilkinson model},}\ }\href@noop {} {\bibfield
  {journal} {\bibinfo  {journal} {Phys. Rev. E}\ }\textbf {\bibinfo {volume}
  {94}},\ \bibinfo {pages} {042314} (\bibinfo {year} {2016})}\BibitemShut
  {NoStop}%
\bibitem [{\citenamefont {Bertrand}\ \emph {et~al.}(2019)\citenamefont
  {Bertrand}, \citenamefont {Chatenay},\ and\ \citenamefont
  {Voituriez}}]{bertrand2019nonlinear}%
  \BibitemOpen
  \bibfield  {author} {\bibinfo {author} {\bibfnamefont {T.}~\bibnamefont
  {Bertrand}}, \bibinfo {author} {\bibfnamefont {D.}~\bibnamefont {Chatenay}},
  \ and\ \bibinfo {author} {\bibfnamefont {R.}~\bibnamefont {Voituriez}},\
  }\bibfield  {title} {\enquote {\bibinfo {title} {Nonlinear diffusion and
  hyperuniformity from poisson representation in systems with interaction
  mediated dynamics},}\ }\href@noop {} {\bibfield  {journal} {\bibinfo
  {journal} {New J. Phys.}\ }\textbf {\bibinfo {volume} {21}},\ \bibinfo
  {pages} {123048} (\bibinfo {year} {2019})}\BibitemShut {NoStop}%
\bibitem [{\citenamefont {Goldfriend}\ \emph {et~al.}(2017)\citenamefont
  {Goldfriend}, \citenamefont {Diamant},\ and\ \citenamefont
  {Witten}}]{goldfriend2017screening}%
  \BibitemOpen
  \bibfield  {author} {\bibinfo {author} {\bibfnamefont {T.}~\bibnamefont
  {Goldfriend}}, \bibinfo {author} {\bibfnamefont {H.}~\bibnamefont {Diamant}},
  \ and\ \bibinfo {author} {\bibfnamefont {T.~A.}\ \bibnamefont {Witten}},\
  }\bibfield  {title} {\enquote {\bibinfo {title} {Screening, hyperuniformity,
  and instability in the sedimentation of irregular objects},}\ }\href@noop {}
  {\bibfield  {journal} {\bibinfo  {journal} {Phys. Rev. Lett.}\ }\textbf
  {\bibinfo {volume} {118}},\ \bibinfo {pages} {158005} (\bibinfo {year}
  {2017})}\BibitemShut {NoStop}%
\bibitem [{\citenamefont {Chen}\ \emph {et~al.}(2024)\citenamefont {Chen},
  \citenamefont {Lei}, \citenamefont {Xiang}, \citenamefont {Duan},
  \citenamefont {Peng},\ and\ \citenamefont {Zhang}}]{hepeng2024}%
  \BibitemOpen
  \bibfield  {author} {\bibinfo {author} {\bibfnamefont {J.}~\bibnamefont
  {Chen}}, \bibinfo {author} {\bibfnamefont {X.}~\bibnamefont {Lei}}, \bibinfo
  {author} {\bibfnamefont {Y.}~\bibnamefont {Xiang}}, \bibinfo {author}
  {\bibfnamefont {M.}~\bibnamefont {Duan}}, \bibinfo {author} {\bibfnamefont
  {X.}~\bibnamefont {Peng}}, \ and\ \bibinfo {author} {\bibfnamefont {H.~P.}\
  \bibnamefont {Zhang}},\ }\bibfield  {title} {\enquote {\bibinfo {title}
  {Emergent chirality and hyperuniformity in an active mixture with
  nonreciprocal interactions},}\ }\href@noop {} {\bibfield  {journal} {\bibinfo
   {journal} {Phys. Rev. Lett.}\ }\textbf {\bibinfo {volume} {132}},\ \bibinfo
  {pages} {118301} (\bibinfo {year} {2024})}\BibitemShut {NoStop}%
\bibitem [{\citenamefont {Jack}\ \emph {et~al.}(2015)\citenamefont {Jack},
  \citenamefont {Thompson},\ and\ \citenamefont
  {Sollich}}]{jack2015hyperuniformity}%
  \BibitemOpen
  \bibfield  {author} {\bibinfo {author} {\bibfnamefont {R.~L.}\ \bibnamefont
  {Jack}}, \bibinfo {author} {\bibfnamefont {I.~R.}\ \bibnamefont {Thompson}},
  \ and\ \bibinfo {author} {\bibfnamefont {P.}~\bibnamefont {Sollich}},\
  }\bibfield  {title} {\enquote {\bibinfo {title} {Hyperuniformity and phase
  separation in biased ensembles of trajectories for diffusive systems},}\
  }\href@noop {} {\bibfield  {journal} {\bibinfo  {journal} {Phys. Rev. Lett.}\
  }\textbf {\bibinfo {volume} {114}},\ \bibinfo {pages} {060601} (\bibinfo
  {year} {2015})}\BibitemShut {NoStop}%
\bibitem [{\citenamefont {Karevski}\ and\ \citenamefont
  {Sch{\"u}tz}(2017)}]{karevski2017conformal}%
  \BibitemOpen
  \bibfield  {author} {\bibinfo {author} {\bibfnamefont {D.}~\bibnamefont
  {Karevski}}\ and\ \bibinfo {author} {\bibfnamefont {GM}~\bibnamefont
  {Sch{\"u}tz}},\ }\bibfield  {title} {\enquote {\bibinfo {title} {Conformal
  invariance in driven diffusive systems at high currents},}\ }\href@noop {}
  {\bibfield  {journal} {\bibinfo  {journal} {Phys. Rev. Lett.}\ }\textbf
  {\bibinfo {volume} {118}},\ \bibinfo {pages} {030601} (\bibinfo {year}
  {2017})}\BibitemShut {NoStop}%
\bibitem [{\citenamefont {Garrahan}\ and\ \citenamefont
  {Pollmann}(2024)}]{garrahan2024topological}%
  \BibitemOpen
  \bibfield  {author} {\bibinfo {author} {\bibfnamefont {J.~P.}\ \bibnamefont
  {Garrahan}}\ and\ \bibinfo {author} {\bibfnamefont {F.}~\bibnamefont
  {Pollmann}},\ }\bibfield  {title} {\enquote {\bibinfo {title} {Topological
  phases in the dynamics of the simple exclusion process},}\ }\href@noop {}
  {\bibfield  {journal} {\bibinfo  {journal} {Phys. Rev. E}\ }\textbf {\bibinfo
  {volume} {109}},\ \bibinfo {pages} {L032105} (\bibinfo {year}
  {2024})}\BibitemShut {NoStop}%
\bibitem [{\citenamefont {Torquato}\ \emph {et~al.}(2008)\citenamefont
  {Torquato}, \citenamefont {Scardicchio},\ and\ \citenamefont
  {Zachary}}]{torquato2008point}%
  \BibitemOpen
  \bibfield  {author} {\bibinfo {author} {\bibfnamefont {S.}~\bibnamefont
  {Torquato}}, \bibinfo {author} {\bibfnamefont {A.}~\bibnamefont
  {Scardicchio}}, \ and\ \bibinfo {author} {\bibfnamefont {C.~E.}\ \bibnamefont
  {Zachary}},\ }\bibfield  {title} {\enquote {\bibinfo {title} {Point processes
  in arbitrary dimension from fermionic gases, random matrix theory, and number
  theory},}\ }\href@noop {} {\bibfield  {journal} {\bibinfo  {journal} {J.
  Stat. Mech.: Theory Exp.}\ }\textbf {\bibinfo {volume} {2008}},\ \bibinfo
  {pages} {P11019} (\bibinfo {year} {2008})}\BibitemShut {NoStop}%
\bibitem [{\citenamefont {Carollo}\ \emph {et~al.}(2017)\citenamefont
  {Carollo}, \citenamefont {Garrahan}, \citenamefont {Lesanovsky},\ and\
  \citenamefont {P{\'e}rez-Espigares}}]{carollo2017fluctuating}%
  \BibitemOpen
  \bibfield  {author} {\bibinfo {author} {\bibfnamefont {F.}~\bibnamefont
  {Carollo}}, \bibinfo {author} {\bibfnamefont {J.~P.}\ \bibnamefont
  {Garrahan}}, \bibinfo {author} {\bibfnamefont {I.}~\bibnamefont
  {Lesanovsky}}, \ and\ \bibinfo {author} {\bibfnamefont {C.}~\bibnamefont
  {P{\'e}rez-Espigares}},\ }\bibfield  {title} {\enquote {\bibinfo {title}
  {Fluctuating hydrodynamics, current fluctuations, and hyperuniformity in
  boundary-driven open quantum chains},}\ }\href@noop {} {\bibfield  {journal}
  {\bibinfo  {journal} {Physical Review E}\ }\textbf {\bibinfo {volume} {96}},\
  \bibinfo {pages} {052118} (\bibinfo {year} {2017})}\BibitemShut {NoStop}%
\bibitem [{\citenamefont {Sakai}\ \emph {et~al.}(2022)\citenamefont {Sakai},
  \citenamefont {Arita},\ and\ \citenamefont {Ohtsuki}}]{sakai2022quantum}%
  \BibitemOpen
  \bibfield  {author} {\bibinfo {author} {\bibfnamefont {S.}~\bibnamefont
  {Sakai}}, \bibinfo {author} {\bibfnamefont {R.}~\bibnamefont {Arita}}, \ and\
  \bibinfo {author} {\bibfnamefont {T.}~\bibnamefont {Ohtsuki}},\ }\bibfield
  {title} {\enquote {\bibinfo {title} {Quantum phase transition between
  hyperuniform density distributions},}\ }\href@noop {} {\bibfield  {journal}
  {\bibinfo  {journal} {Phys. Rev. Res.}\ }\textbf {\bibinfo {volume} {4}},\
  \bibinfo {pages} {033241} (\bibinfo {year} {2022})}\BibitemShut {NoStop}%
\bibitem [{\citenamefont {Cort{\'e}}\ \emph {et~al.}(2008)\citenamefont
  {Cort{\'e}}, \citenamefont {Chaikin}, \citenamefont {Gollub},\ and\
  \citenamefont {Pine}}]{corte2008random}%
  \BibitemOpen
  \bibfield  {author} {\bibinfo {author} {\bibfnamefont {L.}~\bibnamefont
  {Cort{\'e}}}, \bibinfo {author} {\bibfnamefont {P.~M.}\ \bibnamefont
  {Chaikin}}, \bibinfo {author} {\bibfnamefont {J.~P.}\ \bibnamefont {Gollub}},
  \ and\ \bibinfo {author} {\bibfnamefont {D.~J.}\ \bibnamefont {Pine}},\
  }\bibfield  {title} {\enquote {\bibinfo {title} {Random organization in
  periodically driven systems},}\ }\href@noop {} {\bibfield  {journal}
  {\bibinfo  {journal} {Nat. Phys.}\ }\textbf {\bibinfo {volume} {4}},\
  \bibinfo {pages} {420--424} (\bibinfo {year} {2008})}\BibitemShut {NoStop}%
\bibitem [{\citenamefont {Hexner}\ and\ \citenamefont
  {Levine}(2015)}]{hexner2015hyperuniformity}%
  \BibitemOpen
  \bibfield  {author} {\bibinfo {author} {\bibfnamefont {D.}~\bibnamefont
  {Hexner}}\ and\ \bibinfo {author} {\bibfnamefont {D.}~\bibnamefont
  {Levine}},\ }\bibfield  {title} {\enquote {\bibinfo {title} {Hyperuniformity
  of critical absorbing states},}\ }\href@noop {} {\bibfield  {journal}
  {\bibinfo  {journal} {Phys. Rev. Lett.}\ }\textbf {\bibinfo {volume} {114}},\
  \bibinfo {pages} {110602} (\bibinfo {year} {2015})}\BibitemShut {NoStop}%
\bibitem [{\citenamefont {Pine}\ \emph {et~al.}(2005)\citenamefont {Pine},
  \citenamefont {Gollub}, \citenamefont {Brady},\ and\ \citenamefont
  {Leshansky}}]{pine2005chaos}%
  \BibitemOpen
  \bibfield  {author} {\bibinfo {author} {\bibfnamefont {D.~J.}\ \bibnamefont
  {Pine}}, \bibinfo {author} {\bibfnamefont {J.~P.}\ \bibnamefont {Gollub}},
  \bibinfo {author} {\bibfnamefont {J.~F.}\ \bibnamefont {Brady}}, \ and\
  \bibinfo {author} {\bibfnamefont {A.~M.}\ \bibnamefont {Leshansky}},\
  }\bibfield  {title} {\enquote {\bibinfo {title} {Chaos and threshold for
  irreversibility in sheared suspensions},}\ }\href@noop {} {\bibfield
  {journal} {\bibinfo  {journal} {Nature}\ }\textbf {\bibinfo {volume} {438}},\
  \bibinfo {pages} {997--1000} (\bibinfo {year} {2005})}\BibitemShut {NoStop}%
\bibitem [{\citenamefont {Schrenk}\ and\ \citenamefont
  {Frenkel}(2015)}]{frenkelnonergodic}%
  \BibitemOpen
  \bibfield  {author} {\bibinfo {author} {\bibfnamefont {K.~J.}\ \bibnamefont
  {Schrenk}}\ and\ \bibinfo {author} {\bibfnamefont {D.}~\bibnamefont
  {Frenkel}},\ }\bibfield  {title} {\enquote {\bibinfo {title} {{Communication:
  Evidence for non-ergodicity in quiescent states of periodically sheared
  suspensions}},}\ }\href@noop {} {\bibfield  {journal} {\bibinfo  {journal}
  {J. Chem. Phys}\ }\textbf {\bibinfo {volume} {143}},\ \bibinfo {pages}
  {241103} (\bibinfo {year} {2015})}\BibitemShut {NoStop}%
\bibitem [{\citenamefont {Menon}\ and\ \citenamefont
  {Ramaswamy}(2009)}]{menon2009universality}%
  \BibitemOpen
  \bibfield  {author} {\bibinfo {author} {\bibfnamefont {G.~I.}\ \bibnamefont
  {Menon}}\ and\ \bibinfo {author} {\bibfnamefont {S.}~\bibnamefont
  {Ramaswamy}},\ }\bibfield  {title} {\enquote {\bibinfo {title} {Universality
  class of the reversible-irreversible transition in sheared suspensions},}\
  }\href@noop {} {\bibfield  {journal} {\bibinfo  {journal} {Phys. Rev. E}\
  }\textbf {\bibinfo {volume} {79}},\ \bibinfo {pages} {061108} (\bibinfo
  {year} {2009})}\BibitemShut {NoStop}%
\bibitem [{\citenamefont {Rossi}\ \emph {et~al.}(2000)\citenamefont {Rossi},
  \citenamefont {Pastor-Satorras},\ and\ \citenamefont
  {Vespignani}}]{rossi2000universality}%
  \BibitemOpen
  \bibfield  {author} {\bibinfo {author} {\bibfnamefont {M.}~\bibnamefont
  {Rossi}}, \bibinfo {author} {\bibfnamefont {R.}~\bibnamefont
  {Pastor-Satorras}}, \ and\ \bibinfo {author} {\bibfnamefont {A.}~\bibnamefont
  {Vespignani}},\ }\bibfield  {title} {\enquote {\bibinfo {title} {Universality
  class of absorbing phase transitions with a conserved field},}\ }\href@noop
  {} {\bibfield  {journal} {\bibinfo  {journal} {Phys. Rev. Lett.}\ }\textbf
  {\bibinfo {volume} {85}},\ \bibinfo {pages} {1803} (\bibinfo {year}
  {2000})}\BibitemShut {NoStop}%
\bibitem [{\citenamefont {Henkel}\ \emph {et~al.}(2008)\citenamefont {Henkel},
  \citenamefont {Hinrichsen},\ and\ \citenamefont
  {L{\"u}beck}}]{henkel2008non}%
  \BibitemOpen
  \bibfield  {author} {\bibinfo {author} {\bibfnamefont {M.}~\bibnamefont
  {Henkel}}, \bibinfo {author} {\bibfnamefont {H.}~\bibnamefont {Hinrichsen}},
  \ and\ \bibinfo {author} {\bibfnamefont {S.}~\bibnamefont {L{\"u}beck}},\
  }\href@noop {} {\emph {\bibinfo {title} {Non-equilibrium phase transitions:
  volume 1: absorbing phase transitions}}}\ (\bibinfo  {publisher} {Springer
  Science \& Business Media},\ \bibinfo {year} {2008})\BibitemShut {NoStop}%
\bibitem [{\citenamefont {Ma}\ and\ \citenamefont
  {Torquato}(2019)}]{ma2019hyperuniformity}%
  \BibitemOpen
  \bibfield  {author} {\bibinfo {author} {\bibfnamefont {Z.}~\bibnamefont
  {Ma}}\ and\ \bibinfo {author} {\bibfnamefont {S.}~\bibnamefont {Torquato}},\
  }\bibfield  {title} {\enquote {\bibinfo {title} {Hyperuniformity of
  generalized random organization models},}\ }\href@noop {} {\bibfield
  {journal} {\bibinfo  {journal} {Phys. Rev. E}\ }\textbf {\bibinfo {volume}
  {99}},\ \bibinfo {pages} {022115} (\bibinfo {year} {2019})}\BibitemShut
  {NoStop}%
\bibitem [{\citenamefont {Hexner}\ \emph {et~al.}(2017)\citenamefont {Hexner},
  \citenamefont {Chaikin},\ and\ \citenamefont {Levine}}]{hexner2017enhanced}%
  \BibitemOpen
  \bibfield  {author} {\bibinfo {author} {\bibfnamefont {D.}~\bibnamefont
  {Hexner}}, \bibinfo {author} {\bibfnamefont {P.~M.}\ \bibnamefont {Chaikin}},
  \ and\ \bibinfo {author} {\bibfnamefont {D.}~\bibnamefont {Levine}},\
  }\bibfield  {title} {\enquote {\bibinfo {title} {Enhanced hyperuniformity
  from random reorganization},}\ }\href@noop {} {\bibfield  {journal} {\bibinfo
   {journal} {Proc. Natl. Acad. Sci. U.S.A.}\ }\textbf {\bibinfo {volume}
  {114}},\ \bibinfo {pages} {4294--4299} (\bibinfo {year} {2017})}\BibitemShut
  {NoStop}%
\bibitem [{\citenamefont {Mari}\ \emph {et~al.}(2022)\citenamefont {Mari},
  \citenamefont {Bertin},\ and\ \citenamefont {Nardini}}]{mari2022absorbing}%
  \BibitemOpen
  \bibfield  {author} {\bibinfo {author} {\bibfnamefont {R.}~\bibnamefont
  {Mari}}, \bibinfo {author} {\bibfnamefont {E.}~\bibnamefont {Bertin}}, \ and\
  \bibinfo {author} {\bibfnamefont {C.}~\bibnamefont {Nardini}},\ }\bibfield
  {title} {\enquote {\bibinfo {title} {Absorbing phase transitions in systems
  with mediated interactions},}\ }\href@noop {} {\bibfield  {journal} {\bibinfo
   {journal} {Phys. Rev. E}\ }\textbf {\bibinfo {volume} {105}},\ \bibinfo
  {pages} {L032602} (\bibinfo {year} {2022})}\BibitemShut {NoStop}%
\bibitem [{\citenamefont {Wilken}\ \emph {et~al.}(2021)\citenamefont {Wilken},
  \citenamefont {Guerra}, \citenamefont {Levine},\ and\ \citenamefont
  {Chaikin}}]{wilken2021random}%
  \BibitemOpen
  \bibfield  {author} {\bibinfo {author} {\bibfnamefont {S.}~\bibnamefont
  {Wilken}}, \bibinfo {author} {\bibfnamefont {R.~E.}\ \bibnamefont {Guerra}},
  \bibinfo {author} {\bibfnamefont {D.}~\bibnamefont {Levine}}, \ and\ \bibinfo
  {author} {\bibfnamefont {P.~M.}\ \bibnamefont {Chaikin}},\ }\bibfield
  {title} {\enquote {\bibinfo {title} {Random close packing as a dynamical
  phase transition},}\ }\href@noop {} {\bibfield  {journal} {\bibinfo
  {journal} {Phys. Rev. Lett.}\ }\textbf {\bibinfo {volume} {127}},\ \bibinfo
  {pages} {038002} (\bibinfo {year} {2021})}\BibitemShut {NoStop}%
\bibitem [{\citenamefont {Ni}\ \emph {et~al.}(2013)\citenamefont {Ni},
  \citenamefont {Cohen~Stuart},\ and\ \citenamefont {Dijkstra}}]{ni_rcp_2013}%
  \BibitemOpen
  \bibfield  {author} {\bibinfo {author} {\bibfnamefont {R.}~\bibnamefont
  {Ni}}, \bibinfo {author} {\bibfnamefont {M.~A.}\ \bibnamefont
  {Cohen~Stuart}}, \ and\ \bibinfo {author} {\bibfnamefont {M.}~\bibnamefont
  {Dijkstra}},\ }\bibfield  {title} {\enquote {\bibinfo {title} {Pushing the
  glass transition towards random close packing using self-propelled hard
  spheres},}\ }\href@noop {} {\bibfield  {journal} {\bibinfo  {journal} {Nat.
  Commun.}\ }\textbf {\bibinfo {volume} {4}},\ \bibinfo {pages} {2704}
  (\bibinfo {year} {2013})}\BibitemShut {NoStop}%
\bibitem [{\citenamefont {Wilken}\ \emph
  {et~al.}(2023{\natexlab{a}})\citenamefont {Wilken}, \citenamefont {Guo},
  \citenamefont {Levine},\ and\ \citenamefont {Chaikin}}]{rcpjamming}%
  \BibitemOpen
  \bibfield  {author} {\bibinfo {author} {\bibfnamefont {S.}~\bibnamefont
  {Wilken}}, \bibinfo {author} {\bibfnamefont {A.~Z.}\ \bibnamefont {Guo}},
  \bibinfo {author} {\bibfnamefont {D.}~\bibnamefont {Levine}}, \ and\ \bibinfo
  {author} {\bibfnamefont {P.~M.}\ \bibnamefont {Chaikin}},\ }\bibfield
  {title} {\enquote {\bibinfo {title} {Dynamical approach to the jamming
  problem},}\ }\href@noop {} {\bibfield  {journal} {\bibinfo  {journal} {Phys.
  Rev. Lett.}\ }\textbf {\bibinfo {volume} {131}},\ \bibinfo {pages} {238202}
  (\bibinfo {year} {2023}{\natexlab{a}})}\BibitemShut {NoStop}%
\bibitem [{\citenamefont {v.~Wijland}(2002)}]{van2002universality}%
  \BibitemOpen
  \bibfield  {author} {\bibinfo {author} {\bibfnamefont {F.}~\bibnamefont
  {v.~Wijland}},\ }\bibfield  {title} {\enquote {\bibinfo {title} {Universality
  class of nonequilibrium phase transitions with infinitely many absorbing
  states},}\ }\href@noop {} {\bibfield  {journal} {\bibinfo  {journal} {Phys.
  Rev. Lett.}\ }\textbf {\bibinfo {volume} {89}},\ \bibinfo {pages} {190602}
  (\bibinfo {year} {2002})}\BibitemShut {NoStop}%
\bibitem [{\citenamefont {v.~Wijland}(2003)}]{wijland2003infinitely}%
  \BibitemOpen
  \bibfield  {author} {\bibinfo {author} {\bibfnamefont {F.}~\bibnamefont
  {v.~Wijland}},\ }\bibfield  {title} {\enquote {\bibinfo {title}
  {Infinitely-many absorbing-state nonequilibrium phase transitions},}\
  }\href@noop {} {\bibfield  {journal} {\bibinfo  {journal} {Braz. J. Phys}\
  }\textbf {\bibinfo {volume} {33}},\ \bibinfo {pages} {551--556} (\bibinfo
  {year} {2003})}\BibitemShut {NoStop}%
\bibitem [{\citenamefont {Doussal}\ and\ \citenamefont
  {Wiese}(2015)}]{le2015exact}%
  \BibitemOpen
  \bibfield  {author} {\bibinfo {author} {\bibfnamefont {P.~L.}\ \bibnamefont
  {Doussal}}\ and\ \bibinfo {author} {\bibfnamefont {K.~J.}\ \bibnamefont
  {Wiese}},\ }\bibfield  {title} {\enquote {\bibinfo {title} {Exact mapping of
  the stochastic field theory for manna sandpiles to interfaces in random
  media},}\ }\href@noop {} {\bibfield  {journal} {\bibinfo  {journal} {Phys.
  Rev. Lett.}\ }\textbf {\bibinfo {volume} {114}},\ \bibinfo {pages} {110601}
  (\bibinfo {year} {2015})}\BibitemShut {NoStop}%
\bibitem [{\citenamefont {Ma}\ \emph {et~al.}(2023)\citenamefont {Ma},
  \citenamefont {Pausch},\ and\ \citenamefont {Cates}}]{ma2023theory}%
  \BibitemOpen
  \bibfield  {author} {\bibinfo {author} {\bibfnamefont {X.}~\bibnamefont
  {Ma}}, \bibinfo {author} {\bibfnamefont {J.}~\bibnamefont {Pausch}}, \ and\
  \bibinfo {author} {\bibfnamefont {M.~E.}\ \bibnamefont {Cates}},\ }\bibfield
  {title} {\enquote {\bibinfo {title} {Theory of hyperuniformity at the
  absorbing state transition},}\ }\href@noop {} {\bibfield  {journal} {\bibinfo
   {journal} {arXiv preprint arXiv:2310.17391}\ } (\bibinfo {year}
  {2023})}\BibitemShut {NoStop}%
\bibitem [{\citenamefont {Wiese}(2024)}]{joerg2024hyperuniformity}%
  \BibitemOpen
  \bibfield  {author} {\bibinfo {author} {\bibfnamefont {K.~J.}\ \bibnamefont
  {Wiese}},\ }\bibfield  {title} {\enquote {\bibinfo {title} {Hyperuniformity
  in the manna model, conserved directed percolation and depinning},}\
  }\href@noop {} {\bibfield  {journal} {\bibinfo  {journal} {arXiv e-prints
  arxiv:2401.09123}\ } (\bibinfo {year} {2024})}\BibitemShut {NoStop}%
\bibitem [{\citenamefont {Wiese}(2022)}]{Wiese_2022}%
  \BibitemOpen
  \bibfield  {author} {\bibinfo {author} {\bibfnamefont {K.~J.}\ \bibnamefont
  {Wiese}},\ }\bibfield  {title} {\enquote {\bibinfo {title} {Theory and
  experiments for disordered elastic manifolds, depinning, avalanches, and
  sandpiles},}\ }\href@noop {} {\bibfield  {journal} {\bibinfo  {journal} {Rep.
  Prog. Phys.}\ }\textbf {\bibinfo {volume} {85}},\ \bibinfo {pages} {086502}
  (\bibinfo {year} {2022})}\BibitemShut {NoStop}%
\bibitem [{\citenamefont {Shapira}\ and\ \citenamefont
  {Wiese}(2023)}]{Shapira_2023}%
  \BibitemOpen
  \bibfield  {author} {\bibinfo {author} {\bibfnamefont {A.}~\bibnamefont
  {Shapira}}\ and\ \bibinfo {author} {\bibfnamefont {K.~J.}\ \bibnamefont
  {Wiese}},\ }\bibfield  {title} {\enquote {\bibinfo {title} {Anchored advected
  interfaces, oslo model, and roughness at depinning},}\ }\href@noop {}
  {\bibfield  {journal} {\bibinfo  {journal} {J. Stat. Mech.: Theory Exp.}\
  }\textbf {\bibinfo {volume} {2023}},\ \bibinfo {pages} {063202} (\bibinfo
  {year} {2023})}\BibitemShut {NoStop}%
\bibitem [{\citenamefont {Rosso}\ \emph {et~al.}(2003)\citenamefont {Rosso},
  \citenamefont {Hartmann},\ and\ \citenamefont {Krauth}}]{rosso2003}%
  \BibitemOpen
  \bibfield  {author} {\bibinfo {author} {\bibfnamefont {A.}~\bibnamefont
  {Rosso}}, \bibinfo {author} {\bibfnamefont {A.~K.}\ \bibnamefont {Hartmann}},
  \ and\ \bibinfo {author} {\bibfnamefont {W.}~\bibnamefont {Krauth}},\
  }\bibfield  {title} {\enquote {\bibinfo {title} {Depinning of elastic
  manifolds},}\ }\href@noop {} {\bibfield  {journal} {\bibinfo  {journal}
  {Phys. Rev. E}\ }\textbf {\bibinfo {volume} {67}},\ \bibinfo {pages} {021602}
  (\bibinfo {year} {2003})}\BibitemShut {NoStop}%
\bibitem [{\citenamefont {Wang}\ \emph {et~al.}(2018)\citenamefont {Wang},
  \citenamefont {Schwarz},\ and\ \citenamefont
  {Paulsen}}]{wang2018hyperuniformity}%
  \BibitemOpen
  \bibfield  {author} {\bibinfo {author} {\bibfnamefont {J.}~\bibnamefont
  {Wang}}, \bibinfo {author} {\bibfnamefont {J.~M.}\ \bibnamefont {Schwarz}}, \
  and\ \bibinfo {author} {\bibfnamefont {J.~D.}\ \bibnamefont {Paulsen}},\
  }\bibfield  {title} {\enquote {\bibinfo {title} {Hyperuniformity with no fine
  tuning in sheared sedimenting suspensions},}\ }\href@noop {} {\bibfield
  {journal} {\bibinfo  {journal} {Nat. Comm.}\ }\textbf {\bibinfo {volume}
  {9}},\ \bibinfo {pages} {2836} (\bibinfo {year} {2018})}\BibitemShut
  {NoStop}%
\bibitem [{\citenamefont {Cort{\'e}}\ \emph {et~al.}(2009)\citenamefont
  {Cort{\'e}}, \citenamefont {Gerbode}, \citenamefont {Man},\ and\
  \citenamefont {Pine}}]{corte2009self}%
  \BibitemOpen
  \bibfield  {author} {\bibinfo {author} {\bibfnamefont {L.}~\bibnamefont
  {Cort{\'e}}}, \bibinfo {author} {\bibfnamefont {S.~J.}\ \bibnamefont
  {Gerbode}}, \bibinfo {author} {\bibfnamefont {W.}~\bibnamefont {Man}}, \ and\
  \bibinfo {author} {\bibfnamefont {D.~J.}\ \bibnamefont {Pine}},\ }\bibfield
  {title} {\enquote {\bibinfo {title} {Self-organized criticality in sheared
  suspensions},}\ }\href@noop {} {\bibfield  {journal} {\bibinfo  {journal}
  {Phys. Rev. Lett.}\ }\textbf {\bibinfo {volume} {103}},\ \bibinfo {pages}
  {248301} (\bibinfo {year} {2009})}\BibitemShut {NoStop}%
\bibitem [{\citenamefont {Hexner}\ and\ \citenamefont
  {Levine}(2017)}]{hexner2017noise}%
  \BibitemOpen
  \bibfield  {author} {\bibinfo {author} {\bibfnamefont {D.}~\bibnamefont
  {Hexner}}\ and\ \bibinfo {author} {\bibfnamefont {D.}~\bibnamefont
  {Levine}},\ }\bibfield  {title} {\enquote {\bibinfo {title} {Noise,
  diffusion, and hyperuniformity},}\ }\href@noop {} {\bibfield  {journal}
  {\bibinfo  {journal} {Phys. Rev. Lett.}\ }\textbf {\bibinfo {volume} {118}},\
  \bibinfo {pages} {020601} (\bibinfo {year} {2017})}\BibitemShut {NoStop}%
\bibitem [{\citenamefont {Lei}\ \emph {et~al.}(2023)\citenamefont {Lei},
  \citenamefont {Zheng},\ and\ \citenamefont {Ni}}]{Lei2023Spherical}%
  \BibitemOpen
  \bibfield  {author} {\bibinfo {author} {\bibfnamefont {Y.}~\bibnamefont
  {Lei}}, \bibinfo {author} {\bibfnamefont {N.}~\bibnamefont {Zheng}}, \ and\
  \bibinfo {author} {\bibfnamefont {R.}~\bibnamefont {Ni}},\ }\bibfield
  {title} {\enquote {\bibinfo {title} {Random organization and non-equilibrium
  hyperuniform fluids on a sphere},}\ }\href@noop {} {\bibfield  {journal}
  {\bibinfo  {journal} {J. Chem. Phys.}\ }\textbf {\bibinfo {volume} {157}},\
  \bibinfo {pages} {081101} (\bibinfo {year} {2023})}\BibitemShut {NoStop}%
\bibitem [{\citenamefont {Lei}\ and\ \citenamefont
  {Ni}(2019)}]{lei2019hydrodynamics}%
  \BibitemOpen
  \bibfield  {author} {\bibinfo {author} {\bibfnamefont {Q.-L.}\ \bibnamefont
  {Lei}}\ and\ \bibinfo {author} {\bibfnamefont {R.}~\bibnamefont {Ni}},\
  }\bibfield  {title} {\enquote {\bibinfo {title} {Hydrodynamics of
  random-organizing hyperuniform fluids},}\ }\href@noop {} {\bibfield
  {journal} {\bibinfo  {journal} {Proc. Natl. Acad. Sci. U.S.A.}\ }\textbf
  {\bibinfo {volume} {116}},\ \bibinfo {pages} {22983--22989} (\bibinfo {year}
  {2019})}\BibitemShut {NoStop}%
\bibitem [{\citenamefont {Tjhung}\ and\ \citenamefont
  {Berthier}(2017)}]{tjhung2017discontinuous}%
  \BibitemOpen
  \bibfield  {author} {\bibinfo {author} {\bibfnamefont {E.}~\bibnamefont
  {Tjhung}}\ and\ \bibinfo {author} {\bibfnamefont {L.}~\bibnamefont
  {Berthier}},\ }\bibfield  {title} {\enquote {\bibinfo {title} {Discontinuous
  fluidization transition in time-correlated assemblies of actively deforming
  particles},}\ }\href@noop {} {\bibfield  {journal} {\bibinfo  {journal}
  {Phys. Rev. E}\ }\textbf {\bibinfo {volume} {96}},\ \bibinfo {pages} {050601}
  (\bibinfo {year} {2017})}\BibitemShut {NoStop}%
\bibitem [{\citenamefont {Brauchart}\ \emph {et~al.}(2019)\citenamefont
  {Brauchart}, \citenamefont {Grabner},\ and\ \citenamefont
  {Kusner}}]{brauchart2019hyperuniform}%
  \BibitemOpen
  \bibfield  {author} {\bibinfo {author} {\bibfnamefont {J.~S.}\ \bibnamefont
  {Brauchart}}, \bibinfo {author} {\bibfnamefont {P.~J.}\ \bibnamefont
  {Grabner}}, \ and\ \bibinfo {author} {\bibfnamefont {W.}~\bibnamefont
  {Kusner}},\ }\bibfield  {title} {\enquote {\bibinfo {title} {Hyperuniform
  point sets on the sphere: Deterministic aspects},}\ }\href@noop {} {\bibfield
   {journal} {\bibinfo  {journal} {Constr. Approx.}\ }\textbf {\bibinfo
  {volume} {50}},\ \bibinfo {pages} {45--61} (\bibinfo {year}
  {2019})}\BibitemShut {NoStop}%
\bibitem [{\citenamefont {Bo{\v{z}}i{\v{c}}}\ and\ \citenamefont
  {{\v{C}}opar}(2019)}]{bovzivc2019spherical}%
  \BibitemOpen
  \bibfield  {author} {\bibinfo {author} {\bibfnamefont {A.~L.}\ \bibnamefont
  {Bo{\v{z}}i{\v{c}}}}\ and\ \bibinfo {author} {\bibfnamefont {S.}~\bibnamefont
  {{\v{C}}opar}},\ }\bibfield  {title} {\enquote {\bibinfo {title} {Spherical
  structure factor and classification of hyperuniformity on the sphere},}\
  }\href@noop {} {\bibfield  {journal} {\bibinfo  {journal} {Phys. Rev. E}\
  }\textbf {\bibinfo {volume} {99}},\ \bibinfo {pages} {032601} (\bibinfo
  {year} {2019})}\BibitemShut {NoStop}%
\bibitem [{\citenamefont {Galliano}\ \emph {et~al.}(2023)\citenamefont
  {Galliano}, \citenamefont {Cates},\ and\ \citenamefont
  {Berthier}}]{berthier2d}%
  \BibitemOpen
  \bibfield  {author} {\bibinfo {author} {\bibfnamefont {L.}~\bibnamefont
  {Galliano}}, \bibinfo {author} {\bibfnamefont {M.~E.}\ \bibnamefont {Cates}},
  \ and\ \bibinfo {author} {\bibfnamefont {L.}~\bibnamefont {Berthier}},\
  }\bibfield  {title} {\enquote {\bibinfo {title} {Two-dimensional crystals far
  from equilibrium},}\ }\href@noop {} {\bibfield  {journal} {\bibinfo
  {journal} {Phys. Rev. Lett.}\ }\textbf {\bibinfo {volume} {131}},\ \bibinfo
  {pages} {047101} (\bibinfo {year} {2023})}\BibitemShut {NoStop}%
\bibitem [{\citenamefont {Maire}\ and\ \citenamefont
  {Plati}(2024)}]{2dquasilong}%
  \BibitemOpen
  \bibfield  {author} {\bibinfo {author} {\bibfnamefont {R.}~\bibnamefont
  {Maire}}\ and\ \bibinfo {author} {\bibfnamefont {A.}~\bibnamefont {Plati}},\
  }\bibfield  {title} {\enquote {\bibinfo {title} {Enhancing (quasi-)long-range
  order in a two-dimensional driven crystal},}\ }\href@noop {} {\bibfield
  {journal} {\bibinfo  {journal} {arXiv preprint arXiv:2405.05621}\ } (\bibinfo
  {year} {2024})}\BibitemShut {NoStop}%
\bibitem [{\citenamefont {Landau}\ \emph {et~al.}(1992)\citenamefont {Landau},
  \citenamefont {Lifshitz}, \citenamefont {Beyer} \emph
  {et~al.}}]{landau1992hydrodynamic}%
  \BibitemOpen
  \bibfield  {author} {\bibinfo {author} {\bibfnamefont {L.}~\bibnamefont
  {Landau}}, \bibinfo {author} {\bibfnamefont {E.}~\bibnamefont {Lifshitz}},
  \bibinfo {author} {\bibfnamefont {R.}~\bibnamefont {Beyer}},  \emph
  {et~al.},\ }\bibfield  {title} {\enquote {\bibinfo {title} {Hydrodynamic
  fluctuations},}\ }in\ \href@noop {} {\emph {\bibinfo {booktitle}
  {Perspectives in Theoretical Physics}}}\ (\bibinfo  {publisher} {Elsevier},\
  \bibinfo {year} {1992})\ pp.\ \bibinfo {pages} {359--361}\BibitemShut
  {NoStop}%
\bibitem [{\citenamefont {Dybiec}\ \emph {et~al.}(2017)\citenamefont {Dybiec},
  \citenamefont {Gudowska-Nowak},\ and\ \citenamefont {Sokolov}}]{ept}%
  \BibitemOpen
  \bibfield  {author} {\bibinfo {author} {\bibfnamefont {B.}~\bibnamefont
  {Dybiec}}, \bibinfo {author} {\bibfnamefont {E.}~\bibnamefont
  {Gudowska-Nowak}}, \ and\ \bibinfo {author} {\bibfnamefont {I.~M.}\
  \bibnamefont {Sokolov}},\ }\bibfield  {title} {\enquote {\bibinfo {title}
  {Underdamped stochastic harmonic oscillator driven by l\'evy noise},}\
  }\href@noop {} {\bibfield  {journal} {\bibinfo  {journal} {Phys. Rev. E}\
  }\textbf {\bibinfo {volume} {96}},\ \bibinfo {pages} {042118} (\bibinfo
  {year} {2017})}\BibitemShut {NoStop}%
\bibitem [{\citenamefont {Lei}\ \emph {et~al.}(2021)\citenamefont {Lei},
  \citenamefont {Hu},\ and\ \citenamefont {Ni}}]{lei2021barrier}%
  \BibitemOpen
  \bibfield  {author} {\bibinfo {author} {\bibfnamefont {Q.-L.}\ \bibnamefont
  {Lei}}, \bibinfo {author} {\bibfnamefont {H.}~\bibnamefont {Hu}}, \ and\
  \bibinfo {author} {\bibfnamefont {R.}~\bibnamefont {Ni}},\ }\bibfield
  {title} {\enquote {\bibinfo {title} {Barrier-controlled nonequilibrium
  criticality in reactive particle systems},}\ }\href@noop {} {\bibfield
  {journal} {\bibinfo  {journal} {Phys. Rev. E}\ }\textbf {\bibinfo {volume}
  {103}},\ \bibinfo {pages} {052607} (\bibinfo {year} {2021})}\BibitemShut
  {NoStop}%
\bibitem [{\citenamefont {Lei}\ and\ \citenamefont
  {Ni}(2023)}]{lei2023howHUfreeze}%
  \BibitemOpen
  \bibfield  {author} {\bibinfo {author} {\bibfnamefont {Y.}~\bibnamefont
  {Lei}}\ and\ \bibinfo {author} {\bibfnamefont {R.}~\bibnamefont {Ni}},\
  }\bibfield  {title} {\enquote {\bibinfo {title} {How does a hyperuniform
  fluid freeze?}}\ }\href@noop {} {\bibfield  {journal} {\bibinfo  {journal}
  {Proc. Natl. Acad. Sci. U.S.A.}\ }\textbf {\bibinfo {volume} {120}},\
  \bibinfo {pages} {e2312866120} (\bibinfo {year} {2023})}\BibitemShut
  {NoStop}%
\bibitem [{\citenamefont {Zhang}\ and\ \citenamefont
  {Snezhko}(2022)}]{zhang2022hyperuniform}%
  \BibitemOpen
  \bibfield  {author} {\bibinfo {author} {\bibfnamefont {B.}~\bibnamefont
  {Zhang}}\ and\ \bibinfo {author} {\bibfnamefont {A.}~\bibnamefont
  {Snezhko}},\ }\bibfield  {title} {\enquote {\bibinfo {title} {Hyperuniform
  active chiral fluids with tunable internal structure},}\ }\href@noop {}
  {\bibfield  {journal} {\bibinfo  {journal} {Phys. Rev. Lett.}\ }\textbf
  {\bibinfo {volume} {128}},\ \bibinfo {pages} {218002} (\bibinfo {year}
  {2022})}\BibitemShut {NoStop}%
\bibitem [{\citenamefont {Ma}\ and\ \citenamefont {Ni}(2022)}]{mazhan2022}%
  \BibitemOpen
  \bibfield  {author} {\bibinfo {author} {\bibfnamefont {Z.}~\bibnamefont
  {Ma}}\ and\ \bibinfo {author} {\bibfnamefont {R.}~\bibnamefont {Ni}},\
  }\bibfield  {title} {\enquote {\bibinfo {title} {Dynamical clustering
  interrupts motility-induced phase separation in chiral active brownian
  particles},}\ }\href@noop {} {\bibfield  {journal} {\bibinfo  {journal} {J.
  Chem. Phys}\ }\textbf {\bibinfo {volume} {156}},\ \bibinfo {pages} {021102}
  (\bibinfo {year} {2022})}\BibitemShut {NoStop}%
\bibitem [{\citenamefont {Kuroda}\ and\ \citenamefont
  {Miyazaki}(2023)}]{kuroda2023microscopic}%
  \BibitemOpen
  \bibfield  {author} {\bibinfo {author} {\bibfnamefont {Y.}~\bibnamefont
  {Kuroda}}\ and\ \bibinfo {author} {\bibfnamefont {K.}~\bibnamefont
  {Miyazaki}},\ }\bibfield  {title} {\enquote {\bibinfo {title} {Microscopic
  theory for hyperuniformity in two-dimensional chiral active fluid},}\
  }\href@noop {} {\bibfield  {journal} {\bibinfo  {journal} {arXiv preprint
  arXiv:2305.06298}\ } (\bibinfo {year} {2023})}\BibitemShut {NoStop}%
\bibitem [{\citenamefont {Kuroda}\ \emph {et~al.}(2024)\citenamefont {Kuroda},
  \citenamefont {Kawasaki},\ and\ \citenamefont {Miyazaki}}]{kuroda2024}%
  \BibitemOpen
  \bibfield  {author} {\bibinfo {author} {\bibfnamefont {Y.}~\bibnamefont
  {Kuroda}}, \bibinfo {author} {\bibfnamefont {T.}~\bibnamefont {Kawasaki}}, \
  and\ \bibinfo {author} {\bibfnamefont {K.}~\bibnamefont {Miyazaki}},\
  }\bibfield  {title} {\enquote {\bibinfo {title} {Long-range translational
  order and hyperuniformity in two-dimensional chiral active crystal},}\
  }\href@noop {} {\bibfield  {journal} {\bibinfo  {journal} {arXiv preprint
  arXiv:2402.19192}\ } (\bibinfo {year} {2024})}\BibitemShut {NoStop}%
\bibitem [{\citenamefont {Ma}\ and\ \citenamefont
  {Torquato}(2017)}]{ma2017random}%
  \BibitemOpen
  \bibfield  {author} {\bibinfo {author} {\bibfnamefont {Z.}~\bibnamefont
  {Ma}}\ and\ \bibinfo {author} {\bibfnamefont {S.}~\bibnamefont {Torquato}},\
  }\bibfield  {title} {\enquote {\bibinfo {title} {Random scalar fields and
  hyperuniformity},}\ }\href@noop {} {\bibfield  {journal} {\bibinfo  {journal}
  {J. Appl. Phys.}\ }\textbf {\bibinfo {volume} {121}} (\bibinfo {year}
  {2017})}\BibitemShut {NoStop}%
\bibitem [{\citenamefont {Shimizu}\ and\ \citenamefont
  {Tanaka}(2015)}]{shimizu2015novel}%
  \BibitemOpen
  \bibfield  {author} {\bibinfo {author} {\bibfnamefont {R.}~\bibnamefont
  {Shimizu}}\ and\ \bibinfo {author} {\bibfnamefont {H.}~\bibnamefont
  {Tanaka}},\ }\bibfield  {title} {\enquote {\bibinfo {title} {A novel
  coarsening mechanism of droplets in immiscible fluid mixtures},}\ }\href@noop
  {} {\bibfield  {journal} {\bibinfo  {journal} {Nat. Comm.}\ }\textbf
  {\bibinfo {volume} {6}},\ \bibinfo {pages} {7407} (\bibinfo {year}
  {2015})}\BibitemShut {NoStop}%
\bibitem [{\citenamefont {Rundman}\ and\ \citenamefont
  {Hilliard}(1967)}]{rundman1967early}%
  \BibitemOpen
  \bibfield  {author} {\bibinfo {author} {\bibfnamefont {K.}~\bibnamefont
  {Rundman}}\ and\ \bibinfo {author} {\bibfnamefont {J.}~\bibnamefont
  {Hilliard}},\ }\bibfield  {title} {\enquote {\bibinfo {title} {Early stages
  of spinodal decomposition in an aluminum-zinc alloy},}\ }\href@noop {}
  {\bibfield  {journal} {\bibinfo  {journal} {Acta. Metall.}\ }\textbf
  {\bibinfo {volume} {15}},\ \bibinfo {pages} {1025--1033} (\bibinfo {year}
  {1967})}\BibitemShut {NoStop}%
\bibitem [{\citenamefont {Langer}(1971)}]{langer1971theory}%
  \BibitemOpen
  \bibfield  {author} {\bibinfo {author} {\bibfnamefont {J.~S.}\ \bibnamefont
  {Langer}},\ }\bibfield  {title} {\enquote {\bibinfo {title} {Theory of
  spinodal decomposition in alloys},}\ }\href@noop {} {\bibfield  {journal}
  {\bibinfo  {journal} {Ann. Phys.}\ }\textbf {\bibinfo {volume} {65}},\
  \bibinfo {pages} {53--86} (\bibinfo {year} {1971})}\BibitemShut {NoStop}%
\bibitem [{\citenamefont {Findik}(2012)}]{findik2012improvements}%
  \BibitemOpen
  \bibfield  {author} {\bibinfo {author} {\bibfnamefont {F.}~\bibnamefont
  {Findik}},\ }\bibfield  {title} {\enquote {\bibinfo {title} {Improvements in
  spinodal alloys from past to present},}\ }\href@noop {} {\bibfield  {journal}
  {\bibinfo  {journal} {Mater. Des.}\ }\textbf {\bibinfo {volume} {42}},\
  \bibinfo {pages} {131--146} (\bibinfo {year} {2012})}\BibitemShut {NoStop}%
\bibitem [{\citenamefont {de~Gennes}(1980)}]{de1980dynamics}%
  \BibitemOpen
  \bibfield  {author} {\bibinfo {author} {\bibfnamefont {P.~G.}\ \bibnamefont
  {de~Gennes}},\ }\bibfield  {title} {\enquote {\bibinfo {title} {Dynamics of
  fluctuations and spinodal decomposition in polymer blends},}\ }\href@noop {}
  {\bibfield  {journal} {\bibinfo  {journal} {J. Chem. Phys}\ }\textbf
  {\bibinfo {volume} {72}},\ \bibinfo {pages} {4756--4763} (\bibinfo {year}
  {1980})}\BibitemShut {NoStop}%
\bibitem [{\citenamefont {Bruder}\ and\ \citenamefont
  {Brenn}(1992)}]{bruder1992spinodal}%
  \BibitemOpen
  \bibfield  {author} {\bibinfo {author} {\bibfnamefont {F.}~\bibnamefont
  {Bruder}}\ and\ \bibinfo {author} {\bibfnamefont {R.}~\bibnamefont {Brenn}},\
  }\bibfield  {title} {\enquote {\bibinfo {title} {Spinodal decomposition in
  thin films of a polymer blend},}\ }\href@noop {} {\bibfield  {journal}
  {\bibinfo  {journal} {Phys. Rev. Lett.}\ }\textbf {\bibinfo {volume} {69}},\
  \bibinfo {pages} {624} (\bibinfo {year} {1992})}\BibitemShut {NoStop}%
\bibitem [{\citenamefont {Laradji}\ \emph {et~al.}(1996)\citenamefont
  {Laradji}, \citenamefont {Mouritsen},\ and\ \citenamefont
  {Toxvaerd}}]{laradji1996spinodal}%
  \BibitemOpen
  \bibfield  {author} {\bibinfo {author} {\bibfnamefont {M.}~\bibnamefont
  {Laradji}}, \bibinfo {author} {\bibfnamefont {O.~G.}\ \bibnamefont
  {Mouritsen}}, \ and\ \bibinfo {author} {\bibfnamefont {S.}~\bibnamefont
  {Toxvaerd}},\ }\bibfield  {title} {\enquote {\bibinfo {title} {Spinodal
  decomposition in multicomponent fluid mixtures: A molecular dynamics
  study},}\ }\href@noop {} {\bibfield  {journal} {\bibinfo  {journal} {Phys.
  Rev. E}\ }\textbf {\bibinfo {volume} {53}},\ \bibinfo {pages} {3673}
  (\bibinfo {year} {1996})}\BibitemShut {NoStop}%
\bibitem [{\citenamefont {Tanaka}(1996)}]{tanaka1996coarsening}%
  \BibitemOpen
  \bibfield  {author} {\bibinfo {author} {\bibfnamefont {H.}~\bibnamefont
  {Tanaka}},\ }\bibfield  {title} {\enquote {\bibinfo {title} {Coarsening
  mechanisms of droplet spinodal decomposition in binary fluid mixtures},}\
  }\href@noop {} {\bibfield  {journal} {\bibinfo  {journal} {J. Chem. Phys}\
  }\textbf {\bibinfo {volume} {105}},\ \bibinfo {pages} {10099--10114}
  (\bibinfo {year} {1996})}\BibitemShut {NoStop}%
\bibitem [{\citenamefont {Cahn}\ and\ \citenamefont
  {Hilliard}(1958)}]{cahn1958free}%
  \BibitemOpen
  \bibfield  {author} {\bibinfo {author} {\bibfnamefont {J.~W.}\ \bibnamefont
  {Cahn}}\ and\ \bibinfo {author} {\bibfnamefont {J.~E.}\ \bibnamefont
  {Hilliard}},\ }\bibfield  {title} {\enquote {\bibinfo {title} {Free energy of
  a nonuniform system. i. interfacial free energy},}\ }\href@noop {} {\bibfield
   {journal} {\bibinfo  {journal} {J. Chem. Phys.}\ }\textbf {\bibinfo {volume}
  {28}},\ \bibinfo {pages} {258--267} (\bibinfo {year} {1958})}\BibitemShut
  {NoStop}%
\bibitem [{\citenamefont {Hohenberg}\ and\ \citenamefont
  {Halperin}(1977)}]{hohenberg1977theory}%
  \BibitemOpen
  \bibfield  {author} {\bibinfo {author} {\bibfnamefont {P.~C.}\ \bibnamefont
  {Hohenberg}}\ and\ \bibinfo {author} {\bibfnamefont {B.~I.}\ \bibnamefont
  {Halperin}},\ }\bibfield  {title} {\enquote {\bibinfo {title} {Theory of
  dynamic critical phenomena},}\ }\href@noop {} {\bibfield  {journal} {\bibinfo
   {journal} {Rev. Mod. Phys.}\ }\textbf {\bibinfo {volume} {49}},\ \bibinfo
  {pages} {435} (\bibinfo {year} {1977})}\BibitemShut {NoStop}%
\bibitem [{\citenamefont {Komura}\ \emph {et~al.}(1984)\citenamefont {Komura},
  \citenamefont {Osamura}, \citenamefont {Fujii},\ and\ \citenamefont
  {Takeda}}]{komura1984dynamical}%
  \BibitemOpen
  \bibfield  {author} {\bibinfo {author} {\bibfnamefont {S.}~\bibnamefont
  {Komura}}, \bibinfo {author} {\bibfnamefont {K.}~\bibnamefont {Osamura}},
  \bibinfo {author} {\bibfnamefont {H.}~\bibnamefont {Fujii}}, \ and\ \bibinfo
  {author} {\bibfnamefont {T.}~\bibnamefont {Takeda}},\ }\bibfield  {title}
  {\enquote {\bibinfo {title} {Dynamical scaling of the structure function in
  quenched al-zn and al-zn-mg alloys},}\ }\href@noop {} {\bibfield  {journal}
  {\bibinfo  {journal} {Phys. Rev. B}\ }\textbf {\bibinfo {volume} {30}},\
  \bibinfo {pages} {2944} (\bibinfo {year} {1984})}\BibitemShut {NoStop}%
\bibitem [{\citenamefont {Katano}\ and\ \citenamefont
  {Iizumi}(1984)}]{katano1984crossover}%
  \BibitemOpen
  \bibfield  {author} {\bibinfo {author} {\bibfnamefont {S.}~\bibnamefont
  {Katano}}\ and\ \bibinfo {author} {\bibfnamefont {M.}~\bibnamefont
  {Iizumi}},\ }\bibfield  {title} {\enquote {\bibinfo {title} {Crossover
  phenomenon in dynamical scaling of phase separation in fe-cr alloy},}\
  }\href@noop {} {\bibfield  {journal} {\bibinfo  {journal} {Phys. Rev. Lett.}\
  }\textbf {\bibinfo {volume} {52}},\ \bibinfo {pages} {835} (\bibinfo {year}
  {1984})}\BibitemShut {NoStop}%
\bibitem [{\citenamefont {Wiltzius}\ \emph {et~al.}(1988)\citenamefont
  {Wiltzius}, \citenamefont {Bates},\ and\ \citenamefont
  {Heffner}}]{wiltzius1988spinodal}%
  \BibitemOpen
  \bibfield  {author} {\bibinfo {author} {\bibfnamefont {P.}~\bibnamefont
  {Wiltzius}}, \bibinfo {author} {\bibfnamefont {F.~S.}\ \bibnamefont {Bates}},
  \ and\ \bibinfo {author} {\bibfnamefont {W.}~\bibnamefont {Heffner}},\
  }\bibfield  {title} {\enquote {\bibinfo {title} {Spinodal decomposition in
  isotopic polymer mixtures},}\ }\href@noop {} {\bibfield  {journal} {\bibinfo
  {journal} {Phys. Rev. Lett.}\ }\textbf {\bibinfo {volume} {60}},\ \bibinfo
  {pages} {1538} (\bibinfo {year} {1988})}\BibitemShut {NoStop}%
\bibitem [{\citenamefont {Hashimoto}\ \emph {et~al.}(1986)\citenamefont
  {Hashimoto}, \citenamefont {Itakura},\ and\ \citenamefont
  {Hasegawa}}]{hashimoto1986late}%
  \BibitemOpen
  \bibfield  {author} {\bibinfo {author} {\bibfnamefont {T.}~\bibnamefont
  {Hashimoto}}, \bibinfo {author} {\bibfnamefont {M.}~\bibnamefont {Itakura}},
  \ and\ \bibinfo {author} {\bibfnamefont {H.}~\bibnamefont {Hasegawa}},\
  }\bibfield  {title} {\enquote {\bibinfo {title} {Late stage spinodal
  decomposition of a binary polymer mixture. i. critical test of dynamical
  scaling on scattering function},}\ }\href@noop {} {\bibfield  {journal}
  {\bibinfo  {journal} {J. Chem. Phys.}\ }\textbf {\bibinfo {volume} {85}},\
  \bibinfo {pages} {6118--6128} (\bibinfo {year} {1986})}\BibitemShut {NoStop}%
\bibitem [{\citenamefont {Yeung}(1988)}]{yeung1988scaling}%
  \BibitemOpen
  \bibfield  {author} {\bibinfo {author} {\bibfnamefont {C.}~\bibnamefont
  {Yeung}},\ }\bibfield  {title} {\enquote {\bibinfo {title} {Scaling and the
  small-wave-vector limit of the form factor in phase-ordering dynamics},}\
  }\href@noop {} {\bibfield  {journal} {\bibinfo  {journal} {Phys. Rev. Lett.}\
  }\textbf {\bibinfo {volume} {61}},\ \bibinfo {pages} {1135} (\bibinfo {year}
  {1988})}\BibitemShut {NoStop}%
\bibitem [{\citenamefont {Shinozaki}\ and\ \citenamefont
  {Oono}(1993)}]{shinozaki1993spinodal}%
  \BibitemOpen
  \bibfield  {author} {\bibinfo {author} {\bibfnamefont {A.}~\bibnamefont
  {Shinozaki}}\ and\ \bibinfo {author} {\bibfnamefont {Y.}~\bibnamefont
  {Oono}},\ }\bibfield  {title} {\enquote {\bibinfo {title} {Spinodal
  decomposition in 3-space},}\ }\href@noop {} {\bibfield  {journal} {\bibinfo
  {journal} {Phys. Rev. E}\ }\textbf {\bibinfo {volume} {48}},\ \bibinfo
  {pages} {2622} (\bibinfo {year} {1993})}\BibitemShut {NoStop}%
\bibitem [{\citenamefont {Kabrede}\ and\ \citenamefont
  {Hentschke}(2006)}]{kabrede2006spinodal}%
  \BibitemOpen
  \bibfield  {author} {\bibinfo {author} {\bibfnamefont {H.}~\bibnamefont
  {Kabrede}}\ and\ \bibinfo {author} {\bibfnamefont {R.}~\bibnamefont
  {Hentschke}},\ }\bibfield  {title} {\enquote {\bibinfo {title} {Spinodal
  decomposition in a 3d lennard--jones system},}\ }\href@noop {} {\bibfield
  {journal} {\bibinfo  {journal} {Phys. A: Stat. Mech. Appl.}\ }\textbf
  {\bibinfo {volume} {361}},\ \bibinfo {pages} {485--493} (\bibinfo {year}
  {2006})}\BibitemShut {NoStop}%
\bibitem [{\citenamefont {Midya}\ and\ \citenamefont
  {Das}(2020)}]{midya2020kinetics}%
  \BibitemOpen
  \bibfield  {author} {\bibinfo {author} {\bibfnamefont {J.}~\bibnamefont
  {Midya}}\ and\ \bibinfo {author} {\bibfnamefont {S.~K.}\ \bibnamefont
  {Das}},\ }\bibfield  {title} {\enquote {\bibinfo {title} {Kinetics of domain
  growth and aging in a two-dimensional off-lattice system},}\ }\href@noop {}
  {\bibfield  {journal} {\bibinfo  {journal} {Phys. Rev. E}\ }\textbf {\bibinfo
  {volume} {102}},\ \bibinfo {pages} {062119} (\bibinfo {year}
  {2020})}\BibitemShut {NoStop}%
\bibitem [{\citenamefont {Tomita}(1991)}]{tomita1991preservation}%
  \BibitemOpen
  \bibfield  {author} {\bibinfo {author} {\bibfnamefont {H.}~\bibnamefont
  {Tomita}},\ }\bibfield  {title} {\enquote {\bibinfo {title} {Preservation of
  isotropy at the mesoscopic stage of phase separation processes},}\
  }\href@noop {} {\bibfield  {journal} {\bibinfo  {journal} {Prog. Theor.
  Phys.}\ }\textbf {\bibinfo {volume} {85}},\ \bibinfo {pages} {47--56}
  (\bibinfo {year} {1991})}\BibitemShut {NoStop}%
\bibitem [{\citenamefont {Mazenko}(1994)}]{mazenko1994growth}%
  \BibitemOpen
  \bibfield  {author} {\bibinfo {author} {\bibfnamefont {G.~F.}\ \bibnamefont
  {Mazenko}},\ }\bibfield  {title} {\enquote {\bibinfo {title} {Growth kinetics
  for a system with a conserved order parameter},}\ }\href@noop {} {\bibfield
  {journal} {\bibinfo  {journal} {Phys. Rev. E}\ }\textbf {\bibinfo {volume}
  {50}},\ \bibinfo {pages} {3485} (\bibinfo {year} {1994})}\BibitemShut
  {NoStop}%
\bibitem [{\citenamefont {Bray}(2002)}]{bray2002theory}%
  \BibitemOpen
  \bibfield  {author} {\bibinfo {author} {\bibfnamefont {A.~J.}\ \bibnamefont
  {Bray}},\ }\bibfield  {title} {\enquote {\bibinfo {title} {Theory of
  phase-ordering kinetics},}\ }\href@noop {} {\bibfield  {journal} {\bibinfo
  {journal} {Adv. Phys.}\ }\textbf {\bibinfo {volume} {51}},\ \bibinfo {pages}
  {481--587} (\bibinfo {year} {2002})}\BibitemShut {NoStop}%
\bibitem [{\citenamefont {Salvalaglio}\ \emph {et~al.}(2020)\citenamefont
  {Salvalaglio}, \citenamefont {Bouabdellaoui}, \citenamefont {Bollani},
  \citenamefont {Benali}, \citenamefont {Favre}, \citenamefont {Claude},
  \citenamefont {Wenger}, \citenamefont {de~Anna}, \citenamefont {Intonti},
  \citenamefont {Voigt} \emph {et~al.}}]{salvalaglio2020hyperuniform}%
  \BibitemOpen
  \bibfield  {author} {\bibinfo {author} {\bibfnamefont {M.}~\bibnamefont
  {Salvalaglio}}, \bibinfo {author} {\bibfnamefont {M.}~\bibnamefont
  {Bouabdellaoui}}, \bibinfo {author} {\bibfnamefont {M.}~\bibnamefont
  {Bollani}}, \bibinfo {author} {\bibfnamefont {A.}~\bibnamefont {Benali}},
  \bibinfo {author} {\bibfnamefont {L.}~\bibnamefont {Favre}}, \bibinfo
  {author} {\bibfnamefont {J.~B.}\ \bibnamefont {Claude}}, \bibinfo {author}
  {\bibfnamefont {J.}~\bibnamefont {Wenger}}, \bibinfo {author} {\bibfnamefont
  {P.}~\bibnamefont {de~Anna}}, \bibinfo {author} {\bibfnamefont
  {F.}~\bibnamefont {Intonti}}, \bibinfo {author} {\bibfnamefont
  {A.}~\bibnamefont {Voigt}},  \emph {et~al.},\ }\bibfield  {title} {\enquote
  {\bibinfo {title} {Hyperuniform monocrystalline structures by spinodal
  solid-state dewetting},}\ }\href@noop {} {\bibfield  {journal} {\bibinfo
  {journal} {Phys. Rev. Lett.}\ }\textbf {\bibinfo {volume} {125}},\ \bibinfo
  {pages} {126101} (\bibinfo {year} {2020})}\BibitemShut {NoStop}%
\bibitem [{\citenamefont {Wilken}\ \emph
  {et~al.}(2023{\natexlab{b}})\citenamefont {Wilken}, \citenamefont
  {Chaderjian},\ and\ \citenamefont {Saleh}}]{wilken2023spatial}%
  \BibitemOpen
  \bibfield  {author} {\bibinfo {author} {\bibfnamefont {S.}~\bibnamefont
  {Wilken}}, \bibinfo {author} {\bibfnamefont {A.}~\bibnamefont {Chaderjian}},
  \ and\ \bibinfo {author} {\bibfnamefont {O.~A.}\ \bibnamefont {Saleh}},\
  }\bibfield  {title} {\enquote {\bibinfo {title} {Spatial organization of
  phase-separated dna droplets},}\ }\href@noop {} {\bibfield  {journal}
  {\bibinfo  {journal} {Phys. Rev. X}\ }\textbf {\bibinfo {volume} {13}},\
  \bibinfo {pages} {031014} (\bibinfo {year} {2023}{\natexlab{b}})}\BibitemShut
  {NoStop}%
\bibitem [{\citenamefont {Zheng}\ \emph {et~al.}(2023)\citenamefont {Zheng},
  \citenamefont {Klatt},\ and\ \citenamefont {L{\"o}wen}}]{zheng2023universal}%
  \BibitemOpen
  \bibfield  {author} {\bibinfo {author} {\bibfnamefont {Y.}~\bibnamefont
  {Zheng}}, \bibinfo {author} {\bibfnamefont {M.~A.}\ \bibnamefont {Klatt}}, \
  and\ \bibinfo {author} {\bibfnamefont {H.}~\bibnamefont {L{\"o}wen}},\
  }\bibfield  {title} {\enquote {\bibinfo {title} {Universal hyperuniformity in
  active field theories},}\ }\href@noop {} {\bibfield  {journal} {\bibinfo
  {journal} {arXiv preprint arXiv:2310.03107}\ } (\bibinfo {year}
  {2023})}\BibitemShut {NoStop}%
\bibitem [{\citenamefont {Cates}\ and\ \citenamefont
  {Tailleur}(2015)}]{cates_review}%
  \BibitemOpen
  \bibfield  {author} {\bibinfo {author} {\bibfnamefont {M.~E.}\ \bibnamefont
  {Cates}}\ and\ \bibinfo {author} {\bibfnamefont {J.}~\bibnamefont
  {Tailleur}},\ }\bibfield  {title} {\enquote {\bibinfo {title}
  {Motility-induced phase separation},}\ }\href@noop {} {\bibfield  {journal}
  {\bibinfo  {journal} {Annu. Rev. Condens. Matter Phys.}\ }\textbf {\bibinfo
  {volume} {6}},\ \bibinfo {pages} {219--244} (\bibinfo {year}
  {2015})}\BibitemShut {NoStop}%
\bibitem [{\citenamefont {Ma}\ \emph {et~al.}(2020)\citenamefont {Ma},
  \citenamefont {Yang},\ and\ \citenamefont {Ni}}]{mazhan2020}%
  \BibitemOpen
  \bibfield  {author} {\bibinfo {author} {\bibfnamefont {Z.}~\bibnamefont
  {Ma}}, \bibinfo {author} {\bibfnamefont {M.}~\bibnamefont {Yang}}, \ and\
  \bibinfo {author} {\bibfnamefont {R.}~\bibnamefont {Ni}},\ }\bibfield
  {title} {\enquote {\bibinfo {title} {Dynamic assembly of active colloids:
  Theory and simulation},}\ }\href@noop {} {\bibfield  {journal} {\bibinfo
  {journal} {Adv. Theory Simul.}\ }\textbf {\bibinfo {volume} {3}},\ \bibinfo
  {pages} {2000021} (\bibinfo {year} {2020})}\BibitemShut {NoStop}%
\bibitem [{\citenamefont {Speck}\ \emph {et~al.}(2014)\citenamefont {Speck},
  \citenamefont {Bialk{\'e}}, \citenamefont {Menzel},\ and\ \citenamefont
  {L{\"o}wen}}]{speck2014effective}%
  \BibitemOpen
  \bibfield  {author} {\bibinfo {author} {\bibfnamefont {T.}~\bibnamefont
  {Speck}}, \bibinfo {author} {\bibfnamefont {J.}~\bibnamefont {Bialk{\'e}}},
  \bibinfo {author} {\bibfnamefont {A.~M.}\ \bibnamefont {Menzel}}, \ and\
  \bibinfo {author} {\bibfnamefont {H.}~\bibnamefont {L{\"o}wen}},\ }\bibfield
  {title} {\enquote {\bibinfo {title} {Effective cahn-hilliard equation for the
  phase separation of active brownian particles},}\ }\href@noop {} {\bibfield
  {journal} {\bibinfo  {journal} {Phys. Rev. Lett.}\ }\textbf {\bibinfo
  {volume} {112}},\ \bibinfo {pages} {218304} (\bibinfo {year}
  {2014})}\BibitemShut {NoStop}%
\bibitem [{\citenamefont {Wittkowski}\ \emph {et~al.}(2014)\citenamefont
  {Wittkowski}, \citenamefont {Tiribocchi}, \citenamefont {Stenhammar},
  \citenamefont {Allen}, \citenamefont {Marenduzzo},\ and\ \citenamefont
  {Cates}}]{wittkowski2014scalar}%
  \BibitemOpen
  \bibfield  {author} {\bibinfo {author} {\bibfnamefont {R.}~\bibnamefont
  {Wittkowski}}, \bibinfo {author} {\bibfnamefont {A.}~\bibnamefont
  {Tiribocchi}}, \bibinfo {author} {\bibfnamefont {J.}~\bibnamefont
  {Stenhammar}}, \bibinfo {author} {\bibfnamefont {R.~J.}\ \bibnamefont
  {Allen}}, \bibinfo {author} {\bibfnamefont {D.}~\bibnamefont {Marenduzzo}}, \
  and\ \bibinfo {author} {\bibfnamefont {M.~E.}\ \bibnamefont {Cates}},\
  }\bibfield  {title} {\enquote {\bibinfo {title} {Scalar $\varphi$ 4 field
  theory for active-particle phase separation},}\ }\href@noop {} {\bibfield
  {journal} {\bibinfo  {journal} {Nat. Comm.}\ }\textbf {\bibinfo {volume}
  {5}},\ \bibinfo {pages} {4351} (\bibinfo {year} {2014})}\BibitemShut
  {NoStop}%
\bibitem [{\citenamefont {Tjhung}\ \emph {et~al.}(2018)\citenamefont {Tjhung},
  \citenamefont {Nardini},\ and\ \citenamefont {Cates}}]{tjhung2018cluster}%
  \BibitemOpen
  \bibfield  {author} {\bibinfo {author} {\bibfnamefont {E.}~\bibnamefont
  {Tjhung}}, \bibinfo {author} {\bibfnamefont {C.}~\bibnamefont {Nardini}}, \
  and\ \bibinfo {author} {\bibfnamefont {M.~E.}\ \bibnamefont {Cates}},\
  }\bibfield  {title} {\enquote {\bibinfo {title} {Cluster phases and bubbly
  phase separation in active fluids: reversal of the ostwald process},}\
  }\href@noop {} {\bibfield  {journal} {\bibinfo  {journal} {Phys. Rev. X}\
  }\textbf {\bibinfo {volume} {8}},\ \bibinfo {pages} {031080} (\bibinfo {year}
  {2018})}\BibitemShut {NoStop}%
\bibitem [{\citenamefont {Reichhardt}\ \emph {et~al.}(2023)\citenamefont
  {Reichhardt}, \citenamefont {Regev}, \citenamefont {Dahmen}, \citenamefont
  {Okuma},\ and\ \citenamefont {Reichhardt}}]{reichhardt2023reversible}%
  \BibitemOpen
  \bibfield  {author} {\bibinfo {author} {\bibfnamefont {C.}~\bibnamefont
  {Reichhardt}}, \bibinfo {author} {\bibfnamefont {I.}~\bibnamefont {Regev}},
  \bibinfo {author} {\bibfnamefont {K.}~\bibnamefont {Dahmen}}, \bibinfo
  {author} {\bibfnamefont {S.}~\bibnamefont {Okuma}}, \ and\ \bibinfo {author}
  {\bibfnamefont {C.~J.~O.}\ \bibnamefont {Reichhardt}},\ }\bibfield  {title}
  {\enquote {\bibinfo {title} {Reversible to irreversible transitions in
  periodic driven many-body systems and future directions for classical and
  quantum systems},}\ }\href@noop {} {\bibfield  {journal} {\bibinfo  {journal}
  {Phys. Rev. Res.}\ }\textbf {\bibinfo {volume} {5}},\ \bibinfo {pages}
  {021001} (\bibinfo {year} {2023})}\BibitemShut {NoStop}%
\bibitem [{\citenamefont {Jocteur}\ \emph {et~al.}(2024)\citenamefont
  {Jocteur}, \citenamefont {Figueiredo}, \citenamefont {Martens}, \citenamefont
  {Bertin},\ and\ \citenamefont {Mari}}]{jocteur2024yielding}%
  \BibitemOpen
  \bibfield  {author} {\bibinfo {author} {\bibfnamefont {T.}~\bibnamefont
  {Jocteur}}, \bibinfo {author} {\bibfnamefont {S.}~\bibnamefont {Figueiredo}},
  \bibinfo {author} {\bibfnamefont {K.}~\bibnamefont {Martens}}, \bibinfo
  {author} {\bibfnamefont {E.}~\bibnamefont {Bertin}}, \ and\ \bibinfo {author}
  {\bibfnamefont {R.}~\bibnamefont {Mari}},\ }\bibfield  {title} {\enquote
  {\bibinfo {title} {Yielding is an absorbing phase transition with vanishing
  critical fluctuations},}\ }\href@noop {} {\bibfield  {journal} {\bibinfo
  {journal} {Phys. Rev. Lett.}\ }\textbf {\bibinfo {volume} {132}},\ \bibinfo
  {pages} {268203} (\bibinfo {year} {2024})}\BibitemShut {NoStop}%
\bibitem [{\citenamefont {Reichhardt}\ and\ \citenamefont
  {Reichhardt}(2009)}]{reichhardt2009random}%
  \BibitemOpen
  \bibfield  {author} {\bibinfo {author} {\bibfnamefont {C.}~\bibnamefont
  {Reichhardt}}\ and\ \bibinfo {author} {\bibfnamefont {CJ~O.}\ \bibnamefont
  {Reichhardt}},\ }\bibfield  {title} {\enquote {\bibinfo {title} {Random
  organization and plastic depinning},}\ }\href@noop {} {\bibfield  {journal}
  {\bibinfo  {journal} {Phy. Rev. Lett.}\ }\textbf {\bibinfo {volume} {103}},\
  \bibinfo {pages} {168301} (\bibinfo {year} {2009})}\BibitemShut {NoStop}%
\bibitem [{\citenamefont {Yao}\ \emph {et~al.}(2020)\citenamefont {Yao},
  \citenamefont {Nayak}, \citenamefont {Balents},\ and\ \citenamefont
  {Zaletel}}]{yao2020classical}%
  \BibitemOpen
  \bibfield  {author} {\bibinfo {author} {\bibfnamefont {N.~Y.}\ \bibnamefont
  {Yao}}, \bibinfo {author} {\bibfnamefont {C.}~\bibnamefont {Nayak}}, \bibinfo
  {author} {\bibfnamefont {L.}~\bibnamefont {Balents}}, \ and\ \bibinfo
  {author} {\bibfnamefont {M.~P.}\ \bibnamefont {Zaletel}},\ }\bibfield
  {title} {\enquote {\bibinfo {title} {Classical discrete time crystals},}\
  }\href@noop {} {\bibfield  {journal} {\bibinfo  {journal} {Nat. Phys.}\
  }\textbf {\bibinfo {volume} {16}},\ \bibinfo {pages} {438--447} (\bibinfo
  {year} {2020})}\BibitemShut {NoStop}%
\bibitem [{\citenamefont {Ikeda}(2023{\natexlab{a}})}]{ikeda2023correlated}%
  \BibitemOpen
  \bibfield  {author} {\bibinfo {author} {\bibfnamefont {H.}~\bibnamefont
  {Ikeda}},\ }\bibfield  {title} {\enquote {\bibinfo {title} {Correlated noise
  and critical dimensions},}\ }\href@noop {} {\bibfield  {journal} {\bibinfo
  {journal} {Phys. Rev. E}\ }\textbf {\bibinfo {volume} {108}},\ \bibinfo
  {pages} {064119} (\bibinfo {year} {2023}{\natexlab{a}})}\BibitemShut
  {NoStop}%
\bibitem [{\citenamefont {Ikeda}(2023{\natexlab{b}})}]{ikeda2023harmonic}%
  \BibitemOpen
  \bibfield  {author} {\bibinfo {author} {\bibfnamefont {H.}~\bibnamefont
  {Ikeda}},\ }\bibfield  {title} {\enquote {\bibinfo {title} {Harmonic chain
  far from equilibrium: single-file diffusion, long-range order, and
  hyperuniformity},}\ }\href@noop {} {\bibfield  {journal} {\bibinfo  {journal}
  {arXiv preprint arXiv:2309.03155}\ } (\bibinfo {year}
  {2023}{\natexlab{b}})}\BibitemShut {NoStop}%
\bibitem [{\citenamefont {De~Luca}\ \emph {et~al.}(2024)\citenamefont
  {De~Luca}, \citenamefont {Ma}, \citenamefont {Nardini},\ and\ \citenamefont
  {Cates}}]{de2024hyperuniformity}%
  \BibitemOpen
  \bibfield  {author} {\bibinfo {author} {\bibfnamefont {F.}~\bibnamefont
  {De~Luca}}, \bibinfo {author} {\bibfnamefont {X.}~\bibnamefont {Ma}},
  \bibinfo {author} {\bibfnamefont {C.}~\bibnamefont {Nardini}}, \ and\
  \bibinfo {author} {\bibfnamefont {M.~E.}\ \bibnamefont {Cates}},\ }\bibfield
  {title} {\enquote {\bibinfo {title} {Hyperuniformity in phase ordering: the
  roles of activity, noise, and non-constant mobility},}\ }\href@noop {}
  {\bibfield  {journal} {\bibinfo  {journal} {J. Condens. Matter Phys.}\
  }\textbf {\bibinfo {volume} {36}},\ \bibinfo {pages} {405101} (\bibinfo
  {year} {2024})}\BibitemShut {NoStop}%
\bibitem [{\citenamefont {Le~Thien}\ \emph {et~al.}(2017)\citenamefont
  {Le~Thien}, \citenamefont {McDermott}, \citenamefont {Reichhardt},\ and\
  \citenamefont {Reichhardt}}]{le2017enhanced}%
  \BibitemOpen
  \bibfield  {author} {\bibinfo {author} {\bibfnamefont {Q.}~\bibnamefont
  {Le~Thien}}, \bibinfo {author} {\bibfnamefont {D.}~\bibnamefont {McDermott}},
  \bibinfo {author} {\bibfnamefont {CJO}\ \bibnamefont {Reichhardt}}, \ and\
  \bibinfo {author} {\bibfnamefont {C.}~\bibnamefont {Reichhardt}},\ }\bibfield
   {title} {\enquote {\bibinfo {title} {Enhanced pinning for vortices in
  hyperuniform pinning arrays and emergent hyperuniform vortex configurations
  with quenched disorder},}\ }\href@noop {} {\bibfield  {journal} {\bibinfo
  {journal} {Phys. Rev. B}\ }\textbf {\bibinfo {volume} {96}},\ \bibinfo
  {pages} {094516} (\bibinfo {year} {2017})}\BibitemShut {NoStop}%
\bibitem [{\citenamefont {Rumi}\ \emph {et~al.}(2019)\citenamefont {Rumi},
  \citenamefont {Arag{\'o}n~S{\'a}nchez}, \citenamefont {El{\'\i}as},
  \citenamefont {Cortes~Maldonado}, \citenamefont {Puig}, \citenamefont
  {Cejas~Bolecek}, \citenamefont {Nieva}, \citenamefont {Konczykowski},
  \citenamefont {Fasano},\ and\ \citenamefont {Kolton}}]{rumi2019hyperuniform}%
  \BibitemOpen
  \bibfield  {author} {\bibinfo {author} {\bibfnamefont {G.}~\bibnamefont
  {Rumi}}, \bibinfo {author} {\bibfnamefont {J.}~\bibnamefont
  {Arag{\'o}n~S{\'a}nchez}}, \bibinfo {author} {\bibfnamefont {F.}~\bibnamefont
  {El{\'\i}as}}, \bibinfo {author} {\bibfnamefont {R.}~\bibnamefont
  {Cortes~Maldonado}}, \bibinfo {author} {\bibfnamefont {J.}~\bibnamefont
  {Puig}}, \bibinfo {author} {\bibfnamefont {N.~R.}\ \bibnamefont
  {Cejas~Bolecek}}, \bibinfo {author} {\bibfnamefont {G.}~\bibnamefont
  {Nieva}}, \bibinfo {author} {\bibfnamefont {M.}~\bibnamefont {Konczykowski}},
  \bibinfo {author} {\bibfnamefont {Y.}~\bibnamefont {Fasano}}, \ and\ \bibinfo
  {author} {\bibfnamefont {A.~B.}\ \bibnamefont {Kolton}},\ }\bibfield  {title}
  {\enquote {\bibinfo {title} {Hyperuniform vortex patterns at the surface of
  type-ii superconductors},}\ }\href@noop {} {\bibfield  {journal} {\bibinfo
  {journal} {Phys. Rev. Res.}\ }\textbf {\bibinfo {volume} {1}},\ \bibinfo
  {pages} {033057} (\bibinfo {year} {2019})}\BibitemShut {NoStop}%
\bibitem [{\citenamefont {Llorens}\ \emph {et~al.}(2020)\citenamefont
  {Llorens}, \citenamefont {Guillam{\'o}n}, \citenamefont {Serrano},
  \citenamefont {C{\'o}rdoba}, \citenamefont {Ses{\'e}}, \citenamefont
  {De~Teresa}, \citenamefont {Ibarra}, \citenamefont {Vieira}, \citenamefont
  {Ortu{\~n}o},\ and\ \citenamefont {Suderow}}]{llorens2020disordered}%
  \BibitemOpen
  \bibfield  {author} {\bibinfo {author} {\bibfnamefont {J.~B.}\ \bibnamefont
  {Llorens}}, \bibinfo {author} {\bibfnamefont {I.}~\bibnamefont
  {Guillam{\'o}n}}, \bibinfo {author} {\bibfnamefont {I.~G.}\ \bibnamefont
  {Serrano}}, \bibinfo {author} {\bibfnamefont {R.}~\bibnamefont
  {C{\'o}rdoba}}, \bibinfo {author} {\bibfnamefont {J.}~\bibnamefont
  {Ses{\'e}}}, \bibinfo {author} {\bibfnamefont {J.~M.}\ \bibnamefont
  {De~Teresa}}, \bibinfo {author} {\bibfnamefont {M.~R.}\ \bibnamefont
  {Ibarra}}, \bibinfo {author} {\bibfnamefont {S.}~\bibnamefont {Vieira}},
  \bibinfo {author} {\bibfnamefont {M.}~\bibnamefont {Ortu{\~n}o}}, \ and\
  \bibinfo {author} {\bibfnamefont {H.}~\bibnamefont {Suderow}},\ }\bibfield
  {title} {\enquote {\bibinfo {title} {Disordered hyperuniformity in
  superconducting vortex lattices},}\ }\href@noop {} {\bibfield  {journal}
  {\bibinfo  {journal} {Phys. Rev. Res.}\ }\textbf {\bibinfo {volume} {2}},\
  \bibinfo {pages} {033133} (\bibinfo {year} {2020})}\BibitemShut {NoStop}%
\bibitem [{\citenamefont {S{\'a}nchez}\ \emph {et~al.}(2023)\citenamefont
  {S{\'a}nchez}, \citenamefont {Maldonado}, \citenamefont {Amig{\'o}},
  \citenamefont {Nieva}, \citenamefont {Kolton},\ and\ \citenamefont
  {Fasano}}]{sanchez2023disordered}%
  \BibitemOpen
  \bibfield  {author} {\bibinfo {author} {\bibfnamefont {J.~A.}\ \bibnamefont
  {S{\'a}nchez}}, \bibinfo {author} {\bibfnamefont {Ra{\'u}l~C.}\ \bibnamefont
  {Maldonado}}, \bibinfo {author} {\bibfnamefont {M.~L.}\ \bibnamefont
  {Amig{\'o}}}, \bibinfo {author} {\bibfnamefont {G.}~\bibnamefont {Nieva}},
  \bibinfo {author} {\bibfnamefont {A.}~\bibnamefont {Kolton}}, \ and\ \bibinfo
  {author} {\bibfnamefont {Y.}~\bibnamefont {Fasano}},\ }\bibfield  {title}
  {\enquote {\bibinfo {title} {Disordered hyperuniform vortex matter with
  rhombic distortions in fese at low fields},}\ }\href@noop {} {\bibfield
  {journal} {\bibinfo  {journal} {Phys. Rev. B}\ }\textbf {\bibinfo {volume}
  {107}},\ \bibinfo {pages} {094508} (\bibinfo {year} {2023})}\BibitemShut
  {NoStop}%
\end{thebibliography}%

\end{document}